\documentclass[twocolumn,showpacs,preprintnumbers,amsmath,amssymb]{revtex4}
\usepackage{epsfig}
\usepackage{dcolumn}
\usepackage{bm}

\newcommand{\remove}[1]{}

\def\be{\begin{equation}}
\def\ee{\end{equation}}

\newcommand{\beq}{\begin{equation}}
\newcommand{\eeq}{\end{equation}}
\newcommand{\beqa}{\begin{eqnarray}}
\newcommand{\eeqa}{\end{eqnarray}}

\renewcommand{\pl}{\partial}

\newcommand{\vv}{{\bf v}}

\newcommand{\vx}{{\bf x}}

\renewcommand{\vr}{{\bf r}}

\newcommand{\ta}{{\tilde{a}}}
\newcommand{\tchi}{{\tilde{\chi}}}

\newcommand{\tg}{{\tilde{g}}}

\newcommand{\tM}{{\tilde{M}}}
\newcommand{\tp}{{\tilde{p}}}
\newcommand{\tPhi}{{\tilde{\Phi}}}
\newcommand{\tPsi}{{\tilde{\Psi}}}

\newcommand{\trho}{{\tilde{\rho}}}

\newcommand{\ts}{{\tilde{s}}}

\newcommand{\cM}{{\cal M}}

\newcommand{\bea}{\begin{array}}
\newcommand{\ea}{\end{array}}

\begin{document}

\title{Ultra-local models of modified gravity without kinetic term}

\author{Philippe Brax}
\affiliation{Institut de Physique Th\'eorique,\\
CEA, IPhT, F-91191 Gif-sur-Yvette, C\'edex, France\\
CNRS, URA 2306, F-91191 Gif-sur-Yvette, C\'edex, France}
\author{Luca Alberto Rizzo}
\affiliation{Institut de Physique Th\'eorique,\\
CEA, IPhT, F-91191 Gif-sur-Yvette, C\'edex, France\\
CNRS, URA 2306, F-91191 Gif-sur-Yvette, C\'edex, France}
\author{Patrick Valageas}
\affiliation{Institut de Physique Th\'eorique,\\
CEA, IPhT, F-91191 Gif-sur-Yvette, C\'edex, France\\
CNRS, URA 2306, F-91191 Gif-sur-Yvette, C\'edex, France}
\vspace{.2 cm}

\date{\today}
\vspace{.2 cm}

\begin{abstract}

We present a class of modified-gravity theories which we call ultra-local models.
We add a scalar field, with negligible kinetic terms, to the  Einstein-Hilbert action. We also introduce a conformal coupling to matter.
This gives rise to a new screening mechanism
which is not entirely due to the non-linearity of the scalar field potential or the coupling function
but to the absence of the kinetic term. As a result this removes any fifth force between isolated objects
in vacuum. It turns out that these models are similar to chameleon-type theories with a large mass when considered outside
the Compton wave-length but differ on shorter scales.
The predictions of these models only depend on a single free function,
as the potential and the coupling function are degenerate, with an amplitude given by
a parameter $\alpha \lesssim 10^{-6}$, whose magnitude springs from requiring a small
modification of Newton's potential astrophysically and cosmologically. This
singles out a redshift $z_{\alpha} \sim \alpha^{-1/3} \gtrsim 100$ where the
fifth force is the greatest.
The cosmological background follows the $\Lambda$-CDM history within a $10^{-6}$
accuracy, while cosmological perturbations are significantly enhanced (or damped)
on small scales, $k \gtrsim 2 h {\rm Mpc}^{-1}$ at $z=0$. The spherical collapse and
the halo mass function are modified in the same manner. We find that the modifications
of gravity are greater for galactic or sub-galactic structures.
We also present a thermodynamic analysis of the non-linear and inhomogeneous
fifth-force regime where we find that the Universe is not made more inhomogeneous before
$z_\alpha$ when the fifth force dominates, and does not lead to the existence of clumped
matter on extra small scales inside halos for large masses while this possibility exists for
masses $M\lesssim 10^{11} M_\odot$ where the phenomenology of ultra-local models
would be most different from $\Lambda$-CDM.

\end{abstract}

 \vskip2pc

\maketitle

\section{Introduction}
\label{introduction}

Since the discovery of the accelerated expansion of the Universe
\cite{Perlmutter:1998np,Riess:1998cb}, explaining its nature has become a major problem
in modern cosmology. Most of the possible solutions for this problem rely either on the
inclusion of a dark energy component  and/or on modifications of General Relativity (GR)
\cite{Amendola:2012ys}. These alternative theories of gravitation, which go beyond
a simple cosmological constant, usually imply the presence
of at least one additional low-mass scalar field in the theory and  induce the presence of a fifth
force on cosmological scales.
However, the presence of the scalar field must have a very small impact on the dynamics of
the Solar System and on any laboratory test due to very stringent constraints imposed by
observations (e.g. \cite{Will:2001mx}). One possible solution, which was recently explored
in \cite{Khoury:2010xi}, is to construct modified gravity theories with a screening mechanism
that provides convergence to GR in dense environments \cite{Brax2012a,Brax:2013ida}
such as the Solar System.

In this paper we investigate a particular type of modification of gravity with a new screening
mechanism, that we will call ``ultra-local models''. We add to the Einstein-Hilbert
action a scalar field  whose Lagrangian has a zero (or negligible) kinetic
term.  For this reason, the equation of motion for the scalar field of this theory,
which gives the relation between the scalar and the matter fields, is a "constraint" equation
with no time derivatives, contrary to what happens in the case of other scalar tensor theories.
In the  ultra-local models, the scalar field is coupled to the matter field via a non-linear
conformal transformation function of the field itself and depends on the local value of the
matter density. In such a way the fifth force associated to the scalar field is proportional to
the local gradient of the matter density.
This provides an automatic screening mechanism as it implies that there is no fifth force
between isolated compact objects, independently of the parameters of the model.
This ultra-local property ensures that astrophysical systems like the Solar System
are perfectly screened.

Thus, although these models can be seen as the limit of chameleon scenarios with
a scalar field mass or potential that is much greater than its kinetic energy, outside the
Compton wave-length, they differ from the chameleon scenarios on smaller scales.
This gives rise to new features for both the definition of the theory and its phenomenology.
In particular, because the scalar field potential and coupling function are degenerate,
we find it convenient to choose (without loss of generality) a linear potential, so that
the physics arises from the non-linearity of the coupling function.
This is somewhat similar to what happens in the Damour-Polyakov mechanism
\cite{Damour:1994zq}, although here we have no potential and
the field is not attracted to the minimum of the coupling function (the screening does not
arise from the vanishing of the coupling but from its locality).
In contrast, other  screening mechanisms studied in the literature are associated to
non-linearities of the scalar Lagrangian of the scalar field (e.g. chameleon
\citep{Khoury:2003aq}, Vainshtein \cite{Vainshtein:1972sx} or K-mouflage \cite{Brax:2014b})
either in the kinetic terms or the potential.

The non-linear coupling function is the only free function of the theory, which can
be constrained using theoretical and cosmological results. In particular, we require the
coupling function to be severely bounded so that its contribution to the metric potential does
not exceed the Newtonian one, associated with typical cosmological perturbations and
astrophysical objects. These ultra-local models correspond to modified source models \cite{Carroll:2006jn} where the coupling to matter has a magnitude of order $\vert \ln A\vert \lesssim 10^{-6}$ to guarantee that the contribution of modified gravity to Newton's potential is at most of order one.
This implies that the $\Lambda$-CDM expansion history is
recovered up to a $10^{-6}$ accuracy. At the linear level, the scalar field in this theory acts as a
scale and time dependent modification of the growth rate which can either enhance or
diminish it, depending on the shape of the coupling function.
On astrophysical scales, the modification of gravity is the largest on galactic scales
while no effects are expected in the Solar System and on cluster scales.
On the other hand, this is not the case inside halos, in particular for masses
below $10^{12} h^{-1} M_\odot$. The effects of the ultra-local interaction can be so drastic
for smaller masses inside the Navarro-Frenk-White profile that the system can undergo
a thermodynamic phase transition with the presence of small clumps.
We expect that the fifth force on these small scales is eventually screened by the
ultra-locality of the scalar interaction. This would lead to a different landscape of
inhomogenities for small mass objects $M \lesssim 10^{11} M_\odot$ deep inside their cores. A more precise analysis would require numerical simulations and this is left for future work.

The paper is organised as follows.
In section~\ref{sec:Dilaton} we introduce the ultra-local models and in
section~\ref{sec:Equations-of-motion} we study the equations of motion, both in Einstein and
Jordan frames.
In section~\ref{sec:models} we consider some generic constraints on the form of such
theories and we present some explicit models, while in section~\ref{sec:theory-validity}
we study the validity and the self-consistency of the theory.
We study the evolution of the cosmological background in
section~\ref{sec:background-quantities}, of cosmological linear perturbations in
section~\ref{sec:Linear-perturbations}, the dynamics of the spherical collapse in
section~\ref{sec:Spherical-collapse} and the halo mass function in
section~\ref{sec:Halo-mass-function}.
We consider the screening properties of the theory, from clusters of galaxies down to
the Solar System and the Earth, in section \ref{sec:halos}.
We investigate the formation of non-linear structures and the fifth-force non-linear regime
in section~\ref{sec:structures-thermo-analysis}, considering the stability of both
cosmological and astrophysical inhomogeneities.
In section~\ref{sec:alpha} we study the dependence of the previous results under the
variation of the free parameter $\alpha$ of the coupling function and in
section~\ref{sec:Comparison} we compare the ultra-local models to other modified
gravity theories.
Finally we conclude in section~\ref{sec:Conclusions}.

\section{Scalar-field model with negligible kinetic term}
\label{sec:Dilaton}

We consider scalar field models with actions of the form
\beqa
S & = & \int d^4 x \; \sqrt{-\tg} \left[ \frac{\tilde{M}_{\rm Pl}^2}{2} \tilde{R}
+ \tilde{\cal L}_{\varphi}(\varphi) \right]
 \nonumber \\
&& + \int d^4 x \; \sqrt{-g} \, {\cal L}_{\rm m}(\psi^{(i)}_{\rm m},g_{\mu\nu}) ,
\label{S-def}
\eeqa
where the various matter fields follow the Jordan-frame metric $g_{\mu\nu}$,
with determinant $g$, which is related to the Einstein-frame metric $\tg_{\mu\nu}$,
with determinant $\tg$, by \cite{Vainshtein:1972sx}
\beq
g_{\mu\nu} = A^2(\varphi) \tg_{\mu\nu} .
\label{conformal-g-tg}
\eeq
In this paper, we investigate models where the scalar field Lagrangian is
dominated by its potential term, so that we write
\beq
\tilde{\cal L}_{\varphi}(\varphi) = - V(\varphi) ,
\label{L-phi-def}
\eeq
where we set the kinetic term to zero.
Then, assuming that the potential $V(\varphi)$ can be inverted (i.e., that it is
a monotonic function over the range of $\varphi$ of interest), we can make
the change of variable from $\varphi$ to $V$. More precisely, introducing the
characteristic energy scale $\cM^4$ of the potential we define the dimensionless
field $\tilde\chi$ as
\beq
\tilde\chi \equiv - \frac{V(\varphi)}{\cM^4} , \;\;\; \mbox{and} \;\;\;
A(\tilde\chi) \equiv A(\varphi) .
\label{tchi-def}
\eeq
Therefore, in terms of the field $\tilde\chi$ the scalar field Lagrangian and the
conformal metric transformation read as
\beq
\tilde{\cal L}_{\tilde\chi}(\tilde\chi) = \cM^4 \tilde\chi \;\;\; \mbox{and} \;\;\;
g_{\mu\nu} = A^2(\tilde\chi) \tg_{\mu\nu} .
\label{L-chi-def}
\eeq
Thus, these models are fully specified by a single function, $A(\tilde\chi)$, which is
defined from the initial potential $V(\varphi)$ and coupling function $A(\varphi)$
through Eq.(\ref{tchi-def}).
This means that there is a broad degeneracy in the action (\ref{S-def})
as different couples $\{V(\varphi),A(\varphi)\}$ with the same rescaled coupling
$A(\tilde\chi)$ give rise to the same physics. Therefore, in the following we work with
the field $\tilde\chi$ and with Eq.(\ref{L-chi-def}).
The energy scale $\cM^4$ is arbitrary and only defines the normalization of the
field $\tchi$. We can choose without loss of generality $\cM^4>0$ and we
shall typically have $\cM^4 \sim \bar\rho_{\rm de 0}$, where $\bar\rho_{\rm de 0}$
is the mean dark energy density today , if we require the accelerated expansion
of the Universe at low $z$ to be driven by the scalar field potential $V(\varphi)$,
without adding an extra cosmological constant.

\section{Equations of motion}
\label{sec:Equations-of-motion}

Because the matter fields follow the geodesics set by the Jordan frame and
satisfy the usual conservation equations in this frame, we mostly work in the
Jordan frame. This is also the frame that is better suited to make the connection
with observations as atomic physics remains the same throughout cosmic
evolution in this frame \cite{Faraoni1999}.
However, because the gravitational sector is simpler in the Einstein frame,
we first derive the Einstein equations in the Einstein frame, and next translate
these equations in terms of the Jordan tensors.

\subsection{Einstein frame}
\label{sec:Einstein-equations-E}

\subsubsection{Scalar-field and Einstein equations}
\label{sec:scalar-E-E}

The scalar-field Lagrangian (\ref{L-chi-def}) is given in the Einstein frame,
where the equation of motion of the scalar field reads as
\beq
\cM^4 + \tilde{T}  \frac{d\ln A}{d\tchi} = 0 ,
\label{KG-E}
\eeq
where $\tilde{T} = \tilde{T}^{\mu}_{\mu}$ is the trace of the matter
energy-momentum tensor in the Einstein frame.
From the conformal coupling (\ref{L-chi-def}) the energy-momentum tensors in the
Einstein and Jordan frames are related by
\beq
\tilde{T}^{\mu}_{\nu} = A^4 T^{\mu}_{\nu} \;\;\; \mbox{and} \;\;\;
\tilde{T} = A^4 T .
\label{tT-T}
\eeq
As there is no kinetic term in the scalar-field Lagrangian (\ref{L-chi-def}),
the ``Klein-Gordon'' equation (\ref{KG-E}) contains no derivative term and it
is a constraint equation, which gives the field $\tchi(\vx)$ as a function of the
matter density field $\trho(\vx)$.
The energy-momentum tensor of the scalar field also reads as
\beq
\tilde{T}^{\mu}_{\nu(\tchi)} = \cM^4 \tchi \delta^{\mu}_{\nu} ,
\label{T-chi-E}
\eeq
so that the scalar-field energy density and pressure are
\beq
\trho_{\tchi} = - \cM^4 \tchi , \;\;\; \tp_{\tchi} = \cM^4 \tchi = - \trho_{\tchi} .
\label{rho-chi-E}
\eeq
In the Einstein frame, the Einstein equations take their standard form,
$\tilde{G}^{\mu}_{\nu}= \tilde{T}^{\mu}_{\nu}$.

\subsubsection{Cosmological background in the Einstein frame}
\label{sec:background-E}

Using the conformal time $\tau$ and comoving coordinates $\vx$, the background
metrics in both frames are given by
$d\ts^2= \ta^2 (-d\tau^2+d\vx^2)$ and
$d s^2= a^2 (-d\tau^2+d\vx^2)$, with
\beq
a = \bar{A} \ta \;\;\; \mbox{and} \;\;\; d t = \bar{A} d\tilde{t} , \;\;\;
\vr= \bar{A} \tilde{\vr} ,
\label{a-ta-def}
\eeq
where we denote background quantities with a bar.
Thus, the cosmic times and physical distances are different in the two frames.
From Eq.(\ref{tT-T}) the densities and pressures are also related by
\beq
\bar{\tilde\rho} = \bar{A}^4 \bar\rho , \;\;\; \bar{\tilde p} = \bar{A}^4 \bar{p} ,
\label{trho-rho-def}
\eeq
while the Friedmann equation takes the standard form,
\beq
3 \tilde{M}_{\rm Pl}^2 \tilde{\cal H}^2 = \ta^2 ( \bar{\trho} + \bar{\trho}_{\rm rad}
+ \bar{\trho}_{\tchi} ) ,
\label{Friedmann-E}
\eeq
with $\tilde{\cal H} = d\ln\ta/d\tau$ the conformal expansion rate in the Einstein
frame.
From Eq.(\ref{rho-chi-E}) the background scalar field energy density and pressure
are given by
\beq
\bar{\trho}_{\tchi} = - \cM^4 \bar{\tchi} , \;\;\; \bar{\tp}_{\tchi} = \cM^4 \bar{\tchi}
= - \bar{\trho}_{\tchi} .
\label{rho-chi-E-background}
\eeq

\subsubsection{Perturbations in the Einstein frame}
\label{sec:perturbations-E}

Taking into account the perturbations from the homogeneous background,
the Einstein-frame metric reads in the Newtonian gauge as
\beq
d \ts^2= \ta^2 [ - (1+2\tPhi) d\tau^2 + (1-2\tPsi) d\vx^2 ] ,
\label{metric-E}
\eeq
and the Einstein equations yield, at linear order over the metric potentials and in
the quasi-static approximation (for scales much below the Hubble radius),
\beq
\tPhi = \tPsi = \tPsi_{\rm N} \;\;\; \mbox{with} \;\;\;
\frac{\nabla^2}{\ta^2} \tPsi_{\rm N} \equiv \frac{\delta\trho+\delta\trho_{\tchi}}
{2\tM_{\rm Pl}^2} .
\label{Phi-Psi-E}
\eeq
Here we use the non-relativistic limit $v^2 \ll c^2$, so that the gravitational
slip $\tPhi-\tPsi$ vanishes, and $\delta\trho=\trho-\bar{\trho}$ and
$\delta\trho_{\tchi}=\trho_{\tchi}-\bar{\trho}_{\tchi}$ are the matter and scalar-field
density fluctuations.
in particular, we have
\beq
\delta\trho_{\tchi} = - \cM^4 \delta\tchi .
\label{delta-trhochi-def}
\eeq

\subsection{Jordan frame}
\label{sec:Jordan}

\subsubsection{Cosmological background in the Jordan frame}
\label{sec:background-J}

From Eq.(\ref{a-ta-def}) the conformal expansion rates in the two frames are
related by
\beq
\tilde{\cal H} = (1-\epsilon_2) {\cal H} \;\;\; \mbox{with} \;\;\;
\epsilon_2(t) \equiv \frac{d\ln\bar{A}}{d\ln a} ,
\label{eps2-def}
\eeq
while the densities and pressures are related as in Eq.(\ref{trho-rho-def}).
Therefore, the Friedmann equation (\ref{Friedmann-E}) yields
\beq
3 M_{\rm Pl}^2 {\cal H}^2 = (1-\epsilon_2)^{-2} a^2 ( \bar{\rho} + \bar{\rho}_{\rm rad}
+ \bar{\rho}_{\tchi} ) ,
\label{Friedmann-J}
\eeq
where the Jordan-frame Planck mass is
\beq
M_{\rm Pl}^2(t) = \bar{A}^{-2}(t) \, \tM_{\rm Pl}^2 .
\label{Planck-J}
\eeq
Then, we can define an effective dark energy density by
\beq
3 M_{\rm Pl}^2 {\cal H}^2 = a^2 ( \bar{\rho} + \bar{\rho}_{\rm rad}
+ \bar{\rho}_{\rm de} ) ,
\label{Friedmann-J-de}
\eeq
which gives
\beq
\bar{\rho}_{\rm de} = \bar\rho_{\tchi} + \frac{2\epsilon_2-\epsilon_2^2}{(1-\epsilon_2)^2}
( \bar{\rho} + \bar{\rho}_{\rm rad} + \bar{\rho}_{\tchi} ) .
\label{rho-de-def}
\eeq

In the Jordan frame the matter obeys the standard conservation equations,
$\nabla_{\mu} T^{\mu}_{\nu} =0$, and the background matter and radiation densities
evolve as
\beq
\bar\rho = \frac{\bar\rho_0}{a^3} , \;\,\,\,
\bar\rho_{\rm rad} = \frac{\bar\rho_{\rm rad 0}}{a^4} .
\label{rho-a-J}
\eeq
The scalar-field equation of motion (\ref{KG-E}) gives
\beq
\cM^4 = \bar{A}^4 \bar{\rho} \frac{d\ln\bar{A}}{d\bar{\tchi}} \;\;\;
\mbox{and} \;\;\;
\frac{d\bar{\tchi}}{d\tau} = \bar{A}^4 \frac{\bar{\rho}}{\cM^4} \epsilon_2
{\cal H} ,
\label{KG-J-background}
\eeq
hence
\beq
\bar\rho_{\tchi} = - \bar{A}^{-4} \cM^4 \bar{\tchi} , \;\;\;
\frac{d\bar\rho_{\tchi}}{d\tau} = - 4 \epsilon_2 {\cal H} \bar\rho_{\tchi}
- \epsilon_2 {\cal H} \bar\rho .
\label{drho-chi-dtau-J}
\eeq

\subsubsection{Perturbations in the Jordan frame}
\label{sec:Perturbations-J}

In the Jordan frame we write the Newtonian gauge metric as
\beq
d s^2= a^2 [ - (1+2\Phi) d\tau^2 + (1-2\Psi) d\vx^2 ] ,
\label{metric-J}
\eeq
so that the Einstein- and Jordan-frame metric potentials are related by
\beq
1+2\Phi = \frac{A^2}{\bar{A}^2} (1+2\tPhi) , \;\;\;
1-2\Psi = \frac{A^2}{\bar{A}^2} (1-2\tPsi) ,
\label{Phi-J-Phi-E}
\eeq
while the Einstein-frame Newtonian potential (\ref{Phi-Psi-E}) is also the solution
of
\beq
\frac{\nabla^2}{a^2} \tPsi_{\rm N} = \frac{\delta(A^4\rho)+\delta(A^4\rho_{\tchi})}
{2 \bar{A}^4 M_{\rm Pl}^2} .
\label{tPsi-N-E-J}
\eeq

Since we wish the deviations from General Relativity and the $\Lambda$-CDM
cosmology to be small, at most of the order of ten percent,
the potentials $\Phi$ and $\Psi$ cannot deviate too much from the Jordan-frame
Newtonian potential defined by
\beq
\frac{\nabla^2}{a^2} \Psi_{\rm N} \equiv \frac{\delta\rho+\delta\rho_{\tchi}}
{2 M_{\rm Pl}^2} ,
\label{tPsi-N-J}
\eeq
where the scalar field density fluctuations must also remain modest as compared
with the matter density fluctuations.
Therefore, Eqs.(\ref{Phi-J-Phi-E}) and (\ref{tPsi-N-E-J}) lead to the constraints
\beq
\left| \frac{\delta A}{\bar{A}} \right| \lesssim \left| \Psi_{\rm N} \right| , \;\;\;
\left| \delta\rho_{\tchi} \right | \lesssim \left| \delta\rho \right| .
\label{pert-bounds}
\eeq
Then, since $|\Psi_{\rm N}|$ is typically of order $10^{-5}$, we can linearize in
$\delta A$ as we did for the metric potentials, and within a $10^{-5}$ relative accuracy
we obtain
\beq
\Phi = \Psi_{\rm N} + \delta\ln A , \;\;\; \Psi = \Psi_{\rm N} - \delta\ln A ,
\label{Phi-Psi-J}
\eeq
and
\beq
\delta\rho_{\tchi} = - \bar{A}^{-4} \cM^4 \delta{\tchi} .
\label{delta-rho-chi-J}
\eeq

The equation of motion of the scalar field reads as
\beq
\cM^4 = \bar{A}^4 \rho \frac{d\ln A}{d\tchi} .
\label{KG-pert-J}
\eeq

The matter and radiation components obey the standard equations of motion,
which gives for the matter component the continuity and Euler equations
\beq
\frac{\pl\rho}{\pl\tau} + (\vv\cdot\nabla) \rho + ( 3 {\cal H}+\nabla\cdot\vv) \rho = 0 ,
\label{continuity-J}
\eeq
and
\beq
\frac{\pl\vv}{\pl\tau} + (\vv\cdot\nabla)\vv + {\cal H} \vv = -\nabla\Phi .
\label{Euler-J}
\eeq

From Eq.(\ref{Phi-Psi-J}) we have $\nabla\Phi=\nabla\Psi_{\rm N}+\nabla\ln A$,
and the scalar-field equation (\ref{KG-pert-J}) gives
\beq
\nabla\ln A = \frac{d\ln A}{d\tchi} \nabla\tchi = \frac{\cM^4}{\bar{A}^4\rho}
\nabla\tchi ,
\label{grad-lnA}
\eeq
so that the Euler equation (\ref{Euler-J}) also reads as
\beq
\frac{\pl\vv}{\pl\tau} + (\vv\cdot\nabla)\vv + {\cal H} \vv = -\nabla\Psi_{\rm N}
- \frac{\nabla p_A}{\rho}  ,
\label{Euler-1-J}
\eeq
with
\beq
p_A = \frac{\cM^4 c^2}{\bar{A}^4} \tchi ,
\label{pA-def}
\eeq
where we explicitly wrote the factor $c^2$.

Thus, in terms of the matter dynamics, the scalar-field or modified-gravity
effects appear through two factors, a) the modification of the Poisson equation
(\ref{tPsi-N-J}), because of the additional source associated with the scalar-field
energy density fluctuations and of the time dependence of the Jordan-frame
Planck mass, and b) the new pressure term $p_A$ in the Euler equation
(\ref{Euler-1-J}).
This pressure $p_A$ corresponds to a polytropic equation of state, as it only
depends on the matter density (the sum of cold dark matter and baryons).

\subsubsection{Linear regime in the Jordan frame}
\label{sec:linear-J}

On large scales or at early times we may linearize the equations of motion.
Expanding the coupling function $A(\tchi)$ as
\beq
\ln A(\tchi) = \ln\bar{A} + \sum_{n=1}^{\infty} \frac{\beta_n(t)}{n!} (\delta\tchi)^n ,
\label{beta-n-def}
\eeq
the scalar field equation (\ref{KG-pert-J}) gives at the background and linear
orders
\beq
\cM^4 = \bar{A}^4 \bar\rho \beta_1 , \;\;\;
\delta\tchi = - \frac{\beta_1}{\beta_2} \delta ,
\label{dchi-linear}
\eeq
where we note $\delta\equiv \delta\rho/\bar\rho$ the matter density contrast.
This also yields
\beq
\delta p_A = - \epsilon_1(t) \bar\rho c^2 \delta
\;\;\; \mbox{and} \;\;\;
\delta\rho_{\tchi} = \epsilon_1(t) \bar\rho \delta ,
\label{p_A-rho-chi-L}
\eeq
with
\beq
\epsilon_1(t) \equiv \frac{\beta_1}{\beta_2} \frac{\cM^4}{\bar{A}^4 \bar\rho}
= \frac{\beta_1^2}{\beta_2} = \frac{\epsilon_2}{3-4\epsilon_2} ,
\label{eps1-def}
\eeq
where to obtain the last relation we took the time derivative of the first relation
in (\ref{dchi-linear}) and used the second expression in (\ref{eps2-def}).

The continuity equation (\ref{continuity-J}) reads as
$\pl_{\tau}\delta+\nabla\cdot[(1+\delta\vv]=0$ in terms of the density contrast.
Combining with the Euler equation at linear order, this gives
\beq
\frac{\pl^2\delta}{\pl\tau^2} + {\cal H} \frac{\pl\delta}{\pl\tau}
+ \epsilon_1 c^2 \nabla^2\delta =
\frac{\bar\rho a^2}{2 M_{\rm Pl}^2} ( 1 + \epsilon_1 ) \delta .
\label{delta-evol}
\eeq

As compared with the $\Lambda$-CDM cosmology, the pressure term $\nabla^2\delta$
introduces an explicit scale dependence. Going to Fourier space, the linear growing
modes $D(k,t)$ now depend on the wave number $k$ and obey the evolution equation
\beq
\frac{\pl^2 D}{\pl (\ln a)^2} + \left( 2 + \frac{1}{H^2} \frac{d H}{d t} \right)
\frac{\pl D}{\pl\ln a} - \frac{3\Omega_{\rm m}}{2} (1+\epsilon) D = 0 ,
\label{DL}
\eeq
where $H=d\ln a/d t$ is the Jordan-frame expansion rate (with respect to the
Jordan-frame cosmic time $t$) and the factor $\epsilon(k,t)$, which describes the
deviation from the $\Lambda$-CDM cosmology, is given by
\beq
\epsilon(k,t) = \epsilon_1(t) \left( 1 + \frac{2}{3\Omega_{\rm m}}
\frac{c^2k^2}{a^2H^2} \right) .
\label{eps-def}
\eeq
Thus, the two effects of the scalar field, the contribution to the gravitational potential
of $\delta\rho_{\tchi}$ and the pressure term due to the conformal transformation
between the Einstein and Jordan frames, modify the growth of structures in the same
direction, given by the sign of $\epsilon_1$. A positive $\epsilon_1$ gives a scale
dependent amplification of the gravitational force and an acceleration of gravitational
clustering.
The $k$-dependent pressure term dominates when $ck/aH>1$, that is, on sub-horizon
scales. Moreover, we have $(ck/aH)^2 \sim 10^7$ today at scales of about
$1 \, h^{-1}$Mpc. Therefore, we must have
\beq
| \epsilon_1 | \lesssim 10^{-7}
\label{epsilon1-bound1}
\eeq
to ensure that the growth of large-scale structures is not too significantly modified.
This also ensures that the first condition in (\ref{pert-bounds}) is satisfied on cosmological
scales.
Moreover, the fluctuations of the scalar field energy density
in the Poisson equations are negligible and $\epsilon_2$ is very small, of order
$10^{-7}$, from the last relation in Eq.(\ref{eps1-def}).

\section{Explicit models}
\label{sec:models}

\subsection{Constraints}
\label{sec:constraints}

\subsubsection{Small parameter $\alpha$}
\label{sec:small-alpha}

In usual scalar-field models with a kinetic term, the Klein-Gordon equation for the scalar
field that corresponds to Eq.(\ref{KG-pert-J}) contains a derivative term
$\nabla^2\varphi$, which suppresses the fluctuations of the scalar field on small scales.
This mechanism is absent in our case and the scalar field $\tchi$ only follows the
variations of the local matter density.
However, we wish the fluctuations of $\ln A$ to remain small
and of order $10^{-6}$ from cosmological scales down to astrophysical objects such as
stars and planets, to comply with the first constraint in Eq.(\ref{pert-bounds}) and to
ensure that the metric potentials remain close the Newtonian potential.
Because the density varies by many orders of magnitude from the intergalactic medium
to the atmospheres and cores of stars and planets, and to the typical densities found
in the laboratory on Earth, and we cannot rely on the small-scale suppression due to
derivative terms, the function $\ln A$ must be bounded within a small interval over its
full domain,
\beq
| \ln A | \lesssim 10^{-6} , \;\;\; \mbox{hence} \;\;\;
| A - 1 | \lesssim 10^{-6} .
\label{lnA-bound}
\eeq
Therefore, the conformal factor $A$ always remains very close to unity
(we can renormalize $A$ by a constant multiplicative factor without loss of generality).
On the other hand, from Eq.(\ref{KG-pert-J}) we have
$d\ln A/d\tchi = \cM^4/\bar{A}^4\rho$, hence
\beq
\frac{d\ln A}{d\tchi} > 0 ,
\label{dlnA-dchi-positive}
\eeq
and
\beq
\rho \rightarrow 0 : \;\;\; \frac{d\ln A}{d\tchi} \rightarrow +\infty , \;\;\;
\rho \rightarrow \infty : \;\;\; \frac{d\ln A}{d\tchi} \rightarrow 0 .
\label{dlnA-dchi-rho}
\eeq

The small range of the function $A(\tchi)$ in Eq.(\ref{lnA-bound}) also implies
that the Jordan-frame Planck mass (\ref{Planck-J}) does not vary by more
than $10^{-6}$. This ensures that the bounds on the variation with time of Newton's
constant obtained from the BBN constraints \cite{Uzan:2010pm,Alvarez2007}
or the Lunar Ranging measurements \cite{Williams:2004qba} are satisfied.
It also means that at the background level the Einstein and Jordan frames are
identical up to $10^{-6}$.

The small range of $\ln A$ also leads to a small amplitude for the factor $\epsilon_2$
defined in Eq.(\ref{eps2-def}), of order $10^{-6}$. In fact, from Eqs.(\ref{epsilon1-bound1})
and (\ref{eps1-def}) we have seen that we also require $\epsilon_1$ and $\epsilon_2$
of order $10^{-7}$, so that both constraints give about the same condition
(\ref{lnA-bound}) on the coupling function $A(\tchi)$.
Then, we recover a standard $\Lambda$-CDM cosmology up to this
order. Indeed, with $\epsilon_2 \simeq 0$ we recover the usual Friedmann equation
in Eq.(\ref{Friedmann-J}), the dark energy density $\bar\rho_{\rm de}$ is almost
identical to the scalar field energy density $\bar\rho_{\tchi}$ in Eq.(\ref{rho-de-def}),
and the latter is almost constant at low $z$ from Eq.(\ref{drho-chi-dtau-J}).
From Eq.(\ref{drho-chi-dtau-J}) we find that the value of the scalar field today must
satisfy
\beq
\bar\rho_{\rm de 0} = - \bar{A}_0^{-4} \cM^4 \bar{\tchi}_0 \simeq - \cM^4 \bar{\tchi}_0 ,
\label{chi0-rhode0}
\eeq
if the scalar field drives the accelerated expansion of the Universe at low $z$ without
an additional cosmological constant.
In particular, this implies $\bar{\tchi}_0<0$.
Finally, we must check that $\epsilon_1$, defined in Eq.(\ref{eps1-def}), remains
small, as in Eq.(\ref{epsilon1-bound1}), and vanishes at high redshift if we wish
to recover the standard clustering growth in the early matter era.

In the following we use the approximation $\bar{A}\simeq 1$ to simplify the expressions
and we present several explicit models for the coupling function $A(\tchi)$ that satisfy
the conditions (\ref{lnA-bound})-(\ref{chi0-rhode0}).
In particular, the equation of motion of the scalar field (\ref{KG-pert-J}) becomes
\beq
\frac{d\ln A}{d\tchi} = \frac{{\cal M}^4}{\rho} ,
\label{KG-A=1}
\eeq
which implicitly defines the functions $\tilde\chi(\rho)$ and $\ln A(\rho)$ for
each coupling function $\ln A(\tilde\chi)$.
To obtain a unique and well-defined solution $\tchi(\rho)$ and $A(\rho)$
to the scalar-field equation (\ref{KG-A=1}), we require that $d\ln A/d\tchi$
be a monotonic function that goes from $0$ to $+\infty$ over a range of $\tchi$,
which will define the domain of the scalar field values.
Then $\tchi(\rho)$ and $\ln A(\rho)$, defined by the values that are solutions
of Eq.(\ref{KG-A=1}) for a given $\rho$, are also monotonic functions of $\rho$.

\subsubsection{Derived characteristic density $\rho_{\alpha}$ and redshift $z_{\alpha}$}
\label{sec:rhoc}

From Eq.(\ref{lnA-bound}) we write
\beq
\ln A(\tilde\chi) = \alpha \, \lambda(\tilde\chi) , \;\;\; \alpha \lesssim 10^{-6} ,
\label{lambda-def}
\ee
where $\alpha$ is a small parameter that ensures the condition (\ref{lnA-bound})
is satisfied, whereas $\lambda(\tilde\chi)$ is a bounded function of order unity
and $\tilde\chi$ is also typically of order unity.
Then, the equation of motion (\ref{KG-A=1}) reads as
\beq
\frac{d\lambda}{d\tilde\chi} = \frac{1}{\hat\rho} \;\;\; \mbox{with} \;\;\;
\hat\rho = \frac{\alpha\rho}{{\cal M}^4} .
\label{rho-hat-def}
\eeq
This implicitly defines the functions $\lambda(\hat\rho)$ and $\tilde\chi(\hat\rho)$,
from the value of $\tilde\chi$ that solves Eq.(\ref{rho-hat-def}) for a given density.
The changes of variables $\ln A \rightarrow \lambda$ and $\rho \rightarrow \hat\rho$
have removed the explicit parameters
${\cal M}^4 \sim \bar\rho_{\rm de0}$ and $\alpha \lesssim 10^{-6}$, so that
the functions $\lambda(\tilde\chi)$, $\lambda(\hat\rho)$ and $\tilde\chi(\hat\rho)$
do not involve small nor large parameters.
Therefore, in addition to the density ${\cal M}^4 \sim \bar\rho_{\rm de0}$,
which is associated with the current dark energy density from Eq.(\ref{chi0-rhode0}),
these models automatically introduce another higher density scale $\rho_{\alpha}$
given by
\be
\rho_{\alpha} = \frac{{\cal M}^4}{\alpha} \sim \frac{\bar\rho_{\rm de0}}{\alpha}
\gtrsim 10^6 \; \bar\rho_{\rm de0} .
\label{rho-alpha-def}
\ee
This implies that, from the point of view of the coupling function $\ln A$,
the low-redshift mean density of the Universe is within its very low density
regime.
Moreover, we can expect a cosmological transition between low-density and high-density
regimes at the redshift $z_{\alpha}$ where $\bar\rho \sim \rho_{\alpha}$,
which corresponds to
\beq
a_{\alpha} \sim \alpha^{1/3} \lesssim 0.01 , \;\;\;
z_{\alpha} \sim \alpha^{-1/3} \gtrsim 100 .
\label{zalpha-def}
\eeq

\subsection{Model (I): $\tchi$ is a bounded increasing function of $\rho$}
\label{sec:model-I}%

We first consider the case where $\tchi(\rho)$ is a monotonic increasing function
of $\rho$, with $\tchi_- < \tchi < \tchi_+$. From Eq.(\ref{dlnA-dchi-rho}) we find
that $d\ln A/d\tchi$ must decrease from $+\infty$ to $0$ as $\tchi$ grows
from $\tchi_-$ to $\tchi_+$.
Moreover, the boundary $\tchi_-$ will correspond to the late dark energy era while
the boundary $\tchi_+$ will correspond to the early matter era.
From Eq.(\ref{chi0-rhode0}) we have $\tchi_-<0$ and to avoid introducing another parameter
we can take $\tchi_+=0$, which corresponds to a vanishing dark energy density at
early times from Eq.(\ref{drho-chi-dtau-J}) (but we could also take any finite value,
or an infinite boundary $\tchi_+ \rightarrow +\infty$ that is reached sufficiently slowly
to ensure that the dark energy component is subdominant at high redshift).
A simple model that obeys these properties and the constraint
(\ref{lnA-bound}), which also reads as Eq.(\ref{lambda-def}), is
\beq
\mbox{model (I):} \;\;\; -1 < \tchi < 0 , \;\;\; \ln A = \alpha \sqrt{1-\tchi^2} ,
\label{model-I-def}
\eeq
with
\beq
\alpha > 0 , \;\;\; \alpha \sim 10^{-6} .
\label{alpha-I-def}
\eeq
Here we set $\tchi_-$ to $-1$ without loss of generality, as this merely defines
the normalization of $\cM^4$ and $\alpha$.
Instead of the square root we could have chosen a more general exponent,
$\ln A = \alpha (1-\tchi^2)^{\nu}$ with $0<\nu<1$, but $\nu=1/2$ simplifies the
numerical computations.
Then, the scalar-field equation (\ref{KG-A=1}) gives
\beq
\tchi(\rho) = - \left( 1 + \frac{\alpha^2 \rho^2}{\cM^8} \right)^{-1/2} ,
\label{chi-rho-model-I}
\eeq
\beq
\ln A(\rho) = \alpha \left( 1 + \frac{\cM^8}{\alpha^2\rho^2} \right)^{-1/2} ,
\label{lnA-rho-model-I}
\eeq
and
\beq
p_A(\rho) = - \cM^4 c^2 \left( 1 + \frac{\alpha^2 \rho^2}{\cM^8} \right)^{-1/2} .
\label{pA-rho-model-I}
\eeq
We recover the fact that the system depends on the density through the dimensionless
ratio $\hat\rho=\rho/\rho_{\alpha}$ introduced in Eqs.(\ref{rho-hat-def})-(\ref{rho-alpha-def}).
In terms of the scale factor $a(t)$, using $\bar\rho=\bar\rho_0/a^3$, this gives
(at leading order)
\beq
\epsilon_2(a) = 3 \epsilon_1(a) \;\;\; \mbox{and} \;\;\;
\epsilon_1(a) = - \alpha \frac{\alpha\bar\rho/\cM^4}
{(1+\alpha^2\bar\rho^2/\cM^8)^{3/2}}
\label{eps1-2-I}
\eeq
and we can check that $|\epsilon_2| = 3 |\epsilon_1| \lesssim \alpha  \ll 1$ at all
redshifts, while we have $\bar{\tchi}_0 \simeq -1$ and
$\cM^4 \simeq \bar\rho_{\rm de 0}$. At low redshift, $z\simeq 0$, we actually have
$|\epsilon_2| = 3|\epsilon_1| \sim \alpha^2$, which is much smaller than the maximum
value of order $\alpha$ that is reached at a redshift $z_{\alpha} \sim \alpha^{-1/3}$.
Therefore, in this model the modification to the growth of large-scale structures
is the greatest at high redshifts, $z \sim z_{\alpha}$, much before the dark energy era.

As explained in Sec.~\ref{sec:Dilaton}, this choice of $A(\tchi)$ corresponds to an
infinite number of couples $\{V(\varphi),A(\varphi)\}$. In particular, from
Eq.(\ref{tchi-def}) this corresponds for instance to
\beq
\mbox{(Ia) :} \;\;\; V(\varphi) = \cM^4 \sqrt{1-\left( \frac{\beta\varphi}{\alpha M_{\rm Pl}} \right)^2 } ,
\;\;\; A(\varphi) = e^{\beta\varphi/M_{\rm Pl}}
\label{Ia-def}
\eeq
with $\beta>0$ and $0<\beta\varphi/M_{\rm Pl}<\alpha$, where we assumed an exponential
coupling function $A(\varphi)$, or to
\beq
\mbox{(Ib) :} \;\;\; V(\varphi) = \cM^4 e^{-\gamma\varphi/M_{\rm Pl}} ,
\;\;\; A(\varphi) = e^{\alpha\sqrt{1-e^{-2\gamma\varphi/M_{\rm Pl}}}} ,
\label{Ib-def}
\eeq
with $\gamma>0$ and $0<\varphi<+\infty$, where we assumed an exponential potential
$V(\varphi)$.

\subsection{Model (II): $\tchi$ is a bounded decreasing function of $\rho$}
\label{sec:model-II}

We next consider the case where $\tchi$ is a monotonic decreasing function of $\rho$,
over $\tchi_-<\tchi<\tchi_+$. Thus, $d\ln A/d\tchi$ must increase from 0 to $+\infty$
as $\tchi$ grows from $\tchi_-$ to $\tchi_+$. Now $\tchi_-$ corresponds to the early matter
era whereas $\tchi_+$ corresponds to the late dark energy era, hence $\tchi_+<0$.
A simple choice that satisfies these conditions is
\beqa
\mbox{model (II):} && \;\;\; \tchi_* < \tchi < -1 , \nonumber \\
&& \ln A = - \alpha \sqrt{ (1+\tchi) (1+2\tchi_*-\tchi) } , \;\;\;
\label{model-II-def}
\eeqa
where again $\alpha$ is a small positive parameter as in Eq.(\ref{lambda-def})
and we set $\tchi_-=\tchi_*$ and $\tchi_+=-1$ without loss of generality.
Then, the scalar-field equation (\ref{KG-A=1}) gives
\beq
\tchi(\rho) = \tchi_* - \frac{1+\tchi_*}{\sqrt{ 1+\alpha^2\rho^2/\cM^8}} ,
\label{chi-rho-model-II}
\eeq
\beq
\ln A(\rho) = \alpha \frac{1+\tchi_*}{\sqrt{1+\cM^8/\alpha^2\rho^2}} ,
\label{lnA-rho-model-II}
\eeq
\beq
p_A(\rho) = \cM^4 c^2 \tchi_* - \cM^4 c^2 \frac{1+\tchi_*}
{\sqrt{ 1+\alpha^2\rho^2/\cM^8}} ,
\label{pA-rho-model-II}
\eeq
which again makes explicit the dependence on the dimensionless
ratio $\hat\rho=\rho/\rho_c$ introduced in Eqs.(\ref{rho-hat-def})-(\ref{rho-alpha-def}).
In terms of the scale factor $a(t)$ this gives $\epsilon_2(a) = 3 \epsilon_1(a)$
and
\beq
\epsilon_1(a) = - \alpha (1+\tchi_*) \frac{\alpha\bar\rho/\cM^4}
{(1+\alpha^2\bar\rho^2/\cM^8)^{3/2}} .
\label{eps1-2-II}
\eeq
Again, we can check that $|\epsilon_2| = 3 |\epsilon_1| \lesssim \alpha  \ll 1$ at all
redshifts, with $\bar{\tchi}_0 \simeq -1$ and $\cM^4 \simeq \bar\rho_{\rm de 0}$.
We also have $|\epsilon_2| = 3 |\epsilon_1| \sim \alpha^2$ at low $z$
and the maximum value of order $\alpha$ is reached at a redshift
$z_{\alpha} \sim \alpha^{-1/3}$.

This choice of $A(\tchi)$ corresponds for instance to
\beqa
\mbox{(IIa)} & : & V(\varphi) = - \cM^4 \left[ \tchi_* + \sqrt{ (1+\tchi_*)^2
- \left( \frac{\beta\varphi}{\alpha M_{\rm Pl}} \right)^2 } \right] ,
\nonumber \\
&& A(\varphi) = e^{\beta\varphi/M_{\rm Pl}}
\label{IIa-def}
\eeqa
with $\beta>0$ and $\alpha(1+\tchi_*) < \beta\varphi/M_{\rm Pl} < 0$,
for an exponential coupling, or to
\beqa
\mbox{(IIb)} & : & V(\varphi) = \cM^4 e^{-\gamma\varphi/M_{\rm Pl}} , \nonumber \\
&& \hspace{-0.8cm}A(\varphi) = e^{-\alpha\sqrt{1+2\tchi_*  - 2\tchi_* e^{-\gamma\varphi/M_{\rm Pl}}
- e^{-2\gamma\varphi/M_{\rm Pl}} }} ,
\label{IIb-def}
\eeqa
with $\gamma>0$ and $-\ln(-\tchi_*)<\gamma\varphi/M_{\rm Pl}<0$, for an
exponential potential.

\begin{figure*}
\begin{center}
\epsfxsize=5.8 cm \epsfysize=5.5 cm {\epsfbox{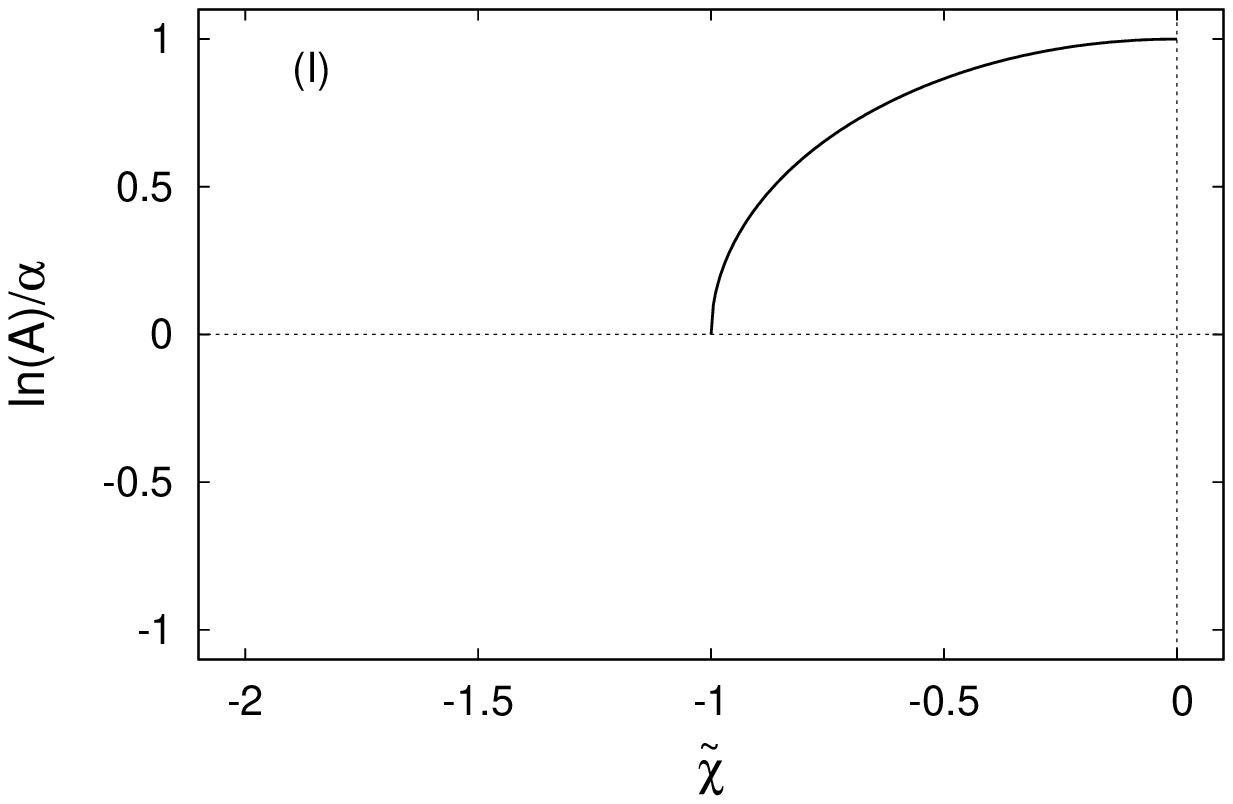}}
\epsfxsize=5.8 cm \epsfysize=5.5 cm {\epsfbox{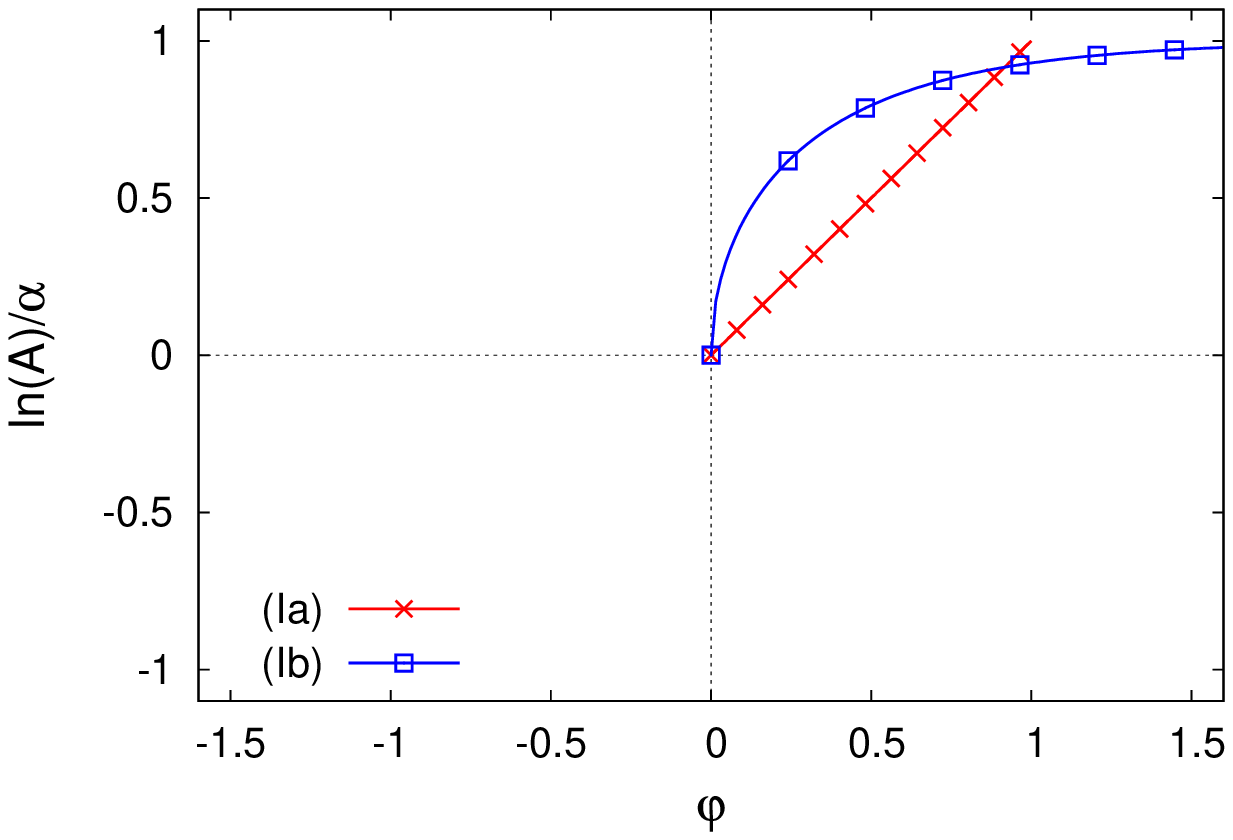}}
\epsfxsize=5.8 cm \epsfysize=5.5 cm {\epsfbox{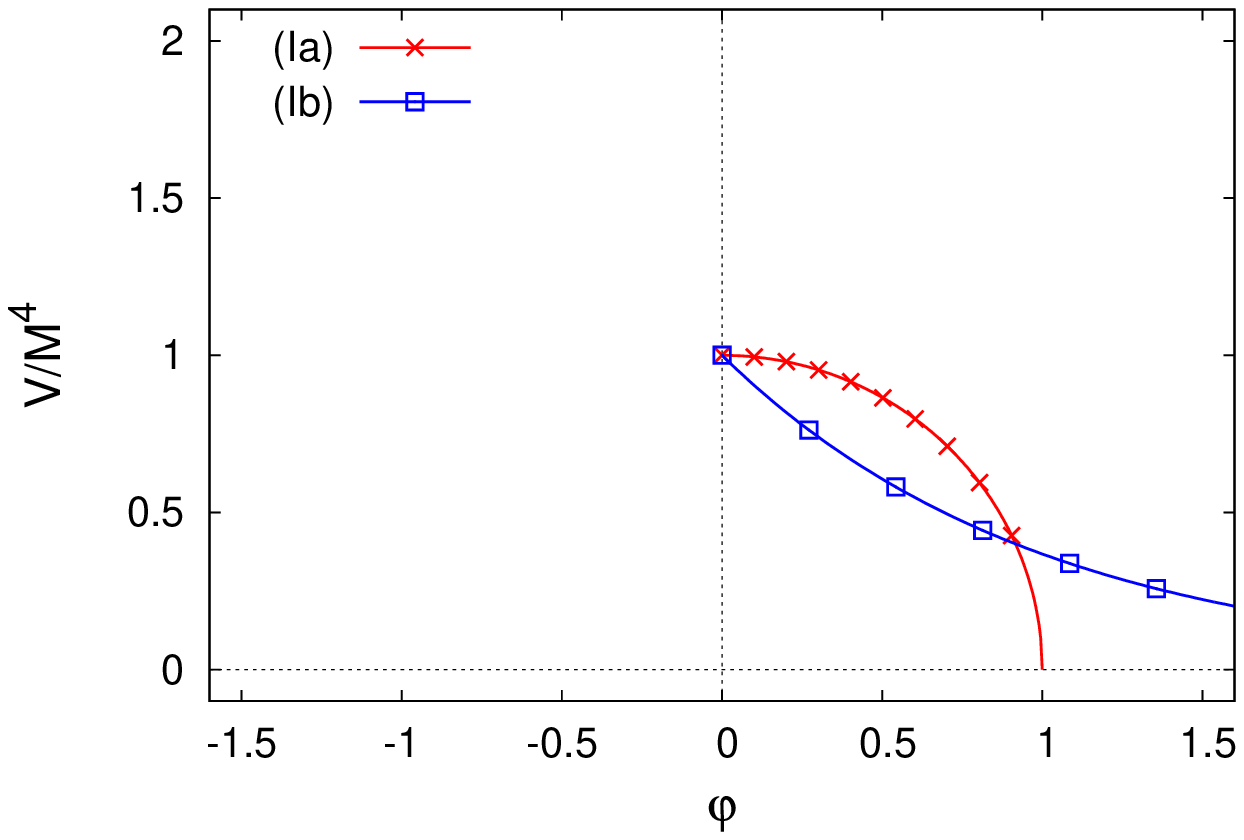}}\\
\epsfxsize=5.8 cm \epsfysize=5.5 cm {\epsfbox{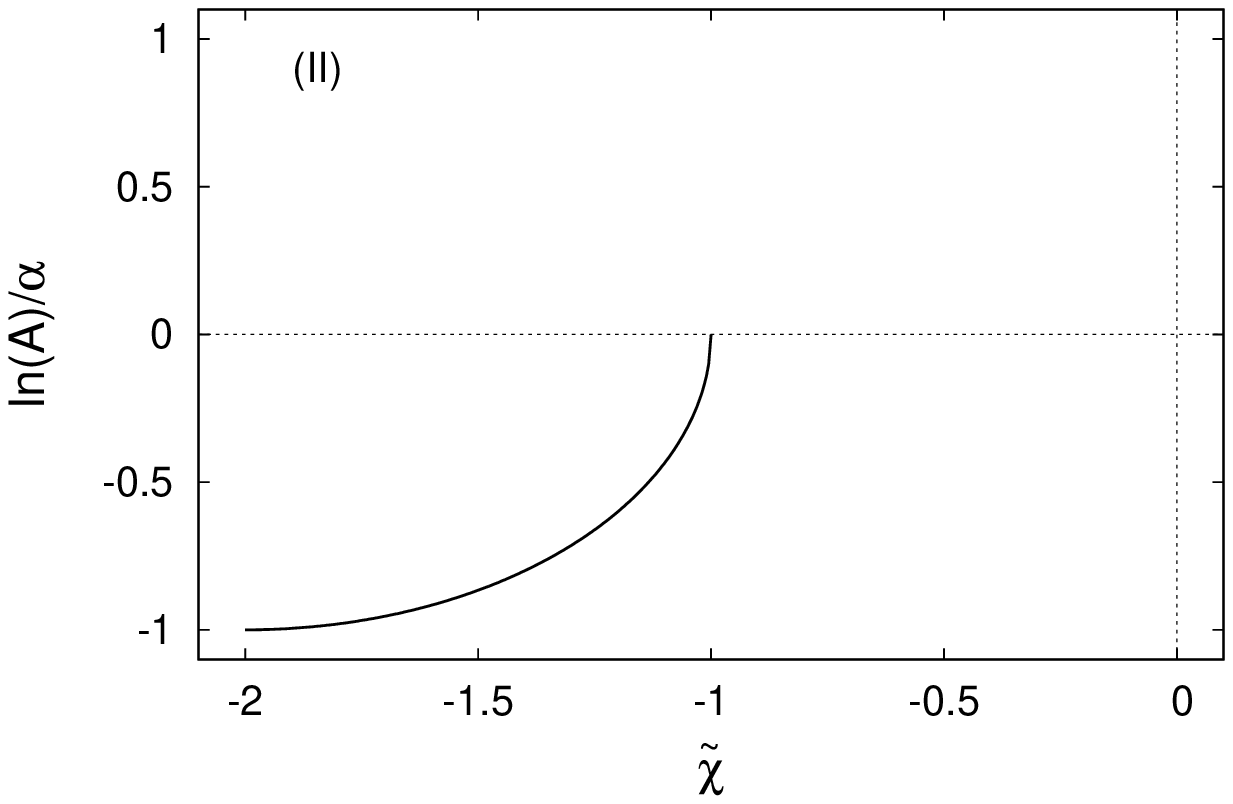}}
\epsfxsize=5.8 cm \epsfysize=5.5 cm {\epsfbox{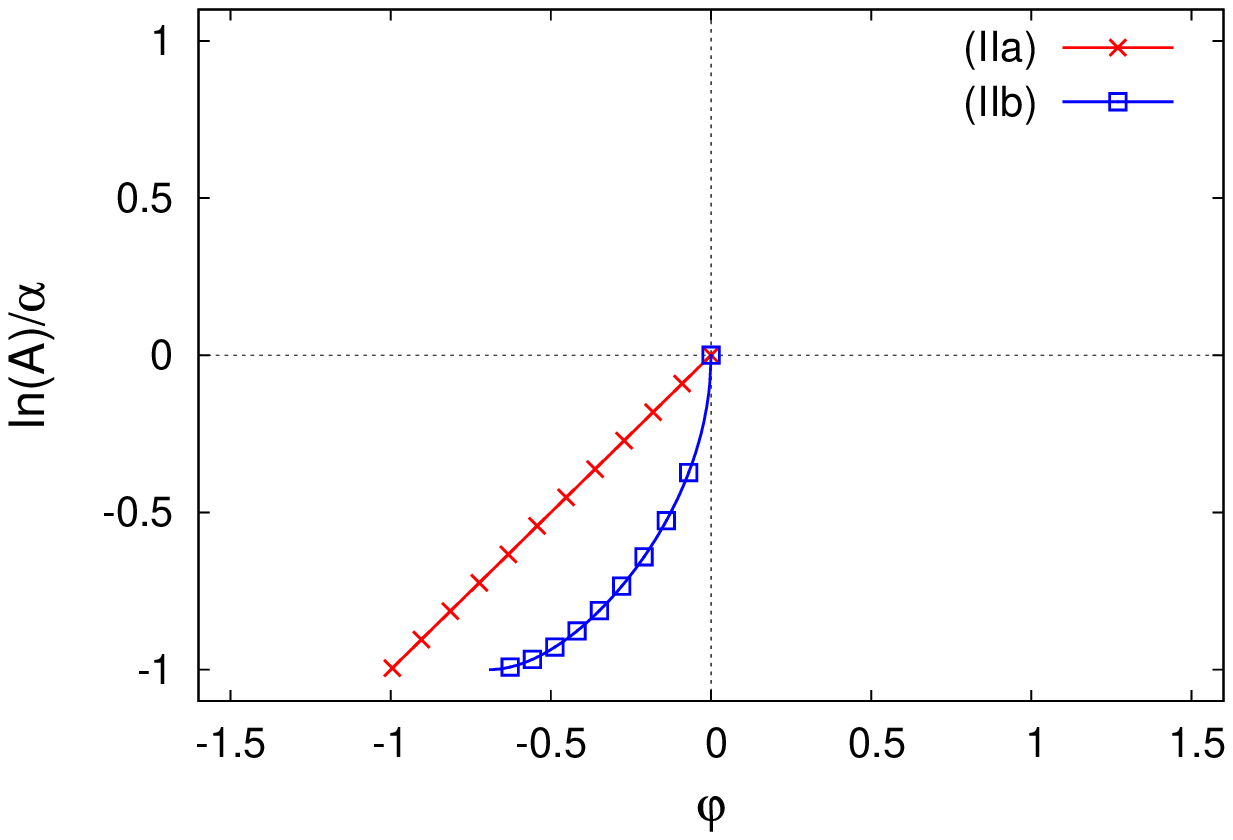}}
\epsfxsize=5.8 cm \epsfysize=5.5 cm {\epsfbox{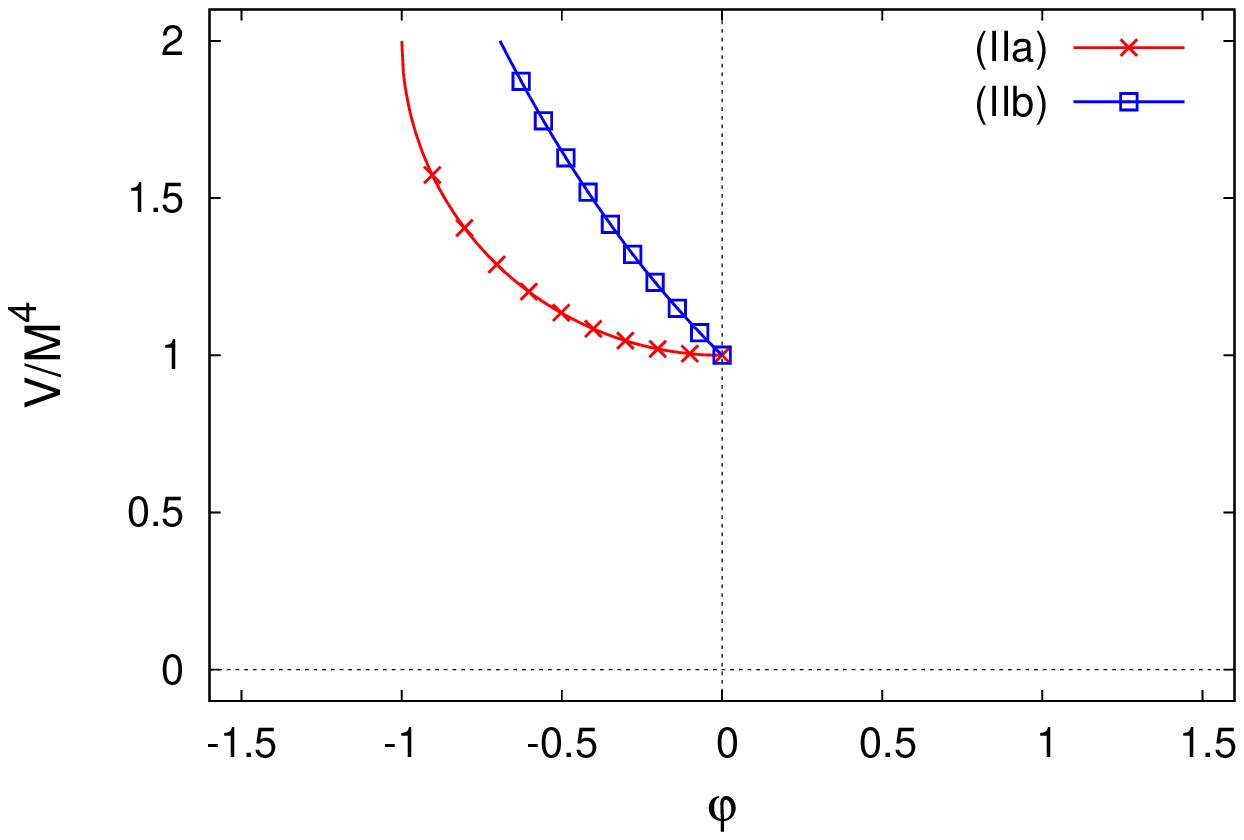}}\\
\epsfxsize=5.8 cm \epsfysize=5.5 cm {\epsfbox{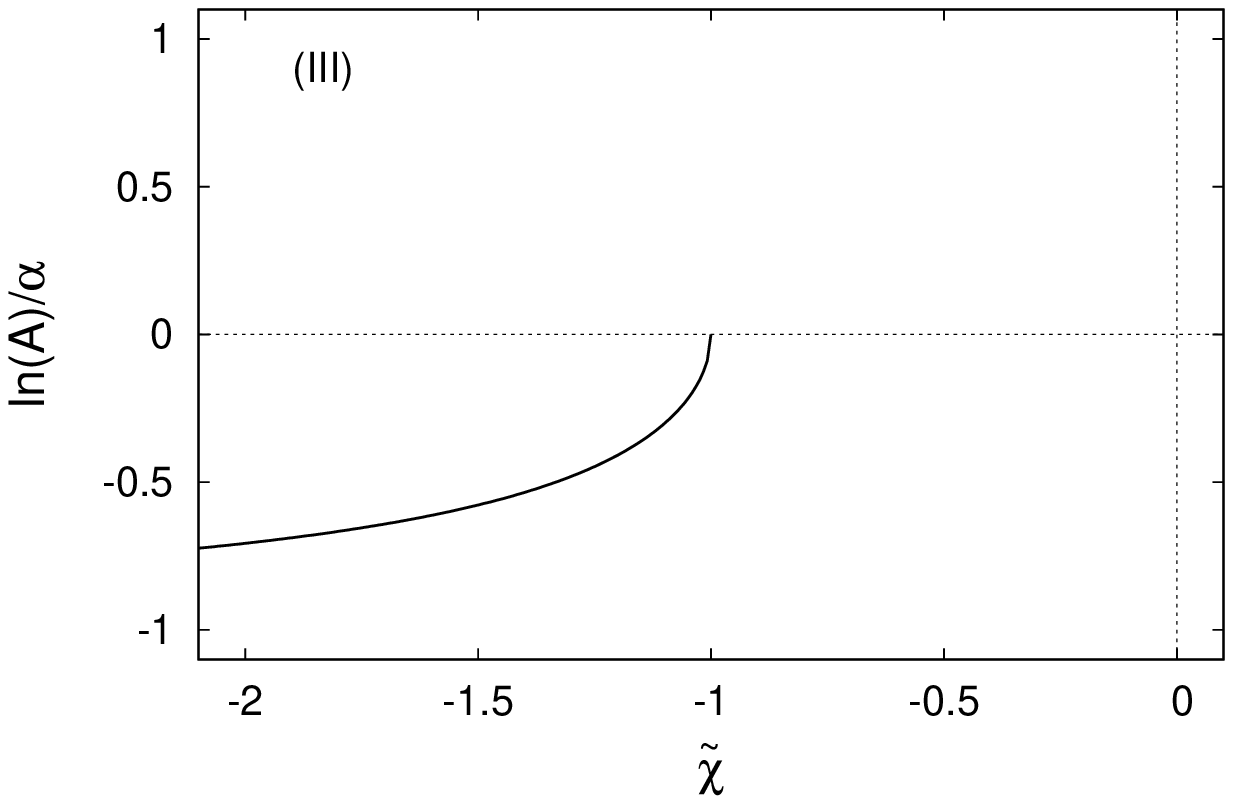}}
\epsfxsize=5.8 cm \epsfysize=5.5 cm {\epsfbox{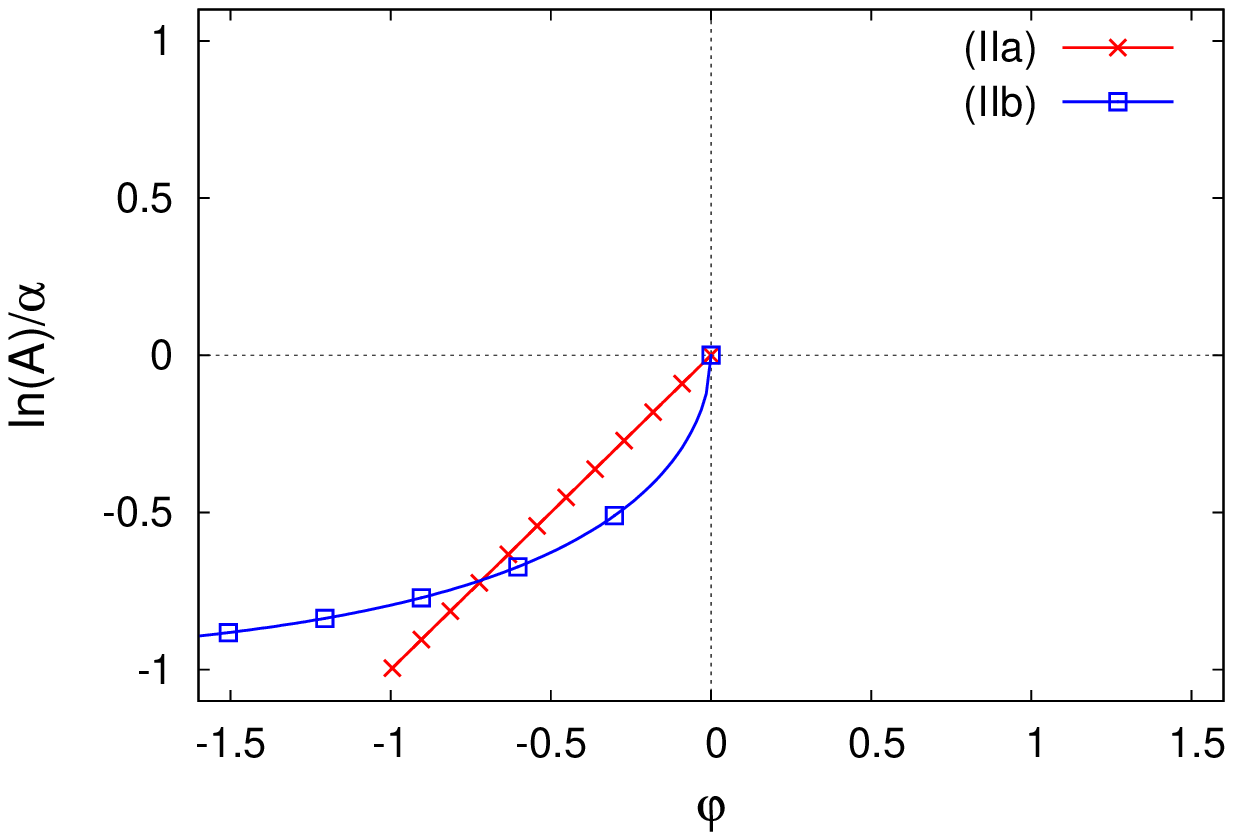}}
\epsfxsize=5.8 cm \epsfysize=5.5 cm {\epsfbox{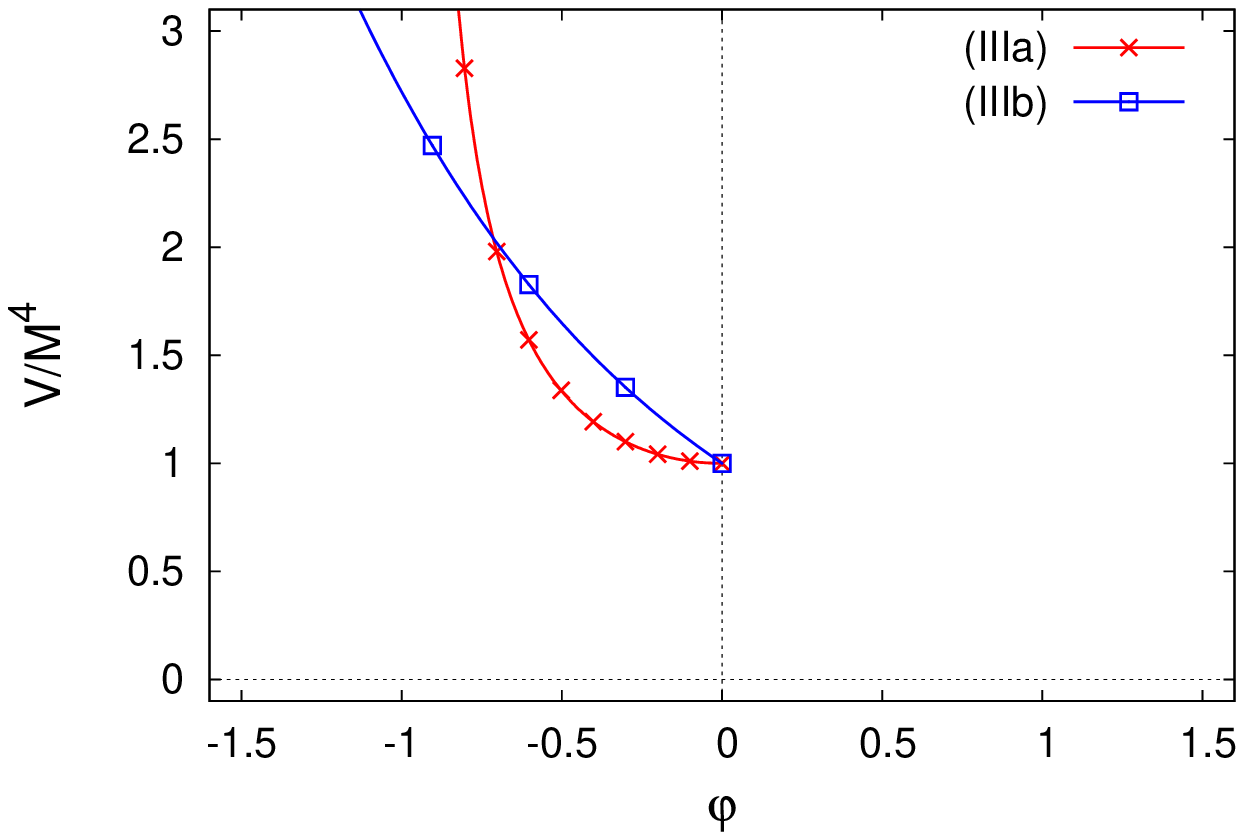}}
\end{center}
\caption{
Coupling functions and scalar field potentials for the model (I) of
Eq.(\ref{model-I-def}) ({\it upper row}), the model (II) of
Eq.(\ref{model-II-def}) ({\it middle row}) and the model (III) of
Eq.(\ref{model-III-def}) ({\it lower row}).
{\it Left column:} coupling function $A(\tchi)$ [the plot shows $\ln(A)/\alpha$].
{\it Middle column:} coupling function $A(\varphi)$ for the examples (a)
(red lines with crosses, in units of $\beta\varphi/\alpha M_{\rm Pl}$),
and (b) (blue lines with squares, in units of $\gamma\varphi/\alpha M_{\rm Pl}$).
{\it Right column:} potential $V(\varphi)/\cM^4$ for the examples (a)
(red lines with crosses, in units of $\beta\varphi/\alpha M_{\rm Pl}$),
and (b) (blue lines with squares, in units of $\gamma\varphi/\alpha M_{\rm Pl}$).
}
\label{fig_AV_1}
\end{figure*}

\subsection{Model (III): $\tchi$ is an unbounded decreasing function of $\rho$}
\label{sec:model-III}

As a variant of the model (II) of Eq.(\ref{model-II-def}), where $\tchi(\rho)$ is
a bounded decreasing function of $\rho$, we can consider the model
\beq
\mbox{model (III):} \;\;\; -\infty < \tchi < -1 , \;\;\; \ln A = - \alpha \sqrt{ 1+\frac{1}{\tchi} } ,
\label{model-III-def}
\eeq
where $\tilde\chi$ is unbounded from below.
This avoids introducing a finite lower bound $\tchi_*$.
Equation (\ref{KG-A=1}) gives
\begin{equation}
\tilde{\chi}^3 \left( \tilde{\chi} + 1 \right) = \frac{\alpha^2 \rho^2}{4 {\cal{M}}^8} ,
\label{KG-equation-model-III}
\end{equation}
which is a fourth-order algebraic equation for $\tilde{\chi}$. We can easily solve it in two different regimes,
namely when $\chi \rightarrow -1$ ($\rho \rightarrow 0)$ and
$\chi \rightarrow - \infty $ ($\rho \rightarrow + \infty$).
In the former case, we obtain
\begin{equation}
\chi \rightarrow -1, \;\;\; \frac{\alpha\rho}{{\cal M}^4} \ll 1 : \;\;\;
\tilde{\chi}(\rho) \simeq - 1 - \frac{\alpha^2 \rho^2}{4{\cal{M}}^8} ,
\label{chi-rho-model-IIIa}
\end{equation}
\begin{equation}
\ln A(\rho) \simeq - \alpha^2 \frac{\rho}{2 {\cal{M}}^4} ,
\label{lnA-chi-model-IIIa}
\end{equation}
\begin{equation}
p_A (\rho) \simeq - {\cal{M}}^4 c^2 - {\cal M}^4 c^2 \frac{\alpha^2 \rho^2}{4 {\cal{M}}^8} .
\label{pressure-rho-model-IIIa}
\end{equation}
At leading order this gives
\begin{equation}
\epsilon_2 = 3 \epsilon_1 \;\; \mbox{with} \;\;
\epsilon_{1}(a) = \frac{\alpha^2 \bar\rho}{2{\cal{M}}^4} ,
\label{eps12-model-IIIa}
\end{equation}
and again we can check that $\epsilon_1(a) \sim \alpha^2$ at low $z$.
On the other hand, in the high-density limit we obtain:
\begin{equation}
\chi \rightarrow - \infty , \;\;\; \frac{\alpha\rho}{{\cal M}^4} \gg 1 : \;\;\;
\tilde{\chi} (\rho) \simeq - \sqrt{\frac{\alpha \rho}{2{\cal{M}}^4}} ,
\label{model-III-large-chi-rho-def}
\end{equation}
\begin{equation}
\ln A (\rho) \simeq - \alpha + \sqrt{\frac{\alpha{\cal{M}}^4}{2\rho}} ,
\label{lnA-rho-model-III-b}
\end{equation}
\begin{equation}
p_A (\rho) \simeq - {\cal{M}}^4 c^2 \sqrt{\frac{\alpha \rho}{2{\cal{M}}^4}} ,
\label{pressure-rho-model-III-b}
\end{equation}
and
\begin{equation}
\epsilon_2 = 3 \epsilon_1 \;\; \mbox{with} \;\;
\epsilon_1 (a) \simeq \sqrt{\frac{\alpha{\cal{M}}^4}{8\rho}} ,
\label{eps12-model-III-b}
\end{equation}
which again shows that $|\epsilon_1(a)| \ll \alpha$ in this limit.

As in the previous cases this coupling function $A(\tilde\chi)$ corresponds to an infinite
number of pairs $\{V(\varphi),A(\varphi)\}$.
In particular, the case of an exponential coupling or an exponential potential are described by
\beq
\mbox{(IIIa)} : \;\; V(\varphi) = \frac{\cM^4}{1- \left( \frac{\beta \varphi}{\alpha M_{\rm Pl}} \right)^2 } ,  \;\;\;
A(\varphi) = e^{\beta\varphi/M_{\rm Pl}}
\label{IIIa-def}
\eeq
with $\beta>0$ and $-\alpha < \beta\varphi/M_{\rm Pl} < 0$, and
\beq
\mbox{(IIIb)} : \;\; V(\varphi) = \cM^4 e^{-\gamma\varphi/M_{\rm Pl}} , \;\;\;
A(\varphi) = e^{-\alpha\sqrt{1 -e^{\gamma\varphi/M_{\rm Pl}}}} ,
\label{IIIb-def}
\eeq
with $\gamma>0$ and $- \infty <\gamma\varphi/M_{\rm Pl}<0$.

\subsection{Common low-redshift and low-density behavior}
\label{sec:models-comparison}

In the following, we consider the case $\alpha=10^{-6}$ and $\chi_*=-2$ for the parameters
that define the models (I), (II) and (III).
In all cases we normalized the low-density limit of the scalar field $\tilde\chi$ to $-1$ and
the derivative $d\ln A/d\tilde\chi$ must go to $+\infty$ in this limit, from Eq.(\ref{KG-A=1}).
For a power-law divergence, $d\ln A/d\tilde\chi \sim \alpha|\tilde\chi+1|^{-\nu}$, with $0<\nu<1$,
this gives $|\tilde\chi+1| \sim (\alpha\rho/{\cal M}^4)^{1/\nu}$ in the low-density regime.
In the explicit models (\ref{model-I-def}), (\ref{model-II-def}) and (\ref{model-III-def}) we have $\nu=1/2$,
so that in all three cases we have:
\beq
\frac{\alpha\rho}{{\cal M}^4} \ll 1 : \;\;\; | \tilde\chi+1 | \sim \left( \frac{\alpha\rho}{{\cal M}^4} \right)^2 ,
\label{chi-low-density}
\eeq
and
\beq
\frac{\alpha\bar\rho}{{\cal M}^4} \ll 1 : \;\;\;
\beta_1 = \frac{{\cal M}^4}{\bar\rho} \sim \frac{\alpha}{|\bar{\tilde\chi}+1|^{1/2}}  ,
\label{beta1-low-density}
\eeq
\beq
| \beta_2 | \sim \frac{\alpha}{|\bar{\tilde\chi}+1|^{3/2}} \sim \alpha^{-2}
\left( \frac{{\cal M}^4}{\bar\rho} \right)^3 ,
\label{beta2-low-density}
\eeq
\beq
| \epsilon_1 | = \frac{\beta_1^2}{| \beta_2 |} \sim
\alpha |\bar{\tilde\chi}+1|^{1/2} \sim \alpha^2
\frac{\bar\rho}{{\cal M}^4} .
\label{eps1-low-density}
\eeq
For future use, writing $d\ln A / d\ln\rho = (d\ln A/d\tilde\chi) (d\tilde\chi/d\ln\rho)$,
we also obtain in this low-density regime
\beq
\frac{\alpha\rho}{{\cal M}^4} \ll 1 : \;\;\;
\left| \frac{d\ln A}{d\ln\rho} \right| \sim \alpha^2 \frac{\rho}{{\cal M}^4} .
\label{dlnAdlnrho-low-density}
\eeq

In models (I) and (II), since the scalar field $\tilde\chi$ has a finite range of order unity,
Eq.(\ref{chi0-rhode0}) implies ${\cal M}^4 \sim \bar\rho_{\rm de 0} \sim \bar\rho_0$.
Therefore, at low redshifts we have from Eq.(\ref{chi-low-density}):
\beq
a \gg a_\alpha \sim \alpha^{1/3} : \;\;\; | \bar{\tilde\chi}+1 | \sim \alpha^2 a^{-6} ,
\label{chib-1-low-z}
\eeq
where we normalized the scale factor to unity today, $a(z=0)=1$.
In particular, we have $|\bar{\tilde\chi}+1| \sim \alpha^2 \sim 10^{-12}$ at low $z$, so that
Eq.(\ref{chi0-rhode0}) implies
\beq
{\cal M}^4 = \bar\rho_{\rm de 0} ,
\label{M4-rhode}
\eeq
up to a $10^{-12}$ accuracy,
and this parameter is completely set by the reference $\Lambda$-CDM cosmology.
Moreover, the dark energy density is almost constant, along with $\bar{\tilde\chi}$,
up to a redshift $z_{\alpha} \sim \alpha^{-1/3} \sim 100$, which means that the background cosmology
cannot be distinguished from the $\Lambda$-CDM reference, in agreement with the analysis
in Sec.~\ref{sec:constraints}.
For the model (III) we also take ${\cal M}^4 = \bar\rho_{\rm de 0}$, which gives the same behaviors.
Then, the scalar-field equation (\ref{KG-A=1}) reads as
\beq
\beta_1(a) = \frac{d\ln \bar{A}}{d\bar{\tilde\chi}}
= \frac{\Omega_{\rm de 0}}{\Omega_{\rm m 0}} a^3 .
\label{beta1-a}
\eeq
Thus, the first derivative of the coupling function, $d\ln A/d\tilde\chi$, at the background level,
is of order unity at low $z$ (despite the prefactor $\alpha$ of $\ln A$, which means that at low
$z$ we are close to the divergence of $d\ln A/d\tilde\chi$) and decreases with redshift
as $(1+z)^{-3}$.

Thus, these models involve two free parameters that must be set to match observations:
the usual dark energy scale ${\cal M}^4=\bar{\rho}_{\rm de 0}$, as in most cosmological models
including $\Lambda$-CDM, and the parameter $\alpha \lesssim 10^{-6}$ that is needed to make
sure that the fifth force never becomes too large as compared with Newtonian gravity.
Of course, this is only an upper bound and we can take $\alpha$ as small as we wish, as we recover
the $\Lambda$-CDM scenario and General Relativity in the limit $\alpha \rightarrow 0$
[where the coupling function $A(\tilde\chi)$ becomes identical to unity and the non-minimal coupling
between matter and the scalar field vanishes].

In Fig.~\ref{fig_AV_1}, we show the coupling and potential functions of the models (I) (upper row),
(II) (middle row) and (III) (lower row).
The left column shows $\lambda(\tilde\chi)=\alpha^{-1} \ln A(\tilde\chi)$ from
Eqs.(\ref{model-I-def}), (\ref{model-II-def}) and (\ref{model-III-def}).
The middle column shows $\lambda(\varphi)=\alpha^{-1} \ln A(\varphi)$ for the variants (a)
and (b).
The right column shows $-\tilde\chi=V(\varphi)/{\cal M}^4$ for the same cases.

In models (Ia,Ib) $\varphi$ is positive whereas in models (IIa,IIb,IIIa,IIIb) it is negative.
It has a finite range in models (Ia) ($0<\beta\varphi/\alpha M_{\rm Pl}<1$),
(IIa) ($1+\tilde\chi_*<\beta\varphi/\alpha M_{\rm Pl}<0$),
(IIb) ($-\ln(-\tilde\chi_*)<\gamma\varphi/M_{\rm Pl}<0$) and
(IIIa) ($-1<\beta\varphi/\alpha M_{\rm Pl}<0$), while it extends from zero to $+\infty$
in model (Ib) and from zero to $-\infty$ in model (IIIb).

In all cases, the late-time dark energy era, $t\rightarrow+\infty$, corresponds to
$\tilde\chi\rightarrow -1$, $\varphi\rightarrow 0$, $V/{\cal M}^4 \rightarrow 1$
(i.e. the end-point at the center of the plots).
It is the maximum of the potential $V(\varphi)$ in model (I) and the minimum
in models (II) and (III).
This low-density limit corresponds to $d\ln A/d\tilde\chi \rightarrow +\infty$,
which implies $(d\ln A/d\varphi) (d\varphi/d V) \rightarrow -\infty$.
In models (a), this is achieved by $d V/d\varphi \rightarrow 0$, while
in models (b) this is achieved by $d\ln A/d\varphi \rightarrow +\infty$.

The early-time or high density limit corresponds to $\tilde\chi \rightarrow 0$ in
model (I), $\tilde\chi\rightarrow -2$ in model (II) and $\tilde\chi\rightarrow -\infty$
in model (III). It also corresponds to $d\ln A/d\tilde\chi \rightarrow 0$,
which implies $(d\ln A/d\varphi) (d\varphi/d V) \rightarrow 0$.
In models (a), this is achieved by $d V/d\varphi \rightarrow -\infty$, while
in models (b) this is achieved by $d\ln A/d\varphi \rightarrow 0$.

\section{Self-consistency and regime of validity of the theory}
\label{sec:theory-validity}

Before we investigate the properties of the models introduced in this paper,
from cosmological to Solar System scales, we consider in this section the self-consistency
and the range of validity of theories defined by the Lagrangian (\ref{L-phi-def}).

\subsection{Stability with respect to a nonzero kinetic term}
\label{sec:Stability}

In the Lagrangian (\ref{L-phi-def}) we set the kinetic term of the scalar field to zero.
However, in realistic scenarios the models studied in this paper may rather correspond
to cases where the scalar field Lagrangian is merely dominated by its potential term
with a negligible but non-zero kinetic term. Then, we must check whether the
solution (\ref{KG-J-background}) obtained in the previous sections remains meaningful
for a small non-zero kinetic term.
Thus, we generalize the Lagrangian (\ref{L-chi-def}) to
\beq
\tilde{\cal L}_{\tilde\chi}(\tilde\chi) = - \frac{\kappa}{2} \tilde\nabla_{\mu} \tilde\chi \tilde\nabla^{\mu}
\tilde\chi + \cM^4 \tilde\chi \;\;\; \mbox{with} \;\;\;
\kappa \rightarrow 0 .
\label{L-chi-def-kinetic}
\eeq
Using $\bar{A} \simeq 1$, so that the Einstein and Jordan frames are identical at the
background level, the equations of motion (\ref{KG-E}) or (\ref{KG-A=1})
of the scalar field generalize to
\beq
- \kappa a^{-4} \frac{\partial}{\partial\tau} \left( a^2 \frac{\partial\tilde\chi}{\partial\tau} \right)
+ \kappa a^{-2} c^2 \nabla^2 \tilde\chi + \cM^4
= \rho \frac{d\ln A}{d \tilde\chi} .
\label{KG-J-kinetic}
\eeq

\subsubsection{Quasi-static cosmological background}
\label{sec:quasi-static-background}

At the background level, considering a scalar field that only depends on time,
we expand the solution of the Klein-Gordon equation (\ref{KG-J-kinetic})
around the solution $\bar{\tilde\chi}_0$ of Eq.(\ref{KG-J-background}), obtained in the
previous sections with a zero kinetic term,
\beq
\bar{\tilde\chi} =  \bar{\tilde\chi}_0 + \bar\phi \;\;\; \mbox{with} \;\;\;
\cM^4 = \bar{\rho} \frac{d\ln A}{d\tchi}(\bar{\tilde\chi}_0) .
\label{tchi0-def}
\eeq
Using the expansion (\ref{beta-n-def}) of $\ln A$ around $\ln A(\bar{\tilde\chi}_0)$, this gives at
linear order over $\bar\phi$,
\beq
\kappa a^{-4} \frac{d}{d\tau} \left( a^2 \frac{d\bar\phi}{d\tau} \right)
+ \bar{\rho} \beta_2 \bar\phi =
- \kappa a^{-4} \frac{d}{d\tau} \left( a^2 \frac{d\bar{\tilde\chi}_0}{d\tau} \right) .
\label{phi-chi0}
\eeq
In the limit $\kappa\rightarrow 0$, the particular solution reads at linear order in $\kappa$ as
\beq
\bar\phi_0 = - \frac{\kappa}{a^4\bar\rho\beta_2} \frac{d}{d\tau}
\left( a^2 \frac{d\bar{\tilde\chi}_0}{d\tau} \right) .
\label{phi0-chi0}
\eeq
As expected, it vanishes in the limit $\kappa\rightarrow 0$.
More precisely, the correction $\bar\phi_0$ is negligible as compared to the quasi-static
solution $\bar{\tilde\chi}_0$ if
\beq
| \kappa | \ll | \beta_2 | \frac{\bar\rho}{H^2} \sim | \beta_2 | M_{\rm Pl}^2 .
\label{kappa-small}
\eeq
From the expressions given in section~\ref{sec:models} and as we will check
in Fig.~\ref{fig_Aa_1} and Eq.(\ref{beta2-a-esp1-a}) below, $\beta_2$ is of
order $\alpha^{-2} \gg 1$ today and decreases at higher redshift until $z_{\alpha} \sim 100$,
where it is of order $\alpha$. At higher redshift it typically remains of order $\alpha$
[because of the prefactor $\alpha$ in the coupling function $\ln A(\tilde\chi)$], or decays
to zero in models such as (III) where $|\bar\tilde\chi|$ is not bounded and goes to
infinity at high $z$.
Thus, for practical purposes the condition (\ref{kappa-small}) is satisfied if
\beq
| \kappa | \ll \alpha M_{\rm Pl}^2 \sim 10^{-6} \; M_{\rm Pl}^2 .
\label{kappa-alpha}
\eeq
For models such as (III), the condition (\ref{kappa-small}) will be violated
at very high $z$, $z \gg z_{\alpha}$, if $\kappa$ does not go to zero.
Then, one must take into account the kinetic terms in the scalar field equation
to obtain the background solution $\bar{\tilde\chi}$.
However, at these high redshifts the scalar field should not play an
important role and our results should be independent of this early-time modification.
On the other hand, in such cases the kinetic prefactor $\kappa$ generically depends
on time, through the factor $(d\varphi/d V)^2$ introduced by the change of
variable (\ref{tchi-def}), and we expect for instance $\kappa$ to decrease as fast as
$\beta_2$, as $1/V^2$, for models where $V$ goes to infinity while $\varphi$ remains
bounded, so that the condition (\ref{kappa-small}) remains satisfied.

So far we have only introduced two parameters in the models, the dimensional
dark-energy density today,
$\bar\rho_{\rm de 0} = {\cal M}^4 \simeq (2.296 \times 10^{-12} {\rm GeV} )^4$ \cite{Planck2015},
and the dimensionless parameter $\alpha \sim 10^{-6}$.
Since ${\cal M}$ is smaller than the reduced Planck mass,
$M_{\rm Pl} \simeq 2.44 \times 10^{18} {\rm GeV}$, by 30 orders of magnitude,
we can see that we do not need to introduce additional small parameters to
satisfy Eq.(\ref{kappa-alpha}). Apart from $\kappa=0$, the choices
$\kappa \sim {\cal M}^2$, $\kappa \sim {\cal M} M_{\rm Pl}$ or
$\kappa \sim {\cal M}^{1/2} M_{\rm Pl}^{3/2}$, satisfy the constraint.

The homogeneous solutions of Eq.(\ref{phi-chi0}) obey
\beq
\frac{d^2\bar\phi}{d\tau^2} + \frac{a^2\bar\rho\beta_2}{\kappa} \bar\phi = 0 ,
\label{phi-homogeneous}
\eeq
in the high-frequency limit (i.e., over time scales much below $1/H$).
From the condition (\ref{kappa-small}) we have $a^2\bar\rho\beta_2/\kappa \gg {\cal H}^2$,
so that the homogeneous solution evolves indeed on time scales much shorter than
the Hubble time.
For the solution $\bar{\tilde\chi}_0$ to be stable the homogeneous solutions
(\ref{phi-homogeneous}) must not show exponential growth but only
fast oscillations, of frequency $\omega \propto \kappa^{-1/2}$. This leads to the constraint
\beq
\frac{\beta_2}{\kappa} > 0 .
\label{beta2-kappa-+}
\eeq

As the field $\tilde\chi$ typically arises through the change of variable (\ref{tchi-def}), the kinetic
coefficient $\kappa$ introduced in Eq.(\ref{L-chi-def-kinetic}) depends on time. However,
its sign is not modified by the change of variable and it is positive for standard well-behaved
models.
Then, the constraint (\ref{beta2-kappa-+}) leads to $\beta_2>0$, which means that
$\ln A(\tilde\chi)$ must be a convex function.
This rules out the model (I) introduced in Sec.~\ref{sec:model-I}. More generally, from the
definition of the coefficients $\beta_1$ and $\beta_2$ and the scalar-field equation
(\ref{KG-A=1}), the condition $\beta_2>0$ implies that $d\rho/d\tilde\chi<0$ and the function
$\tilde\chi(\rho)$ is a monotonic decreasing function of $\rho$.
At the background level, this implies that $\bar{\tilde\chi}$ increases with time, hence
the potential $V(\bar\varphi)$ defined from the change of variable (\ref{tchi-def})
decreases with time.
As expected, the stable case corresponds to scenarios where the background scalar field
$\bar\varphi$ rolls down its potential $V(\bar\varphi)$ [as in models (II) and (III) of
Secs.~\ref{sec:model-II} and \ref{sec:model-III}], whereas the unstable case corresponds
to a background scalar field that climbs up its potential [as in model (I) of Sec.~\ref{sec:model-I}].

Models with $\beta_2<0$ could be made stable, with respect to the classical background
perturbations analyzed here, by choosing a non-standard sign $\kappa<0$ for the small
kinetic term.
However, such models are typically plagued by ghost instabilities, as the kinetic energy is
unbounded from below, unless one sets a high-energy cutoff of the theory at a sufficiently
low energy to tame down these instabilities.
In the following we also present our results for the model (I) of Sec.~\ref{sec:model-I}
to keep this work as general as possible, even though this is unlikely to correspond to
realistic and natural scenarios.

\subsubsection{Cosmological large-scale structures}
\label{sec:stability-large-scale}

To apply the equations of motion derived in Sec.~\ref{sec:Equations-of-motion}
to the formation of large-scale structures, we must also check that the kinetic
term plays no role on these scales.
Thus, we now take into account the Laplacian term in Eq.(\ref{KG-J-kinetic}) and the
perturbations of the scalar field, $\phi=\delta\tilde\chi-\bar{\tilde\chi}$, obey at linear order
\beq
- \kappa a^{-4} \frac{\partial}{\partial\tau} \left( a^2 \frac{\partial\phi}{\partial\tau} \right)
+ \kappa a^{-2} c^2 \nabla^2 \phi - \bar\rho \beta_2 \phi = \beta_1 \delta\rho .
\label{KG-J-kinetic-Laplacian}
\eeq
As for the background case, the time derivatives are negligible when the condition
(\ref{kappa-small}) is satisfied. The spatial Laplacian can be neglected at comoving scale
$1/k$, where $k$ is the wave number of interest, if we have
\beq
| \kappa | \ll \frac{\bar\rho | \beta_2 | a^2}{c^2 k^2} \sim \left( \frac{a H}{c k} \right)^2 | \beta_2 |
M_{\rm Pl}^2 .
\label{kappa-k}
\eeq
This constraint is tighter than the background condition (\ref{kappa-small}) as we
require the theory to remain valid down to sub-horizon scales, $ck/aH \gg 1$.
If we wish to apply the model without kinetic term down to $1 \, h^{-1} {\rm kpc}$,
below the galaxy-halo scale, we must have
\beq
k \sim 1 \; h \, {\rm kpc}^{-1} : \;\;\; \kappa \ll 10^{-19} \; M_{\rm Pl}^2 ,
\label{kappa-k-1kpc}
\eeq
where we used again $|\beta_2| \sim \alpha \sim 10^{-6}$.
We can still choose for instance $\kappa \sim {\cal M}^2$ or $\kappa \sim {\cal M} M_{\rm Pl}$.
Thus, we do not need to introduce a new low-energy scale to build a small-enough kinetic
term that can be neglected for both the background and the cosmological
structures.

Of course, on scales $1 \, h^{-1} {\rm kpc}$ the density and scalar fields are in the nonlinear
regime, which modifies Eq.(\ref{KG-J-kinetic-Laplacian}).
If we expand around the local solution, $\tilde\chi_0[\rho({\vx})]$, the factors
$\bar\rho\beta_2(\bar\rho)$ and $\beta_1(\bar\rho)$ must be replaced by
$\rho\beta_2(\rho)$ and $\beta_1(\rho)$. For models such as (I) and (II), where $\beta_2$
remains of order $\alpha$ at high densities the upper bound in Eq.(\ref{kappa-k-1kpc})
is simply multiplied by a factor $\rho/\bar\rho$. Then, nonlinearities actually loosen up
the constraint (\ref{kappa-k-1kpc}) and the kinetic term in the scalar-field Lagrangian
becomes even more negligible.
In practice, the coefficient $\kappa$ will depend on the local value of the scalar
field, and hence on the local density, but the relatively high upper bound
$(\rho/\bar\rho) 10^{-19} M_{\rm Pl}^2$, as compared with ${\cal M}^2$ or ${\cal M} M_{\rm Pl}$,
suggests that the scale ${\cal M}$ will be sufficient to construct small-enough coefficients
$\kappa$ without introducing additional finely tuned parameters.

\subsection{Small-scale cutoff}
\label{sec:cutoff}

Independently of a possible kinetic term, the local model (\ref{L-chi-def}) considered so far
is not expected to apply down to arbitrarily small scales.
In the previous sections and through most of this paper, we implicitly assume that
we can work with a continuous density field $\rho(\vx)$ defined by some coarse-graining
procedure, instead of a singular field made of Dirac peaks (in the limit of classical point-like
particles) or of isolated density peaks (finite-size classical particles).
Therefore, we assume the models studied in this paper to be effective
theories that only apply beyond some small-scale cutoff $\ell_s$, so that the density field
is defined by a coarse-graining at scale $\ell_s$.

If we consider for instance the mean inter-particle distance, $\lambda = ( m/\rho )^{1/3}$,
we obtain on the Earth, with $\rho \sim 1 \; {\rm g/cm}^{3}$ and $m \sim m_p$, the proton mass,
\beq
\lambda_{\rm Earth} \sim 10^{-8} \; {\rm cm} ,
\label{lambda-Earth}
\eeq
and in the intergalactic medium (IGM), using the mean density of the Universe,
\beq
\lambda_{\rm IGM} \sim \left( \frac{m}{m_p} \right)^{1/3} 100 \; {\rm cm} ,
\label{lambda-intergalactic}
\eeq
where $m$ can be taken as the largest among the proton and the dark matter particle mass.
This typically gives a distance of the order of a meter.
In fact, in our study of the cosmological background and of cosmological structures,
we assume some coarse-graining of the density field on scales at least as large as
$\lambda_{\rm IGM}$, so that we can use the density $\rho$ associated with the
continuum limit.

In terms of energy scales, this corresponds to
$\lambda_{\rm IGM} \sim 1 \; {\rm m} = ( 1.973 \times 10^{-16} {\rm GeV} )^{-1}$.
On the other hand, the mean dark-energy density today is
$\bar\rho_{\rm de 0} = {\cal M}^4 = 2.778 \times 10^{-47} {\rm GeV}^4$, which gives
$\lambda_{\rm IGM} \sim 1 \; {\rm m} \sim 10^4 {\cal M}^{-1}$.
Therefore, the small-scale cutoff, $\ell_s$, which defines the smoothing scale of the density
field in such an effective approach, does not require the introduction of a new fundamental scale.
For instance, it is sufficient to set $\ell_s = \alpha^{-1} {\cal M}^{-1} \sim 100 \, {\rm m}$,
 using the two parameters $\cM$ and $\alpha$ that have already been introduced to
 characterize the model, or $\ell_s = M_{\rm Pl}^{1/2} {\cal M}^{-3/2} \sim 1 \, {\rm A.U.}$,
 using a combination with the Planck mass.

A natural way to introduce a smoothing cutoff on small scales is to have a nonzero
kinetic term in the scalar-field Lagrangian (\ref{L-chi-def}), as considered in
Sec.~\ref{sec:Stability}. Using again the Lagrangian (\ref{L-chi-def-kinetic}),
the fluctuations $\phi$ of the scalar field around the cosmological background
obey Eq.(\ref{KG-J-kinetic-Laplacian}) and the smoothing associated with the Laplacian term
becomes important at the physical scale $\ell_s$ if we have
\beq
| \kappa | \sim \frac{\bar\rho |\beta_2| \ell_s^2}{c^2} \sim \left( \frac{\ell_s H}{c} \right)^2
| \beta_2 | M_{\rm Pl}^ 2 .
\label{kappa-ls}
\eeq
We can check that this constraint is not contradictory with the conditions
(\ref{kappa-small}) and (\ref{kappa-k}), associated with the validity of the solution
without kinetic term for the background and cosmological structures, because
the small-scale cutoff $\ell_s$ can be taken to be much smaller than cosmological scales.
At $z=0$, using $|\beta_2| \sim \alpha^{-2} \sim 10^{12}$, the condition
(\ref{kappa-ls}) reads as
\beq
z=0 : \;\; | \kappa | \sim  \left( \frac{\ell_s}{1 \; {\rm m}} \right)^2  6 \times 10^{-41} \; M_{\rm Pl}^2 ,
\label{kappa-ls-z=0}
\eeq
and for $\ell_s > 1 \; {\rm m}$ we can take for instance $\kappa \sim {\cal M} M_{\rm Pl}$.

Such a Laplacian term is also sufficient to regularize the theory at the atomic scale
(\ref{lambda-Earth}) on the Earth if we have
\beq
| \kappa | > \frac{m_p |\beta_2|_{\rm Earth}}{\lambda_{\rm Earth} c^2} ,
\label{kappa-Earth}
\eeq
where $ |\beta_2|_{\rm Earth}$ is the value obtained for $\rho \sim 1 \; \rm{g . cm}^{-3}$.
Using $|\beta_2|_{\rm Earth} \sim \alpha \sim 10^{-6}$, as appropriate for the
high-density regime (see for instance Fig.~\ref{fig_Aa_1} below), this yields
\beq
{\rm Earth:} \;\;\; | \kappa | > 3 \times 10^{-49} M_{\rm Pl}^2 .
\label{kappa-Earth-Mpl2}
\eeq
This is a looser bound than the cosmological constraint (\ref{kappa-ls-z=0}).
However, because of the change of variable (\ref{tchi-def}) the kinetic prefactor $\kappa$
generically depends on the environment through the local value of the scalar field,
so that usually Eq.(\ref{kappa-Earth-Mpl2}) cannot be directly compared to
Eq.(\ref{kappa-ls-z=0}).
Nevertheless, in any case the estimates (\ref{kappa-ls-z=0}) and (\ref{kappa-Earth-Mpl2})
show that it is not difficult to regularize the theory on very small scales through
a small kinetic term in the scalar-field Lagrangian, without violating the condition
(\ref{kappa-small}).

\section{Evolution of the cosmological background}
\label{sec:background-quantities}

\subsection{Evolution of the background scalar field}
\label{sec:evolution-back-fields}

As explained in section \ref{sec:constraints}, we require the function $\ln A$ to be bounded within
a small interval of order $10^{-6}$, see Eq.(\ref{lnA-bound}), so that contributions of the fifth force
to the metric potentials do not exceed the Newtonian potential by several orders of magnitude.
As pointed out in section \ref{sec:constraints}, this implies that $|\bar{A}-1| \lesssim 10^{-6} \ll 1$.
This implies in turn that we recover a $\Lambda$-CDM cosmology at the background level
up to a $10^{-6}$ accuracy.
We also have $|\epsilon_1|$ and $|\epsilon_2|$ of order $\alpha \sim 10^{-6}$, and
$\bar{\tilde\chi}$ and $\bar\rho_{\rm de}$ are almost constant in the dark energy era,
see Eq.(\ref{drho-chi-dtau-J}).

\begin{figure}
\begin{center}
\epsfxsize=8. cm \epsfysize=5.5 cm {\epsfbox{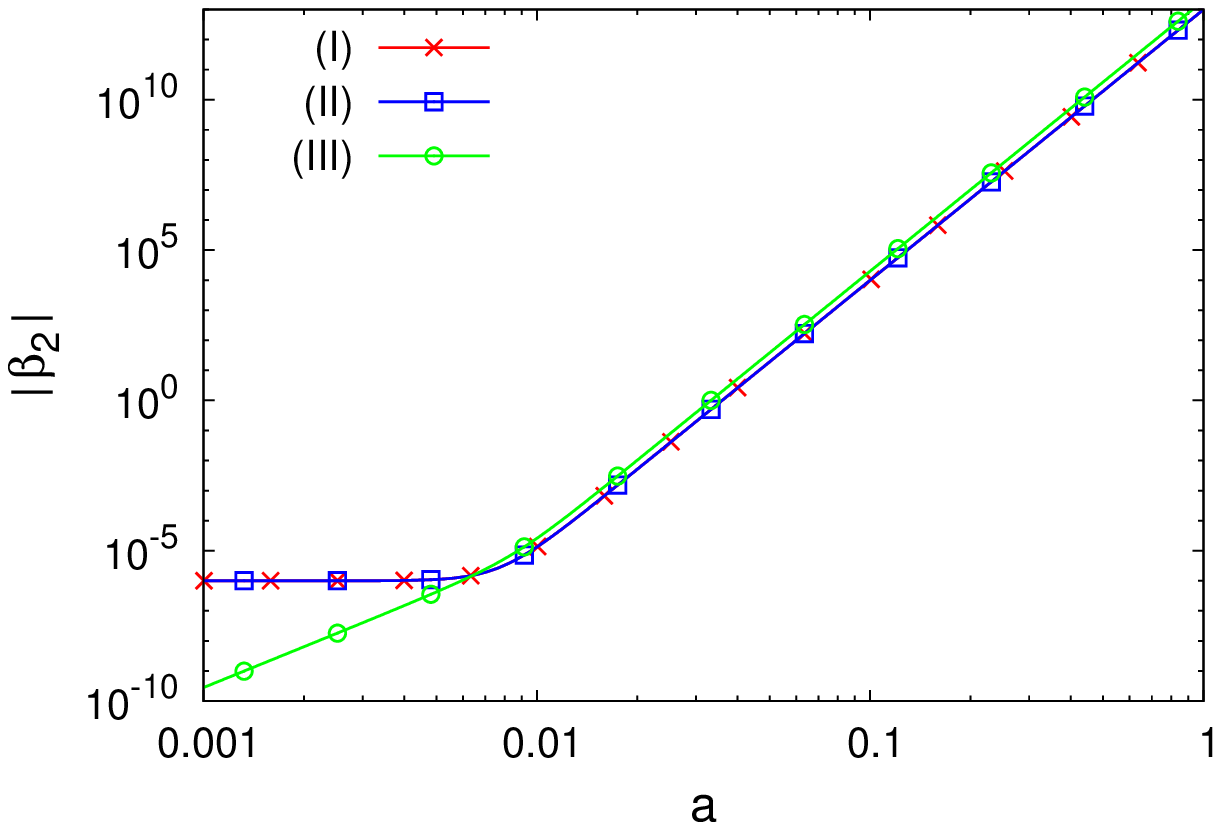}}\\
\epsfxsize=8. cm \epsfysize=5.5 cm {\epsfbox{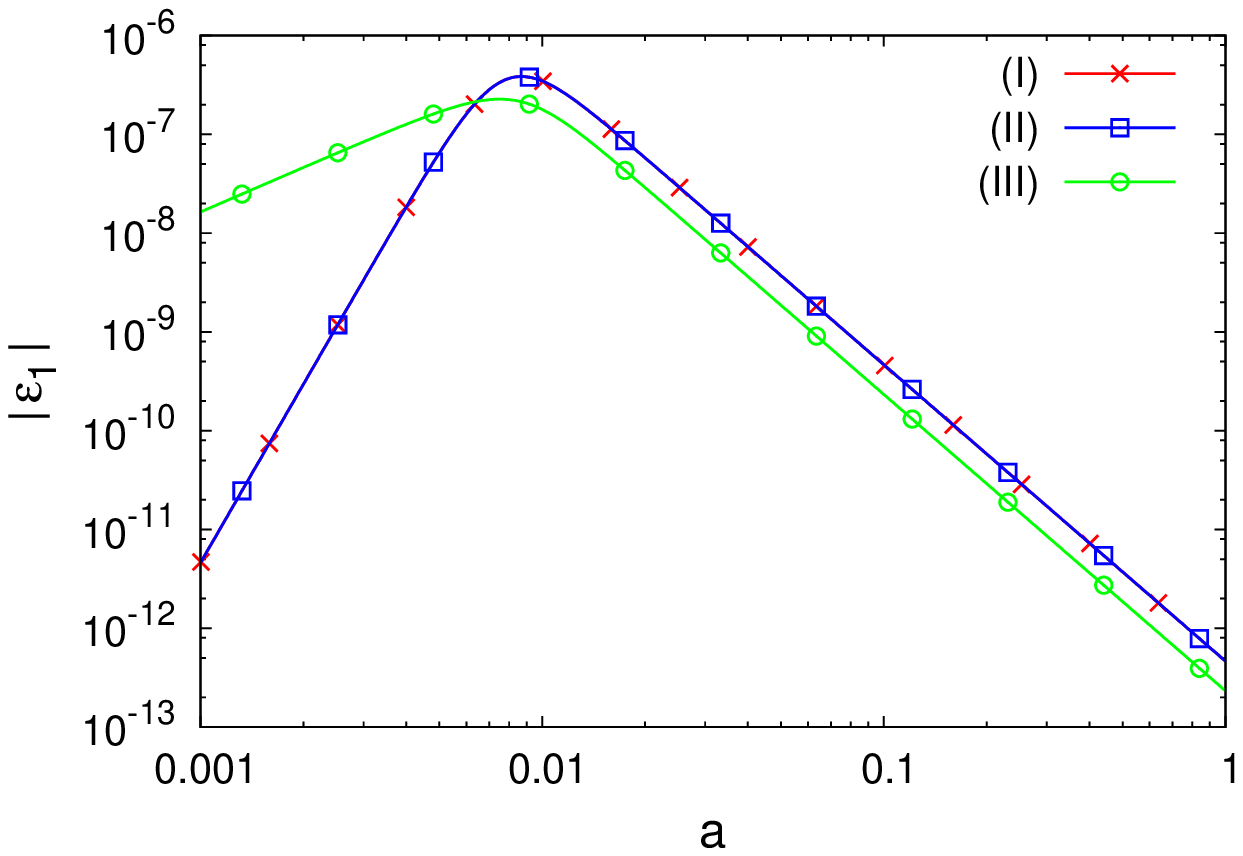}}
\end{center}
\caption{
Absolute value of the factors $\beta_2$ (upper panel) and $\epsilon_1$ (lower panel)
as a function of the scale factor.
We show $|\beta_2|$ and $|\epsilon_1|$ for models (I) (red line with crosses),
(II) (blue line with squares) and (III) (green line with circles);
$\beta_2 <0$ and $\epsilon_1<0$ for model (I); $\beta_2>0$ and $\epsilon_1>0$
for models (II) and (III).
The absolute values $|\beta_2|$ and $|\epsilon_1|$ of models (I) and (II) are equal.
}
\label{fig_Aa_1}
\end{figure}

The factor $\beta_1(a)$ is always positive and decreases with redshift as in Eq.(\ref{beta1-a}),
independently of the details of the coupling function $\ln A(\tilde\chi)$, because it is
directly set by the scalar-field equation of motion (\ref{KG-A=1}).

We show the factors $\beta_2(a)$ and $\epsilon_1(a)$ in Fig.~\ref{fig_Aa_1}.
These factors are positive for models (II) and (III), where the fifth force amplifies
Newtonian gravity, while they are negative for model (I), where the fifth force decreases
Newtonian gravity.
It happens that for our explicit choices (\ref{model-I-def}) and (\ref{model-II-def})
(with $\tilde\chi_*=-2$) the factors $\beta_2$ and $\epsilon_1$ of models (I) and (II) have
the same amplitude, but opposite signs, so that their curves coincide in Fig.~\ref{fig_Aa_1}.

From Eqs.(\ref{beta2-low-density}) and (\ref{eps1-low-density}), we have at low redshifts
\beq
a \gg a_{\alpha} \sim \alpha^{1/3} : \;\;\; | \beta_2 | \sim \alpha^{-2} a^9 , \;\;\;
| \epsilon_1 | \sim  \alpha^2 a^{-3} .
\label{beta2-a-esp1-a}
\eeq
Thus, $|\beta_2|$ is maximum today, with $|\beta_2|_0 \sim \alpha^{-2} \sim 10^{12}$,
 and decreases with redshift until $z_{\alpha} \sim \alpha^{-1/3} \sim 100$, where it is of order
 $| \beta_2 |_{z_{\alpha}} \sim \alpha \sim 10^{-6}$. At higher redshift $|\beta_2|$ typically
 remains of order $\alpha$, or goes to zero with a rate that depends on the details of the model.

The factor $|\epsilon_1|$ reaches a maximum of order $\alpha$ at a high redshift,
$z_{\alpha} \sim \alpha^{-1/3} \sim 100$, and later decays as $a^{-3}$ to reach a value of order
$\alpha^2$ today.
Therefore, the scenarios considered in this paper have the characteristic property that the main
modification to the gravitational dynamics actually occurs at a high redshift $z_{\alpha} \sim 100$,
much before the dark energy era. This is related to the small parameter $\alpha \ll 1$,
in agreement with Eq.(\ref{zalpha-def}) and the characteristic density $\rho_{\alpha}$ of
Eq.(\ref{rho-alpha-def}).
At higher redshift, $z \gg z_{\alpha}$, $|\epsilon_1|$ decreases again, as we have seen that
$\epsilon_1 = \epsilon_2/3$ (in the approximation $\bar{A}\simeq 1$) and
$\epsilon_2 = d\ln \bar{A}/d\ln a$ must vanish because $\bar{A}$ converges to
a constant close to unity at early times.
For models (I) and (II) we have $|\epsilon_1| \sim a^6/\alpha$ while for model (III) we have
$\epsilon_1 \sim \alpha^{1/2} a^{3/2}$.
Thus, the decay of $\epsilon_1$ is much slower at very high redshift for model (III).
Indeed, using the scalar-field equation (\ref{KG-A=1}) we can also write $\epsilon_2$ as
$\epsilon_2 = ({\cal M}^4/\bar\rho) d\tilde\chi/d\ln a$. Then, $d\tilde\chi/d\ln a$
goes to zero at high redshift for models (I) and (II), as $\bar{\tilde\chi}$ converges to a finite value,
whereas $d\tilde\chi/d\ln a$ goes to infinity as $\bar{\tilde\chi}$ goes to $-\infty$.
We can check these behaviors in Fig.~\ref{fig_Aa_1}.

\subsection{Negligible backreaction of small-scale nonlinearities
onto the cosmological background}
\label{sec:backreaction}

We have seen in Sec.~\ref{sec:linear-J} that for small enough $\epsilon_1$ and $\epsilon_2$
the cosmological behavior remains close to the $\Lambda$-CDM scenario at the background
and linear levels.
However, the nonlinearities associated with the scalar field could jeopardize this result.
In this section, we check that the nonlinearity of the scalar-field energy density does not
give rise to a significant backreaction onto the background dynamics.

We have seen that the scalar field energy density reads as
$\rho_{\tilde\chi} = - {\cal M}^4 \tilde\chi$ (using again $\bar{A} \simeq 1$).
Because $\tilde\chi(\rho)$ is a nonlinear function of $\rho$, its volume average
is not identical to the background value $\bar{\tilde\chi} \equiv \tilde\chi(\bar\rho)$.
This implies that the mean Hubble expansion rate over a large volume,
as large as the Hubble radius today, could significantly differ from the
background expansion obtained from the background Friedmann equation
(\ref{Friedmann-J}), especially if the volume average is actually dominated by the
highest-density regions.

In the models described in Sec.~\ref{sec:models}, the background value $\bar{\tilde\chi}$
at low redshift, in the dark energy era, is very close to the value $\tilde\chi(0)=-1$
associated with a zero density, as $|\bar{\tilde\chi}+1| \sim \alpha^2 \ll 1$,
see Eq.(\ref{chib-1-low-z}).
This is because of the small parameter $\alpha$ that was introduced to ensure
a cosmological behavior that is close to the $\Lambda$-CDM predictions.
Then, we simply check that $\langle\tilde\chi\rangle \simeq \tilde\chi(0)$ too, where the
volume average $\langle\tilde\chi\rangle$ is given by
\beq
\langle\tilde\chi\rangle = \int_V \frac{d\vx}{V} \; \tilde\chi(\rho)
= \int_0^{\infty} d\rho \; {\cal P}(\rho) \; \tilde\chi(\rho) .
\label{chi-V-def}
\eeq
Here $V$ is a large volume, with a size of the order of the Hubble radius, while
${\cal P}(\rho)$ is the probability distribution of the density within this volume.
It obeys the two normalization properties:
\beq
\int_0^{\infty} d\rho \; {\cal P}(\rho) = 1 , \;\;\;
\int_0^{\infty} d\rho \; {\cal P}(\rho) \; \rho = \bar\rho .
\label{Prho-norm}
\eeq
For any density threshold $\rho_s>0$, the second property (\ref{Prho-norm}) implies the Bienaym\'e-Tchebychev inequality
\beq
\rho_s > 0 : \;\;\; \int_{\rho_s}^{\infty} d\rho \; {\cal P}(\rho) \leq \frac{\bar\rho}{\rho_s} .
\label{Int-rhos}
\eeq
For monotonic functions $\tilde\chi(\rho)$ we have
$| \langle \tilde\chi \rangle - \tilde\chi(0) | = \int_0^{\infty} d\rho {\cal P}(\rho)
| \tilde\chi(\rho) - \tilde\chi(0) |$. Splitting the integral over the two domains $\rho \leq \rho_s$
and $\rho \geq \rho_s$, and using Eq.(\ref{Int-rhos}), gives
\beq
| \langle \tilde\chi \rangle - \tilde\chi(0) | \leq | \tilde\chi_s - \tilde\chi(0) |
+  \frac{\bar\rho}{\rho_s} | \tilde\chi(\infty) - \tilde\chi(0) | ,
\label{chi-vol-1}
\eeq
where we assumed that $\tilde\chi(\rho)$ is bounded.

Let us first consider the model (I) described in Sec.~\ref{sec:model-I}, where
$\tchi$ is a bounded increasing function of $\rho$.
This gives $\tilde\chi(\infty)=0$, $\tilde\chi_s \simeq -1 + \alpha^2\rho_s^2/2{\cal M}^8$,
for densities $\rho_s \lesssim \bar\rho/\alpha$, and the two terms in
Eq.(\ref{chi-vol-1}) are of the same order for $\rho_s \sim \bar\rho \alpha^{-2/3}$.
This choice provides an upper bound
$| \langle \tilde\chi \rangle - \tilde\chi(0) | \lesssim \alpha^{2/3} \ll 1$.

The model (II) described in Sec.~\ref{sec:model-II}, where
$\tchi$ is a bounded decreasing function of $\rho$, gives similar results and again
$| \langle \tilde\chi \rangle - \tilde\chi(0) | \lesssim \alpha^{2/3} \ll 1$.

The model (III) described in Sec.~\ref{sec:model-III}, where
$\tchi$ is an unbounded decreasing function of $\rho$, remains similar to the model (II).
To handle the infinite range of $\tilde\chi$, we split the integral (\ref{chi-V-def})
over three domains, $[0,\rho_s]$, $[\rho_s,\rho_{\alpha}]$ and $[\rho_{\alpha},+\infty[$, where
$\rho_{\alpha} = {\cal M}^4/\alpha$ is also the density scale where the model departs from
the bounded model (II) and probes the infinite tail (\ref{model-III-large-chi-rho-def}).
The first two terms are of order $\alpha^{2/3}$ as for the model (II), with the same choice
$\rho_s \sim \bar\rho \alpha^{-2/3}$.
Using the Cauchy-Schwarz inequality
$\int_{\rho_{\alpha}}^{\infty} d\rho {\cal P}(\rho) \sqrt{\rho} \leq \bar\rho/\sqrt{\rho_{\alpha}}$,
the last term is found to be of order $\alpha$ at most.

Therefore, in all cases we have
$| \langle \tilde\chi \rangle - \tilde\chi(0) | \lesssim \alpha^{2/3} \ll 1$ and the small-scale
nonlinearities do not produce a significant backreaction onto the overall expansion
rate of the Universe in the dark energy era.

\section{Linear perturbations}
\label{sec:Linear-perturbations}

\subsection{Regime of validity}
\label{sec:linear-validity}

We study in more details the growth of large-scale structures at linear order in this section.
We first investigate the regime of validity of the linear theory.
The standard cosmological linear theory applies to large scales where the matter density
fluctuations $\delta$ are small. This yields the transition scale to nonlinearity
$x_{\delta}^{\rm NL}$ defined by $\sigma^2(x_{\delta}^{\rm NL})=1$, where $\sigma(x)$
is the root-mean-square (rms) density contrast at scale $x$, $\sigma^2=\langle \delta_x^2 \rangle$.
In addition to the perturbative expansion in $\delta$, within the context of the scalar-field
models that we study in this paper the perturbative approach involves an additional expansion
in the scalar field fluctuation $\delta\tilde\chi$.
Then, it could happen that this second expansion has a smaller range of validity,
$x_{\tilde\chi}^{\rm NL}$, so that linear theory applies to a smaller range than in the
$\Lambda$-CDM cosmology.

Therefore, we need to investigate the range of validity of the linear regime for the fifth force.
From the Euler equation (\ref{Euler-J}) and the expression of the metric potential (\ref{Phi-Psi-J}),
the linear approximation is valid for the fifth force as long as we can linearize $\delta\ln A$ in
the density contrast $\delta$.
In Sec.~\ref{sec:linear-J} we obtained the linear regime by expanding $\ln A$ in
$\delta\tilde\chi$
and next solving for $\delta\tilde\chi$ from the scalar field equation (\ref{KG-A=1}).
However, this formulation can underestimate the range of validity of the linear regime for the fifth force.
Indeed, because of the factor $1/\rho$ in the right-hand side a perturbative expansion of
Eq.(\ref{KG-A=1})
in powers of $\delta=(\rho-\bar\rho)/\bar\rho$ cannot extend beyond $|\delta| \sim 1$.
This artificial limitation can be removed at once by writing instead the scalar-field equation as
\beq
\frac{d\tilde\chi}{d\ln A} = \frac{\rho}{{\cal M}^4} .
\label{KG-rho-linear-1}
\eeq
If the function $\tilde\chi(\ln A)$ were quadratic the linear theory would be exact for the fifth force.
In the general case, the range of validity $x_{A}^{\rm NL}$ of the linear theory for the fifth force will be
determined by the nonlinearities of the function $d\tilde\chi/d\ln A$ but it can exceed
$x_{\delta}^{\rm NL}$.

In the models described in Sec.~\ref{sec:models} we have $\ln A = \alpha \lambda(\tilde\chi)$,
where the function $\lambda$ and $\tilde\chi$ are of order unity (or more precisely, do not involve
small or large parameters), whereas $\alpha \lesssim 10^{-6} \ll 1$, as noticed in
Eq.(\ref{lambda-def}).
Then, the scalar-field equation (\ref{KG-rho-linear-1}) reads as
\beq
\frac{d\tilde\chi}{d\lambda} = \frac{\alpha\bar\rho}{{\cal M}^4} (1+\delta) ,
\label{KG-lambda-delta}
\eeq
which we must solve for $\lambda(\delta)$.
In the low-density regime, following the same analysis as for Eq.(\ref{chi-low-density}), we have
in the general case $\ln A \sim \pm \alpha |\tilde\chi+1|^{1-\nu}$, with $0<\nu<1$, and
\beq
\frac{\alpha\rho}{\cM^4} \ll 1 : \;\;\; \tilde\chi \simeq -1 \pm |\lambda|^{1/(1-\nu)} , \;\;\;
\lambda \simeq  \pm \left( \frac{\alpha\rho}{{\cal M}^4} \right)^{(1-\nu)/\nu} .
\label{lambda-rho-nu}
\eeq
Then, for generic $\nu$, at low densities the function $\lambda(\rho)$ can be linearized in
$\delta$ in the range $|\delta| \lesssim 1$.
For the specific case $\nu=1/2$, which corresponds to the models
introduced in Sec.~\ref{sec:models}, the last relation (\ref{lambda-rho-nu}) happens to be linear
so that the linear regime for $\lambda(\rho)$ applies up to $\alpha \rho / {\cal M}^4 \sim 1$,
that is, $\delta \lesssim {\cal M}^4/\alpha\bar{\rho} \sim \alpha^{-1} \gg 1$, which yields
a much greater range.
At high densities, $\alpha\rho/{\cal M}^4 \gg 1$, we have a power-law divergence of the form
$d\tilde\chi/d\lambda \sim |\lambda+\lambda_*|^{-\mu}$, with $\mu>0$.
This yields $\lambda \simeq -\lambda_* \pm (\alpha\rho/{\cal M}^4)^{-1/\mu}$, which can be
linearized for $|\delta| \lesssim 1$.

Therefore, we find that in all cases the regime of validity of the linear regime for the fifth force
is at least as broad as that for the matter fluctuations $\delta$. In the specific case of the models
introduced in in Sec.~\ref{sec:models}, which have a square-root singularity in the low-density
regime, the linear regime for the fifth force applies to a much greater range at low $z$,
$|\delta| \lesssim \alpha^{-1}$.

\subsection{Model (I)}
\label{sec:linear-(I)}

We first consider the case of the model (I) introduced in Eq.(\ref{model-I-def}), where $\tilde\chi$ is an
increasing function of $\rho$. This leads to a negative $\epsilon_1$ and the fifth force decreases
Newtonian gravity.
The linear modes $D_{\pm}(a)$ of the matter density contrast satisfy the evolution equation
(\ref{DL}), where the departure from the $\Lambda$-CDM cosmology only comes from the
factor $\epsilon(k,a)$.
Because $|\epsilon_1| \lesssim \alpha \ll 1$, the factor $1$ in Eq.(\ref{eps-def}) gives a
negligible contribution to $(1+\epsilon)$ and we can write
\beq
\epsilon(k,a) = \epsilon_1(a) \frac{2}{3\Omega_{\rm m}} \left( \frac{ck}{aH} \right)^2 .
\label{eps-k-a}
\eeq
On Hubble scales we have $\epsilon \sim \epsilon_1$, hence $|\epsilon| \lesssim \alpha \ll 1$
and we recover the $\Lambda$-CDM growth of structures.
However, on smaller scales $|\epsilon(k,a)|$ grows as $k^2$ and it reaches unity at a wave number
\beq
k_{\alpha}(a) \simeq \frac{aH}{c\sqrt{|\epsilon_1|}}
\simeq \frac{3\times 10^{-4}}{\sqrt{a|\epsilon_1|}} h \, \rm{Mpc}^{-1} ,
\label{k-alpha-def}
\eeq
where we used $H^2 \propto a^{-3}$ in the matter era.
We have seen in Sec.~\ref{sec:evolution-back-fields} and Fig.~\ref{fig_Aa_1} that $|\epsilon_1|$
is maximum at redshift $z_{\alpha} \sim \alpha^{-1/3}$, with an amplitude
$|\epsilon_1|_{\rm max} \sim \alpha$.
More precisely, from Eq.(\ref{eps1-2-I}) we obtain
\beq
a \ll a_{\alpha} : \;\;\; |\epsilon_1| \sim \alpha^{-1} a^6 , \;\;\;
a \gg a_{\alpha} : \;\;\; |\epsilon_1| \sim  \alpha^2 a^{-3} .
\label{epsilon1-I-scaling}
\eeq
Therefore, $k_{\alpha}(a)$ is minimum at $a \simeq a_{\alpha}$, with
\beq
k_{\alpha}^{\rm min} \equiv k_{\alpha}(a_{\alpha}) \sim 3\times 10^{-4} \alpha^{-2/3}
\, h \, \rm{Mpc}^{-1} ,
\label{k-alpha-min}
\eeq
which yields $k_{\alpha}^{\rm min} \sim 3 \, h \, \rm{Mpc}^{-1}$ for $\alpha=10^{-6}$.
Thus, wave numbers below $k_{\alpha}^{\rm min}$ never probe the fifth force,
while higher wave numbers feel the fifth force over a finite time range,
$[a_-(k),a_+(k)]$, around the scale factor $a_{\alpha}$.
From Eq.(\ref{epsilon1-I-scaling}) we obtain, for $k > k_{\alpha}^{\rm min}$,
\beq
a_-(k) \sim \alpha^{1/7} \left( \frac{ck}{H_0} \right)^{-2/7} , \;\;\;
a_+(k) \sim \alpha \frac{ck}{H_0} .
\label{am-ap-k}
\eeq

In the time interval $[a_-,a_+]$, the factor $(1+\epsilon)$ in the linear evolution equation
(\ref{DL}) is dominated by $\epsilon$ and becomes negative.
This means that the density fluctuations no longer feel an attractive gravity but a pressure-like
force. Then, the linear growing mode $D_+(a)$ stops growing, as in the $\Lambda$-CDM cosmology,
but develops an oscillatory behavior.
In the matter era, the evolution equation (\ref{DL}) simplifies as
\beq
D'' + \frac{1}{2} D' - \frac{3}{2} (1+\epsilon) D = 0 ,
\label{D-matter-era}
\eeq
where we denote with a prime the derivative with respect to $\ln a$.
Rescaling the linear modes as
\beq
D(k,a) = a^{-1/4} \, y(k,a) ,
\label{D-y-a-def}
\eeq
we obtain
\beq
y'' - \left( \frac{25}{16}+\frac{3\epsilon}{2} \right) y = 0 .
\eeq
Then, defining $\omega = \sqrt{-25/16-3\epsilon/2}$, we obtain in the limit $-\epsilon \gg 1$ the
WKB solutions
\beq
y = \frac{c}{\sqrt{\omega}} \cos \left[ \int_{a_-}^{a} \frac{d a}{a} \omega(a) \right]
+ \frac{s}{\sqrt{\omega}} \sin \left[ \int_{a_-}^{a} \frac{d a}{a} \omega(a) \right] ,
\label{WKB}
\eeq
where the coefficients $c$ and $s$ are obtained from the matching at $a_-$.

\begin{figure}
\begin{center}
\epsfxsize=8. cm \epsfysize=5.5 cm {\epsfbox{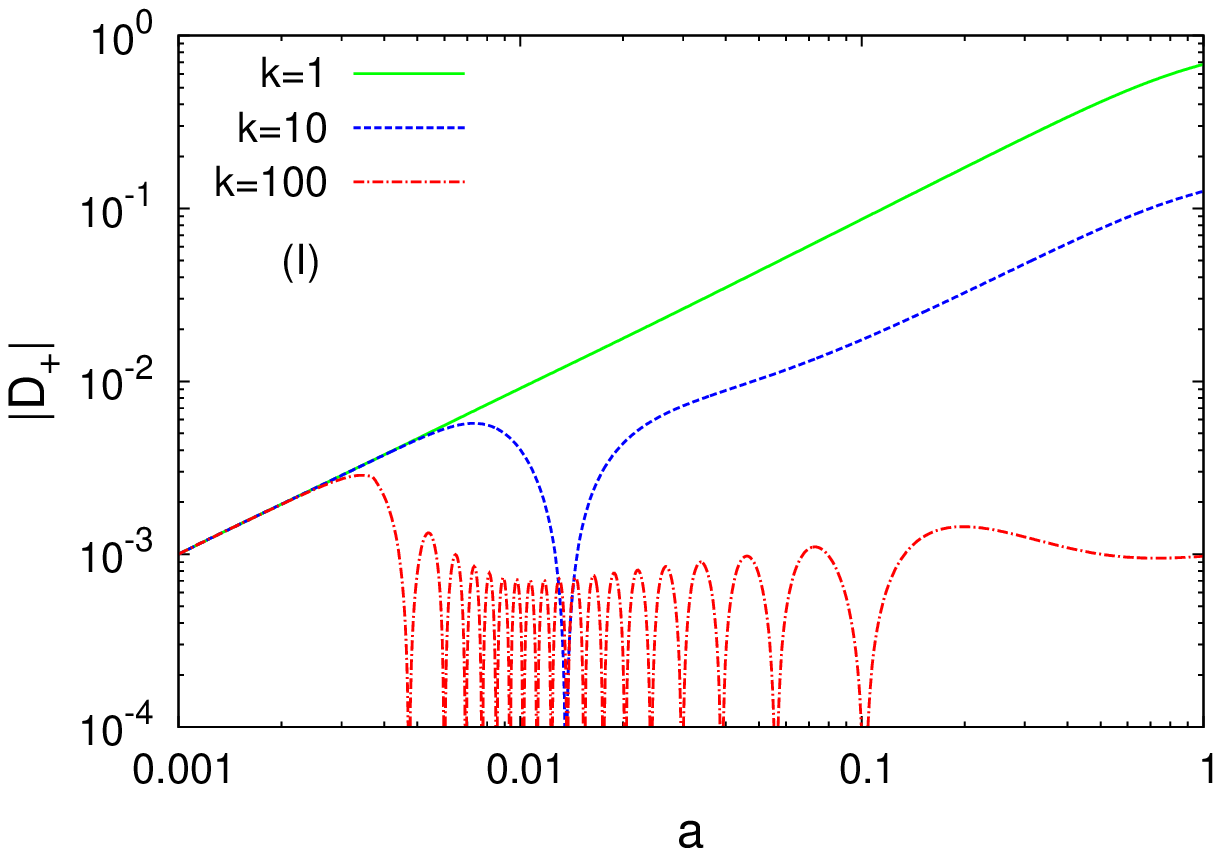}}\\
\epsfxsize=8. cm \epsfysize=5.5 cm {\epsfbox{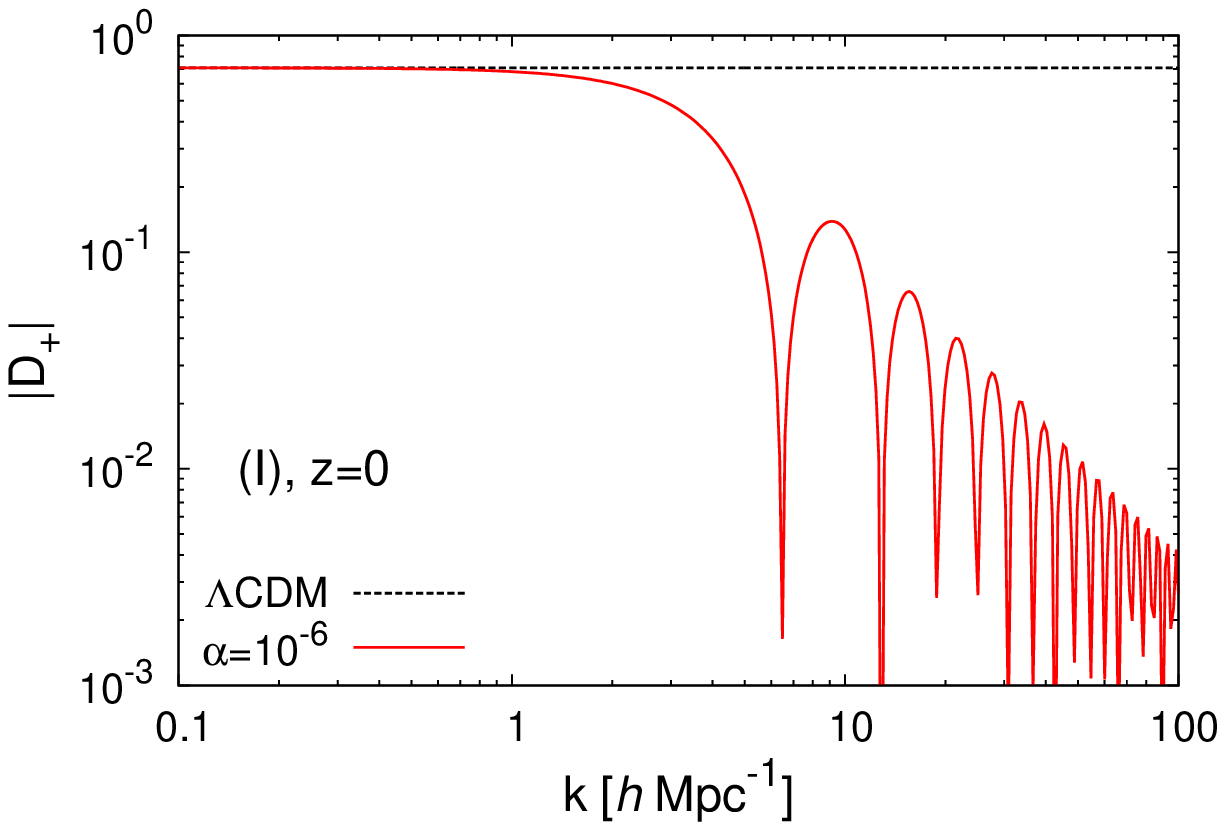}}\\
\epsfxsize=8. cm \epsfysize=5.5 cm {\epsfbox{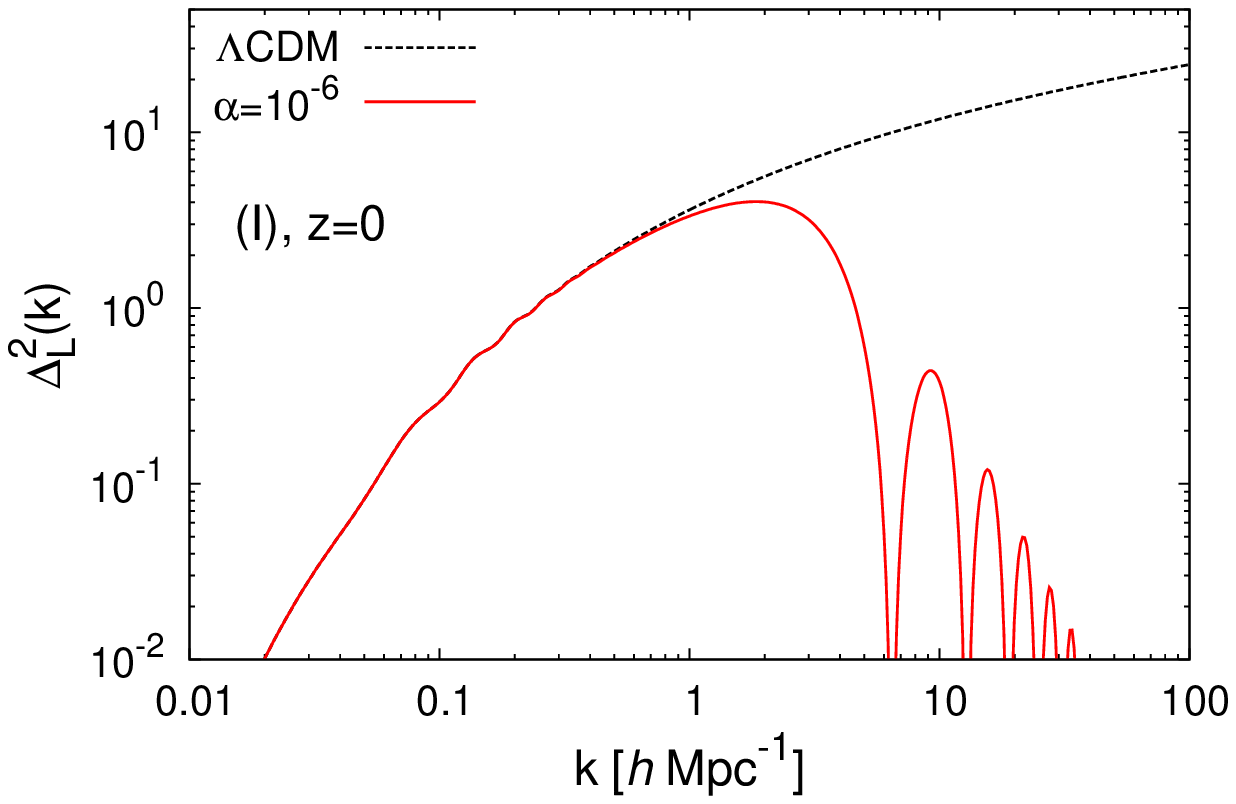}}

\end{center}
\caption{
Linear growing mode $D_+(k,a)$ and logarithmic power spectrum $\Delta^2_L(k,a)$
for the model (I) (we show the absolute value $|D_+|$).
{\it Upper panel:} $D_+(k,a)$ as a function of the scale factor
for $k=1, 10$ and $100 h\rm{Mpc}^{-1}$, from top to bottom.
{\it Middle panel:} $D_+(k,a)$ as a function of the wave number,
at redshift $z=0$. We also show the $\Lambda$-CDM result as the upper dashed line.
{\it Lower panel:} linear logarithmic power spectrum $\Delta^2_L(k,a)$ at redshift $z=0$.
}
\label{fig_Dp_I}
\end{figure}

We show the linear growing mode $D_+(k,a)$ as a function of the scale factor in the upper panel
in Fig.~\ref{fig_Dp_I}.
We can check that we recover the behaviors predicted above.
For $k \lesssim 3 h\rm{Mpc}^{-1}$ the linear growing mode follows the same growth as in the
$\Lambda$-CDM cosmology (which cannot be distinguished from the upper curve in the plot).
At higher $k$ it develops oscillations, in the range $[a_-,a_+]$ around $a_{\alpha} \sim 0.01$.
Because the number of oscillations is not very large in practice we do not need to use the WKB
approximation (\ref{WKB}) and we simply solve the exact evolution equation (\ref{DL}).

We show the dependence on $k$ of the linear growing mode in the middle panel in
Fig.~\ref{fig_Dp_I}.
The oscillatory behavior found for the time evolution at high $k$ gives rise to a decay of the growing
mode at high wave number.
Indeed, for high $k$ the linear mode $D_+(k,a)$ stops growing in the increasingly broader
interval $[a_-,a_+]$, which leads to an increasing delay for $D_+(k,a)$ as compared with the
$\Lambda$-CDM reference.
From the WKB approximation (\ref{WKB}) we can see that $a^{1/4} D_+(k,a)$ has about the
same amplitude at the boundaries $a_-$ and $a_+$, where $\omega \sim 1$, while it decreases
as $1/\sqrt{\omega(a)}$ in-between with a minimum at $a_{\alpha}$.
Therefore, in the matter era after the oscillatory phase, $a> a_+$, we have
$D_+(a) \sim D_+(a_+) a/a_+ \sim D_+(a_-) (a_+/a_-)^{-1/4} (a/a_+) \sim (a_+/a_-)^{-5/4} a$,
where we normalized the $\Lambda$-CDM growing mode as $D_+^{\Lambda\rm CDM}=a$
in the matter era. Thus, at high $k$ the linear growing mode is damped by a factor
$(a_+/a_-)^{-5/4}$.
From Eq.(\ref{am-ap-k}) this gives:
\beq
k > k_{\alpha}^{\rm min} : \;\;\; \frac{D_+(k,a)}{D_+^{\Lambda\rm CDM}(a)} \sim
\left( \frac{k}{k_{\alpha}^{\rm min}} \right)^{-45/28} ,
\label{k-damped-I}
\eeq
which is consistent with the middle panel in Fig.~\ref{fig_Dp_I}.

We display the logarithmic linear power spectrum, $\Delta^2_L(k)= 4 \pi k^3 P_L(k)$, in the
lower panel in Fig.~\ref{fig_Dp_I}.
Its ratio to the $\Lambda$-CDM linear power is given by $(D_+/D_+^{\Lambda\rm CDM})^2$
and shows a steep falloff with oscillations at high $k$, as follows from the middle panel.
The lower panel shows that at $z=0$ the decay of the linear power spectrum appears inside
the nonlinear regime, at $k \gtrsim 2 h\rm{Mpc}^{-1}$, but at higher $z$ it would fall in the linear
regime.

\subsection{Model (II)}
\label{sec:linear-(II)}

We now consider the case of model (II) introduced in Eq.(\ref{model-II-def}), where $\tilde\chi$
is a decreasing function of $\rho$. This leads to a positive $\epsilon_1$ and the fifth force
amplifies Newtonian gravity.
Again, the linear modes $D_{\pm}(a)$ satisfy the evolution equation (\ref{DL}) and the factor
$\epsilon(k,a)$ is given by Eq.(\ref{eps-k-a}). We recover the $\Lambda$-CDM growth on
Hubble scales while $\epsilon$ reaches unity at the wave number $k_{\alpha}(a)$ of
Eq.(\ref{k-alpha-def}).
The amplitude of $\epsilon_1$ verifies the same scalings (\ref{epsilon1-I-scaling}) as for model (I)
and this again defines the minimum wave number $k_{\alpha}^{\rm min}$ of
Eq.(\ref{k-alpha-min}) for which the fifth force ever had a significant impact.
For $k > k_{\alpha}^{\rm min}$ the fifth force is significant in the time interval $[a_-,a_+]$
given by Eq.(\ref{am-ap-k}).

Because $\epsilon>0$ the linear modes do not show oscillations in the range $[a_-,a_+]$ but
faster growth and decay as compared with the $\Lambda$-CDM evolution.
Neglecting the time dependence of $\epsilon$, Eq.(\ref{D-matter-era}) leads to the growing and
decaying modes
\beq
D_{\pm}(a) \sim a^{\gamma_{\pm}} , \;\;\;
\gamma_{\pm} = \frac{\pm \sqrt{25+24\epsilon} -1}{4} .
\label{D-model-II}
\eeq

\begin{figure}
\begin{center}
\epsfxsize=8. cm \epsfysize=5.5 cm {\epsfbox{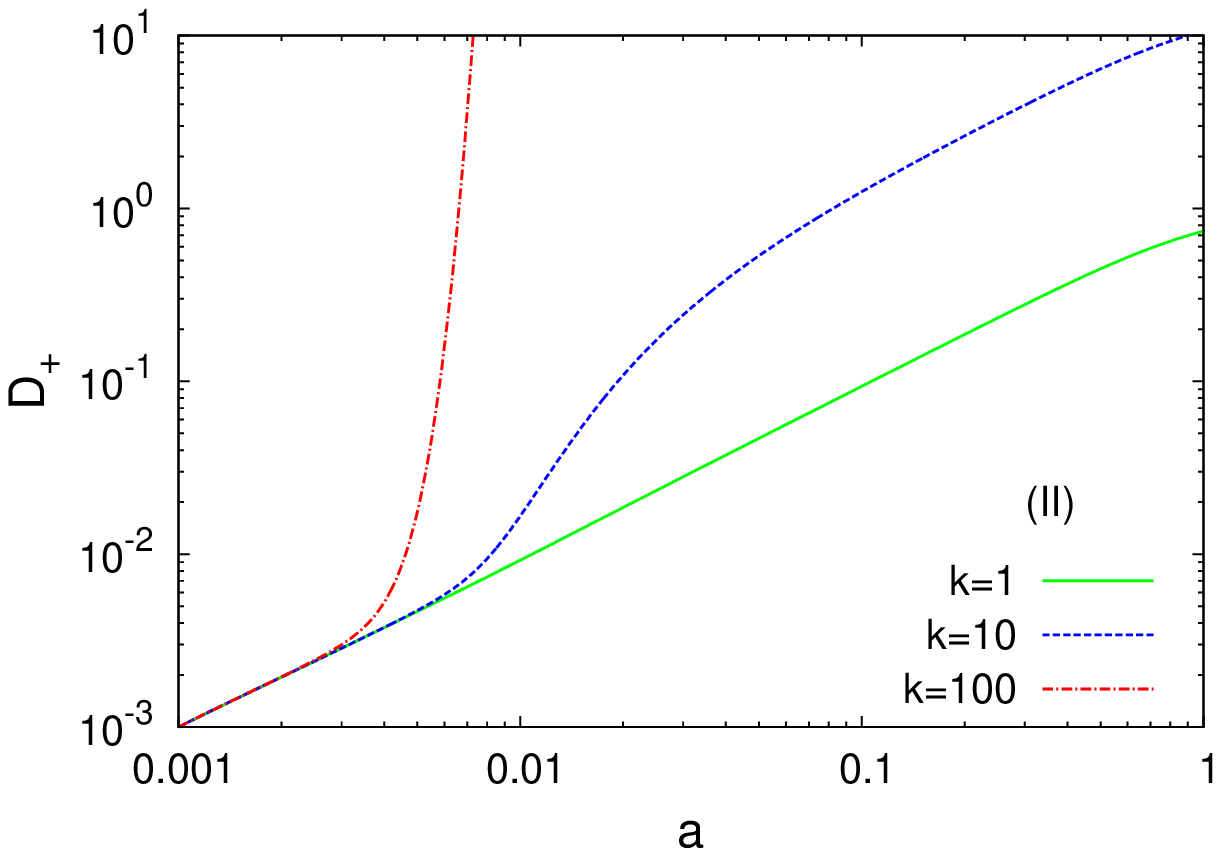}}\\
\epsfxsize=8. cm \epsfysize=5.5 cm {\epsfbox{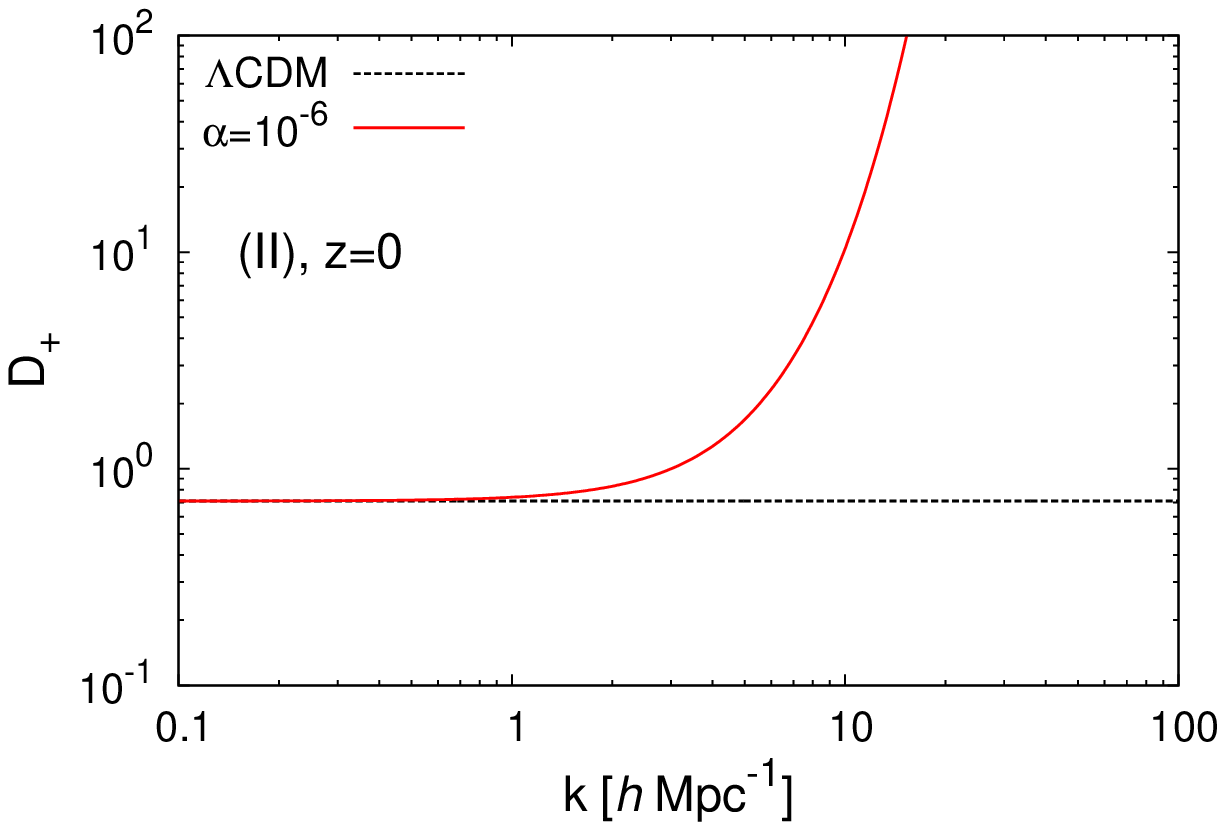}}\\
\epsfxsize=8. cm \epsfysize=5.5 cm {\epsfbox{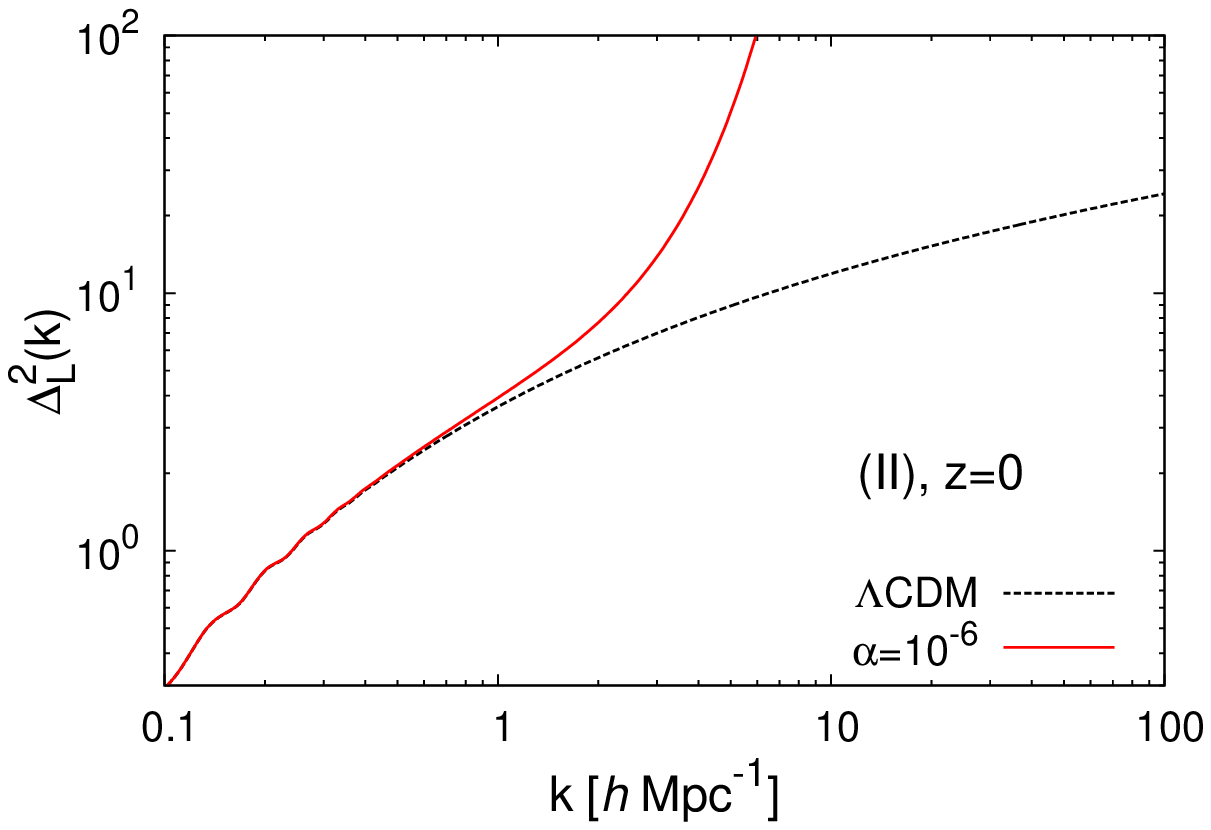}}
\end{center}
\caption{
Linear growing mode $D_+(k,a)$ and logarithmic power spectrum $\Delta^2_L(k,a)$
for the model (II).
{\it Upper panel:} $D_+(k,a)$ as a function of the scale factor
for $k=1, 10$ and $100 h\rm{Mpc}^{-1}$, from bottom to top.
{\it Middle panel:} $D_+(k,a)$ as a function of the wave number,
at redshift $z=0$. We also show the $\Lambda$-CDM result as the lower dashed line.
{\it Lower panel:} linear logarithmic power spectrum $\Delta^2_L(k,a)$ at redshift $z=0$.
}
\label{fig_Dp_II}
\end{figure}

We show our results for the linear growing mode $D_+(k,a)$ and the linear logarithmic
power spectrum $\Delta^2_L(k,a)$ in Fig.~\ref{fig_Dp_II}.
We can see that low wave numbers, $k \lesssim 1 h\rm{Mpc}^{-1}$, follow the same growth
as in the $\Lambda$-CDM cosmology whereas high wave numbers, $k \gtrsim 10 h\rm{Mpc}^{-1}$,
follow a phase of accelerated growth around $a_{\alpha} \sim 0.01$.
This leads to a steep increase of $D_+(k)$ at high $k$, at low redshift.
This means that high wave numbers, $k \gg 10 h\rm{Mpc}^{-1}$, enter the nonlinear regime
at $a \lesssim a_{\alpha}$, much before than in the $\Lambda$-CDM cosmology.
As seen in the lower panel in Fig.~\ref{fig_Dp_II}, at $z=0$ this strong amplification with respect
to $\Lambda$-CDM is restricted to nonlinear scales, but at higher $z$ it would also apply to
scales that would be linear in the $\Lambda$-CDM cosmology.

\subsection{Model (III)}
\label{sec:linear-(III)}

\begin{figure}
\begin{center}
\epsfxsize=8. cm \epsfysize=5.5 cm {\epsfbox{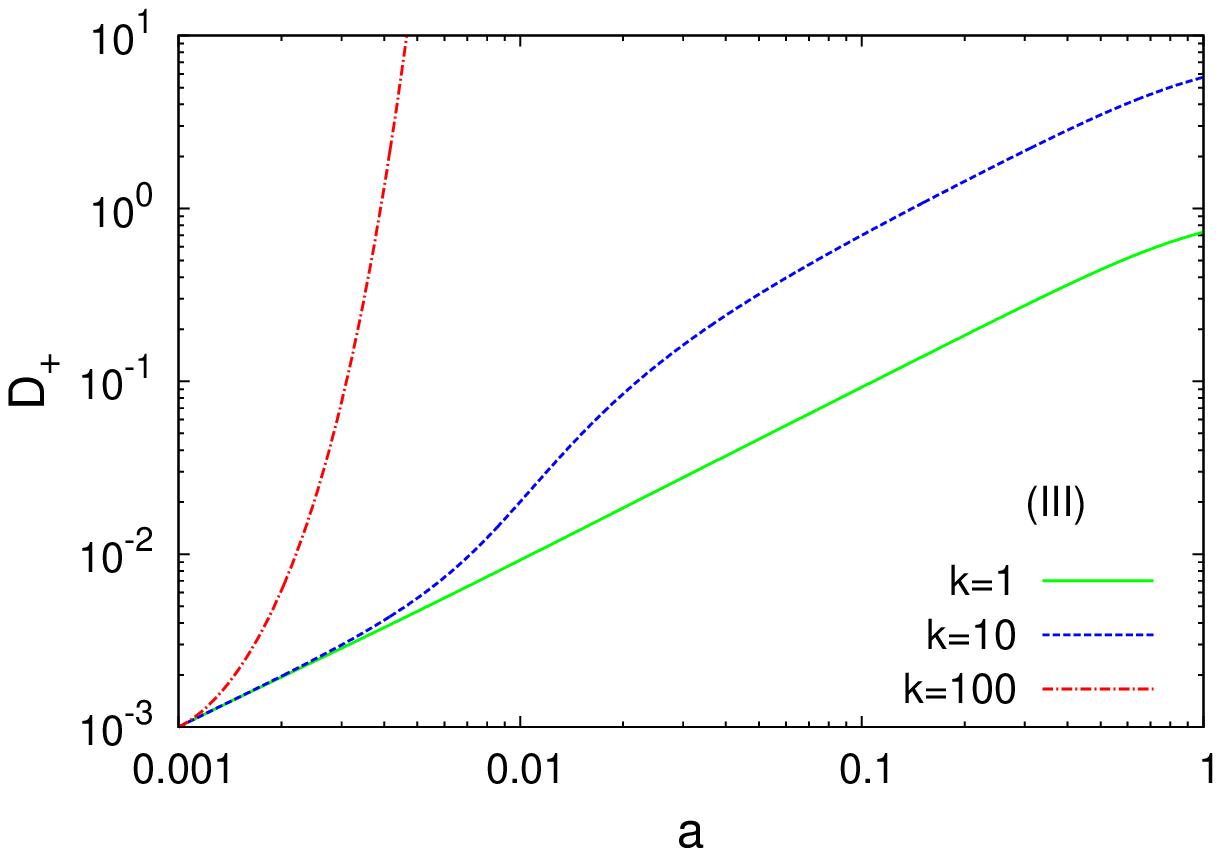}}\\
\epsfxsize=8. cm \epsfysize=5.5 cm {\epsfbox{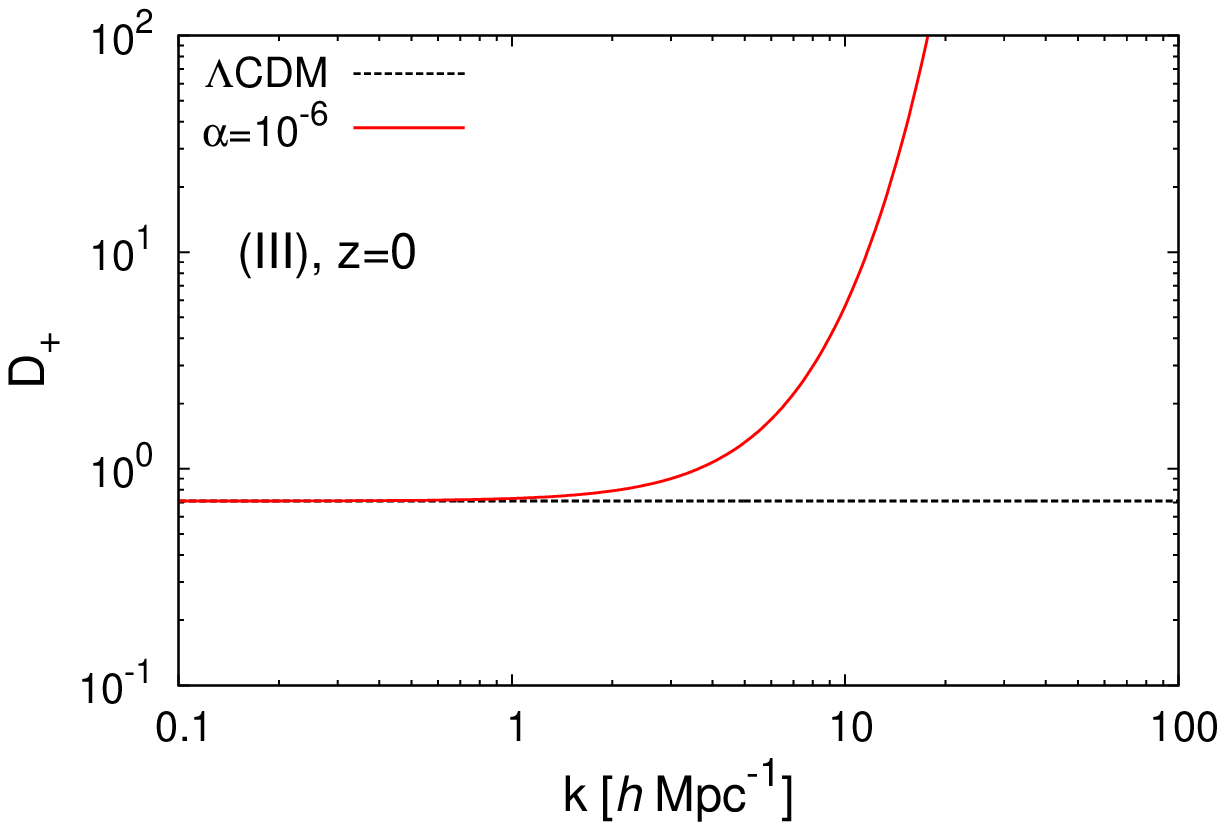}}
\end{center}
\caption{
Linear growing mode $D_+(k,a)$ for the model (III).
{\it Upper panel:} $D_+(k,a)$ as a function of the scale factor
for $k=1, 10$ and $100 h\rm{Mpc}^{-1}$, from bottom to top.
{\it Lower panel:} $D_+(k,a)$ as a function of the wave number,
at redshift $z=0$. We also show the $\Lambda$-CDM result as the lower dashed line.
}
\label{fig_Dp_III}
\end{figure}

The behaviors obtained for the model (III) are similar to those of the model (II), as $\epsilon>0$
and the fifth force accelerates the growth of structures at high $k$, in a time interval
$[a_-(k),a_+(k)]$.
At a given time, the lowest wave number $k_{\alpha}(a)$ where the fifth force is significant is still given by
Eq.(\ref{k-alpha-def}).
The lowest wave number $k_{\alpha}^{\rm min}$ where the fifth force ever played a role
(at $a_{\alpha} \sim \alpha^{1/3}$) is also given by Eq.(\ref{k-alpha-min}).
From Eqs.(\ref{eps12-model-IIIa}) and (\ref{eps12-model-III-b}) we now obtain,
for $k > k_{\alpha}^{\rm min}$,
\beq
a_-(k) \sim \alpha^{-1/5} \left( \frac{ck}{H_0} \right)^{-4/5} , \;\;\;
a_+(k) \sim \alpha \frac{ck}{H_0} .
\label{am-ap-k-III}
\eeq
Thus, the upper boundary $a_+(k)$ behaves as for models (I) and (II), because all three models
have the same low-density or late-time behavior (\ref{chi-low-density}) (up to a sign), but the
lower boundary $a_-(k)$ decreases faster at high $k$. This increases the time span where the
fifth force is dominant and it leads to a stronger impact on the growth of structure at high $k$
than for model (II).
This is due to the slower decrease of $\epsilon_1(a)$ at high redshift found in
Fig.~\ref{fig_Aa_1}.
We show our results for the linear growing mode in Fig.~\ref{fig_Dp_III} and we can check
that we recover these properties. The linear power spectrum is very close to the one
obtained from the model (II) in Fig.~\ref{fig_Dp_II}, hence we do not show it in the figure.

\section{Spherical collapse}
\label{sec:Spherical-collapse}

\subsection{Equation of motion}
\label{sec:eq-spherical}

As can be derived from Eq.\eqref{Euler-J}, on large scales where the
baryonic pressure is negligible, the particle trajectories $\vr(t)$ read as
\beq
\frac{d^2 \vr}{d t^2}  - \frac{1}{a} \frac{d^2 a}{d t^2} \vr = - \nabla_{\vr} \left( \Psi_{\rm N} + \Psi_A \right) ,
\label{trajectory-Jordan}
\eeq
where $\vr=a\vx$ is the physical coordinate, $\nabla_{\vr}=\nabla/a$ the gradient
operator in physical coordinates, and $\Psi_A= c^2 \ln A$ is the fifth force contribution to the metric potential
$\Phi$.
To study the spherical collapse before shell crossing, it is convenient to label each shell
by its Lagrangian radius $q$ or enclosed mass $M$, and to introduce its
normalized radius $y(t)$ by
\beq
y(t) = \frac{r(t)}{a(t) q} \;\;\; \mbox{with} \;\;\;
q = \left(\frac{3M}{4\pi\bar\rho_0}\right)^{1/3} , \;\;\; y(t=0) = 1 .
\label{y-def-Jordan}
\eeq
In particular, the matter density contrast within radius $r(t)$ reads as
\beq
1+ \delta_{<}(r) = y(t)^{-3} .
\label{deltaR-def}
\eeq
Then, Eq.(\ref{trajectory-Jordan}) gives for the evolution of the normalized radius $y$, or
density contrast $\delta_{<}=y^{-3}-1$,
\beq
\frac{d^2 y}{d(\ln a)^2} + \left( 2+\frac{1}{H^2} \frac{d H}{d t} \right)
\frac{d y}{d\ln a} = - \frac{y}{H^2r} \frac{\partial}{\partial r} \left( \Psi_{\rm N} + \Psi_A \right) .
\label{y-lna-Jordan}
\eeq
The Newtonian potential is given by $\Psi_{\rm N} = - {\cal{G}} \delta M / r$, with
$\delta M(<r) = 4\pi \delta_{<}(r) \bar{\rho} r^3/3$, which yields
\begin{equation}
\frac{\partial \Psi_{\rm N}}{\partial r} = \Omega_{\rm m} \frac{H^2 r}{2} \left( y^{-3} - 1 \right) .
\label{derivative-Psi-N}
\end{equation}
The derivative of the fifth force potential reads as
\begin{equation}
\frac{\partial \Psi_A}{\partial r} = c^2 \frac{\partial \ln A }{\partial r}
= \frac{c^2}{r} \frac{d \ln A}{d\ln\rho} \frac{\partial \ln \rho}{\partial \ln r} .
\label{derivative-Psi-A}
\end{equation}
This gives the equation of motion
\beqa
&& \frac{d^2 y}{d(\ln a)^2} + \left( 2+\frac{1}{H^2} \frac{d H}{d t} \right)
\frac{d y}{d\ln a} + \frac{\Omega_{\rm m}}{2} y (y^{-3} - 1) = \nonumber \\
&& - y \left( \frac{c}{Hr} \right)^2 \frac{d\ln A}{d\ln\rho}
\frac{r}{1+\delta} \frac{\partial\delta}{\partial r} .
\label{y-lna-1}
\eeqa
In contrast with the $\Lambda$-CDM case, where the dynamics of different shells are decoupled before
shell crossing, the fifth force introduces a coupling as it depends on the density profile, through the local
density $\rho(r) = \bar\rho (1+\delta(r))$ (which is different from the mean density
$\bar\rho(1+\delta_{<})$ within radius $r$) and its first derivative $\partial\delta/\partial r$.

To obtain a closed expression without solving simultaneously the dynamics of all shells (which would not
be exact at late time when inner shells collapse and cross each other), we use an ansatz for
the density profile.
Following \cite{BraxPV2012,BraxPV2013}, we use the density profile defined by
\beqa
\lefteqn{ \delta(x') = \frac{\delta_{<}(x)}{\sigma^2_{x}} \int_V \frac{d \vx''}{V} \, \xi_L(\vx',\vx'') } \nonumber \\
&& = \frac{\delta_{<}(x)}{\sigma^2_{x}} \int^{+\infty}_{0} d k \, 4 \pi k^2 P_{L}(k)
\tilde{W}(k x) \frac{\sin(kx')}{kx'} .  \;\;\;
\label{density-profile-ansatz}
\eeqa
Here $x(t)=a(t) r(t)$ is the comoving radius of the spherical shell of mass $M$ that we are interested in
while $x'$ is any radius along the profile; $\xi_L$ and $P_L$ are the linear correlation function
and power spectrum of the matter density contrast, $\sigma^2_{x}=\langle \delta_{L<}(x)^2\rangle$
its variance within radius $x$, which defines a sphere of volume $V$; and
$\tilde{W}(k x) =  3 [ \sin(kx) - kx \, \cos(kx) ]/(kx)^3$ the Fourier transform of the 3D top hat of radius $x$.
The choice (\ref{density-profile-ansatz}) corresponds to the typical density profile around a spherical
overdensity of amplitude $\delta_{L<}$ at radius $x$ for a Gaussian field of power spectrum $P_L$.
As the overdensity turns nonlinear the profile should be distorted but we neglect this effect.
The ansatz (\ref{density-profile-ansatz}) allows us to compute the local density contrast $\delta(x)$
and its derivative $\partial\delta/\partial x$ at radius $x$ from $\delta_{<}(x)$ and to close
the equation of motion (\ref{y-lna-1}).

\subsection{Model (I)}
\label{sec:spher-I}

\begin{figure}
\begin{center}
\epsfxsize=8. cm \epsfysize=5.5 cm {\epsfbox{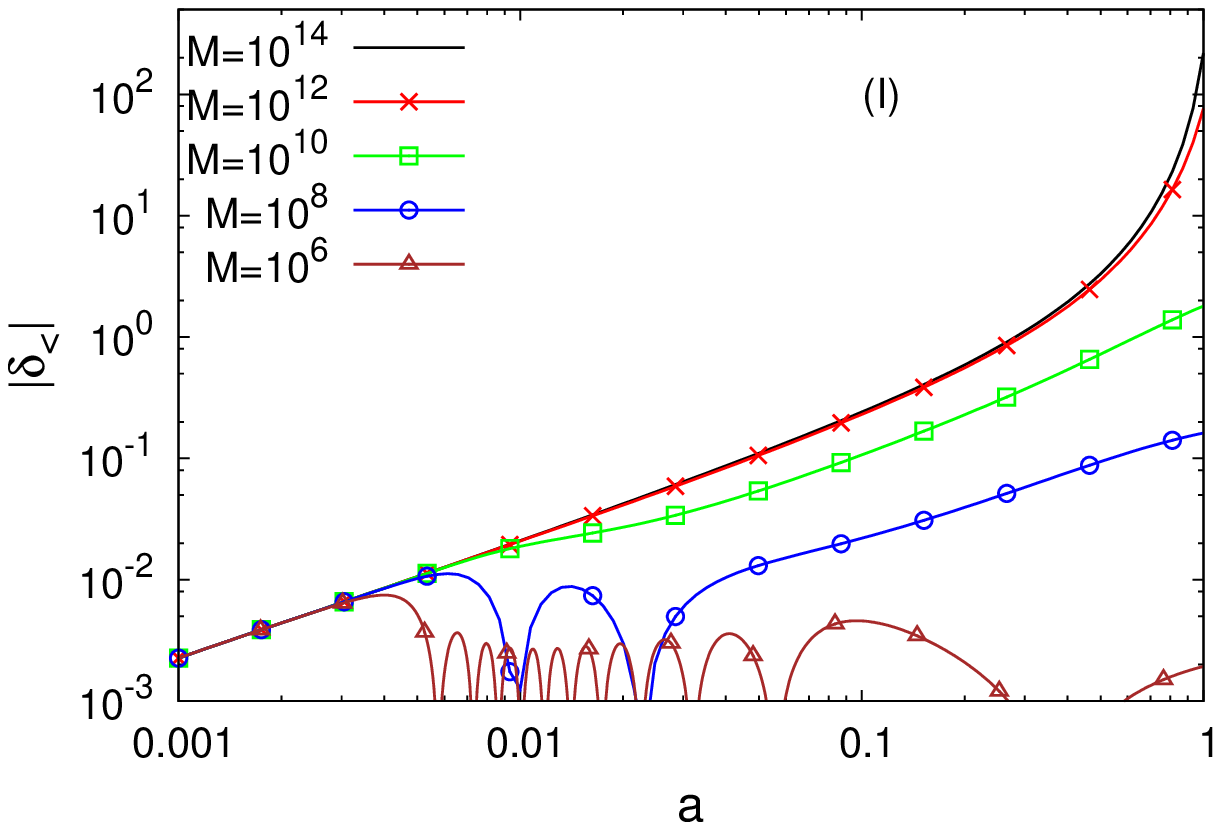}}\\
\epsfxsize=8. cm \epsfysize=5.5 cm {\epsfbox{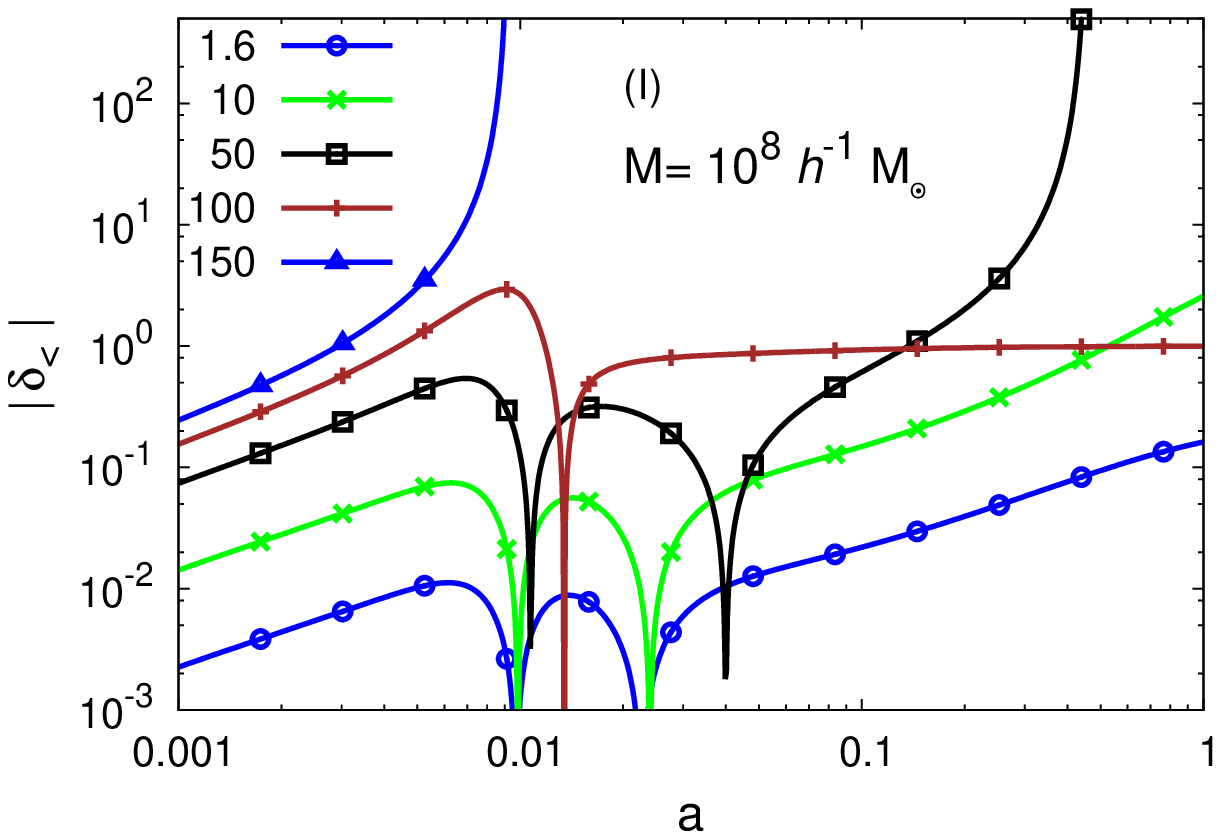}}
\end{center}
\caption{
Time evolution of the nonlinear density contrast $\delta_{<}$ given by the spherical dynamics,
as a function of the scale factor $a$.
{\it Upper panel:} $\delta_{<}(a)$ for several masses, from $M=10^6$ to $10^{14} h^{-1} M_{\odot}$
from bottom to top, with the same initial condition that corresponds to the $\Lambda$-CM
linear density threshold today $\delta_{<L}^{\Lambda \rm -CDM}=1.6$.
{\it Lower panel:} $\delta_{<}(a)$ for several initial conditions, from $\delta_{<L}^{\Lambda \rm -CDM}=1.6$
to $150$ from bottom to top, for the fixed mass $M=10^8 h^{-1} M_{\odot}$.
}
\label{fig_delta_I}
\end{figure}

We show in Fig.~\ref{fig_delta_I} the evolution with time of the nonlinear density contrast within
a shell of mass $M$ given by the spherical dynamics, for the model (I).
In the upper panel, we consider the curves obtained for different masses $M$ with a common
normalization for the linear density contrast $\delta_{<L}$ at a very high redshift, $z_i \gtrsim 10^{3}$.
In the case of the $\Lambda$-CDM cosmology, this corresponds to a linear density contrast today,
at $z=0$, of $\delta_{<L}^{\Lambda \rm -CDM}=1.6$, and to a nonlinear density contrast
$\delta_{<} \simeq 200$, hence to a collapsed and just-virialized halo.
In agreement with the results of Sec.~\ref{sec:linear-(I)} and Fig.~\ref{fig_Dp_I},
we find that for large masses, $M \gtrsim 10^{12} h^{-1} M_{\odot}$, which correspond to large scales,
we remain close to the $\Lambda$-CDM behavior (which cannot be distinguished from the
curves for $M \geq 10^{14} h^{-1} M_{\odot}$), whereas the collapse is delayed for smaller masses.
Because the density contrast is still in the linear regime around $a_{\alpha} \sim 0.01$, where
the fifth force is important (on small scales), the spherical dynamics follows the behavior of the
linear growing mode displayed in Fig.~\ref{fig_Dp_I}.
For large mass it keeps growing as in the $\Lambda$-CDM scenario whereas for small mass it
shows oscillations with an amplitude that is about the same at the end of the oscillatory phase,
$a_+$, as at its beginning, $a_-$. This delays the collapse for small masses and leads to a density
contrast today that is much smaller than $200$.
In fact, because the oscillations imply a change of sign of the density contrast, as was the case
for the linear mode $D_+$, an initially overdense perturbation can come out of the oscillatory phase
as an underdensity, in which case it will never collapse but give rise to a void (neglecting shell crossing).

In the lower panel of Fig.~\ref{fig_delta_I}, we show the spherical dynamics for the fixed mass
$M=10^8 h^{-1} M_{\odot}$ and several values of the initial linear density contrast, which in
the $\Lambda$-CDM cosmology would give rise today to a linear density contrast of
$\delta_{<L}^{\Lambda \rm -CDM}=1.6$ to $150$.
For $\delta_{<L}^{\Lambda \rm -CDM} \lesssim 10$ the dynamics remains in the linear regime until
$z=0$ and the curves are simply a rescaled copy of the result obtained for
$\delta_{<L}^{\Lambda \rm -CDM}=1.6$.
As $\delta_{<}$ only shows two changes of sign (for $M=10^8 h^{-1} M_{\odot}$ the oscillation frequency
is still low) the perturbation comes out of the oscillatory phase as an overdensity, which then resumes
its growth.
Because of the delay of the collapse around $a_{\alpha} \sim 0.01$ the final nonlinear density contrast
does not go much beyond unity at $z=0$.
For the greater initial density contrast $\delta_{<L}^{\Lambda \rm -CDM}=50$,
the overdensity has a higher amplitude at the beginning of the oscillatory phase. It exits with a similar
and positive density contrast and it has time to reach a nonlinear density contrast greater than $200$
before $z=0$.
However, we can see from the curve obtained for $\delta_{<L}^{\Lambda \rm -CDM}=100$
that the final nonlinear density contrast is not a monotonic function of the initial
condition at high initial overdensities, $\delta_{<L}^{\Lambda \rm -CDM}>50$.
Indeed, for the higher initial density $\delta_{<L}^{\Lambda \rm -CDM}=100$
there is a single oscillation,
which implies that the perturbation becomes an underdensity with a nonlinear density contrast
that converges to $-1$ at late time, when the fifth force no longer plays a significant role.
Increasing further the initial density contrast, $\delta_{<L}^{\Lambda \rm -CDM} \gtrsim 150$,
the perturbation collapses before the oscillatory phase and remains highly overdense.

For smaller masses, where the oscillatory phase shows numerous oscillations in the linear regime,
we obtain a similarly non-monotonic behavior as a function of the initial
condition.
In these cases, to obtain a collapsed halo today the overdensity needs to have already collapsed
before the oscillatory phase begins, which leads to a much more stringent condition than for
the $\Lambda$-CDM cosmology. The linear density contrast today, extrapolated from very
early times by the $\Lambda$-CDM growth factor, needs to be greater than about $100$ today,
instead of about $1.6$.

\begin{figure}
\begin{center}
\epsfxsize=8. cm \epsfysize=5.5 cm {\epsfbox{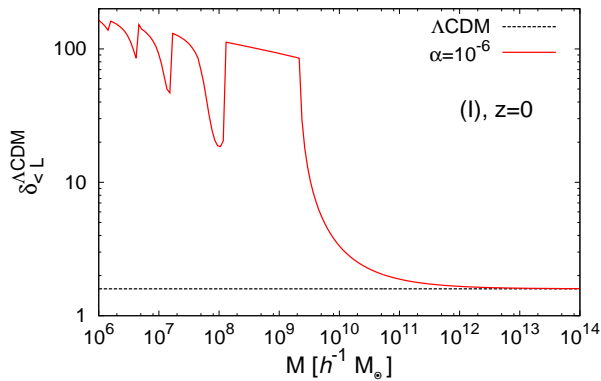}}
\end{center}
\caption{
Initial linear density contrast, as measured by $\delta_{<L}^{\Lambda \rm -CDM}$, that gives rise
to a nonlinear density contrast $\delta_{<}=200$ at $z=0$, as a function of the halo mass $M$.
The lower dashed line is the mass-independent linear density threshold obtained for the
$\Lambda$-CDM cosmology.
}
\label{fig_deltaL_I}
\end{figure}

We show in Fig.~\ref{fig_deltaL_I} the linear density contrast threshold, measured by
$\delta_{<L}^{\Lambda \rm -CDM}$ (i.e., the extrapolation up to $z=0$ of the linear initial
density contrast by the $\Lambda$-CDM growth rate), required to reach a nonlinear density
contrast $\delta_{<}=200$ today.
In  agreement with Fig.~\ref{fig_delta_I}, at large mass we recover the $\Lambda$-CDM
linear density threshold, $\delta_{<L}^{\Lambda \rm -CDM} \simeq 1.6$, whereas at small
mass we obtain a much greater linear density threshold $\delta_{<L}^{\Lambda \rm -CDM} \sim 100$.
We also find a non-monotonic curve, which is due to the oscillation phase and
the complex behavior found in Fig.~\ref{fig_delta_I}.
Moving towards smaller masses, from $M \sim 10^{11}$ down to $M \sim 2 \times 10^9 h^{-1} M_{\odot}$,
the linear density threshold shows a steep rise as it must compensate for the delay around
$a_{\alpha} \sim 0.01$ of structure growth
(this corresponds to the curve $M=10^{10}$ in the upper panel in Fig.~\ref{fig_delta_I}).
The threshold grows until $\delta_{<L}^{\Lambda \rm -CDM} \sim 100$ at
$M \sim 2 \times 10^9 h^{-1} M_{\odot}$, which corresponds to perturbations that have collapsed
just at $a_-$, just before the beginning of the oscillatory phase
(this behavior corresponds to the curve labeled ''150'' in the lower panel in Fig.~\ref{fig_delta_I}).
Next, down to $M \sim 2 \times 10^8 h^{-1} M_{\odot}$ the linear density threshold keeps slowly
increasing as the oscillatory phase expands and $a_-$ decreases
(see the upper panel in Fig.~\ref{fig_delta_I}).
At these masses the oscillatory phase displays a zero and next one change of sign
(so that overdensities emerge as underdensities and never collapse, as for the curve
labeled ``100'' in the lower panel in Fig.~\ref{fig_delta_I}).
At $M \sim 2 \times 10^8 h^{-1} M_{\odot}$ there is a sudden drop in the linear density threshold.
This is because the oscillatory phase now shows two changes of sign, and it is possible for overdensities
that have not yet collapsed before $a_-$ to emerge as overdensities and resume their collapse
(this corresponds to the curve labeled ``50'' in the lower panel in Fig.~\ref{fig_delta_I}).
Moving to lower masses the linear threshold smoothly increases as $a_-$ decreases
(so that the delay grows) until we again reach the plateau around $\sim 100$, and next encounter
a second drop at $M \sim 2 \times 10^7 h^{-1} M_{\odot}$ at the transition from three to four changes
of sign.
The second drop is smaller than the first, because the width of the oscillatory phase has increased
so that it needs a higher initial linear density contrast to eventually reach $\delta_{<}=200$ today.

In any case, the formation of low mass halos, $M \lesssim 2 \times 10^9 h^{-1} M_{\odot}$,
is strongly suppressed as compared with the $\Lambda$-CDM scenario.
In fact, rather than forming in the usual bottom-up hierarchical fashion of CDM models,
low-mass halos may form later in a top-down fashion, by fragmentation of larger-mass halos,
as in Warm Dark Matter (WDM) scenarios.

\subsection{Model (II)}
\label{sec:spher-II}

\begin{figure}
\begin{center}
\epsfxsize=8. cm \epsfysize=5.5 cm {\epsfbox{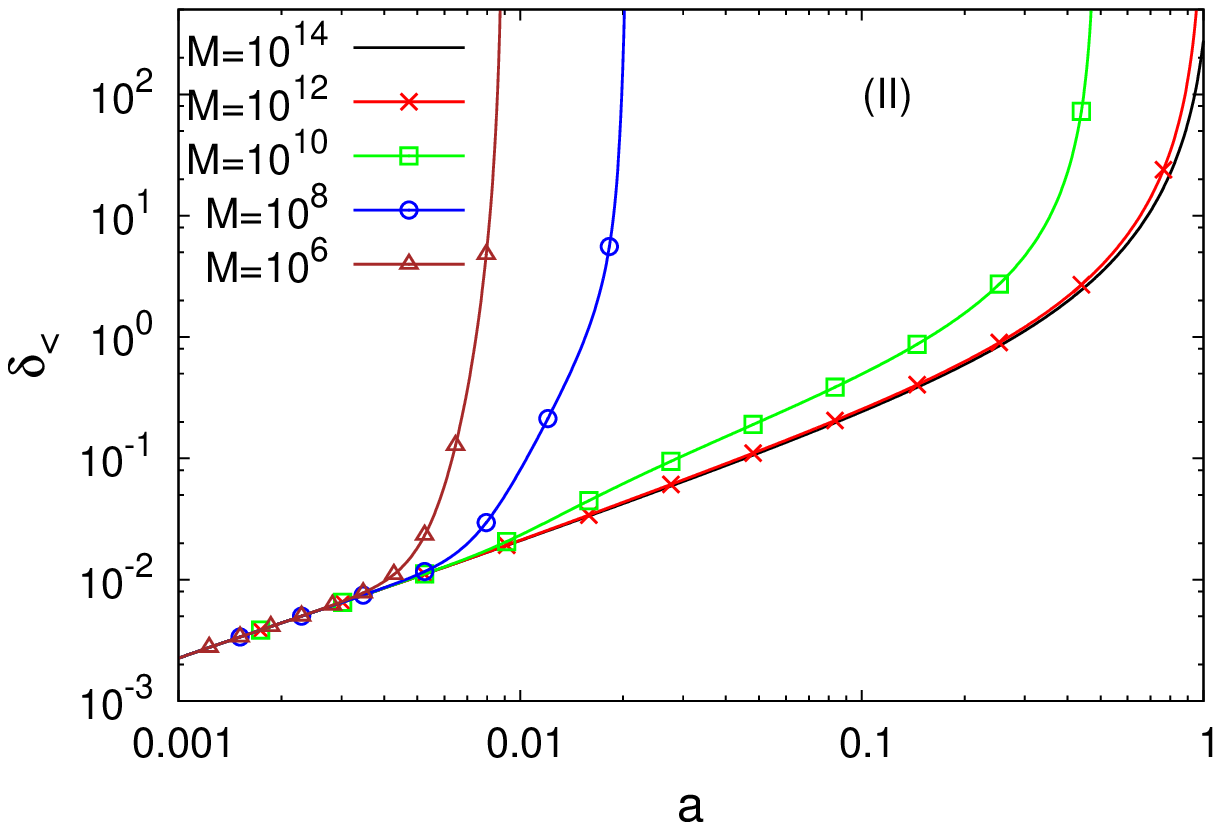}}\\
\epsfxsize=8. cm \epsfysize=5.5 cm {\epsfbox{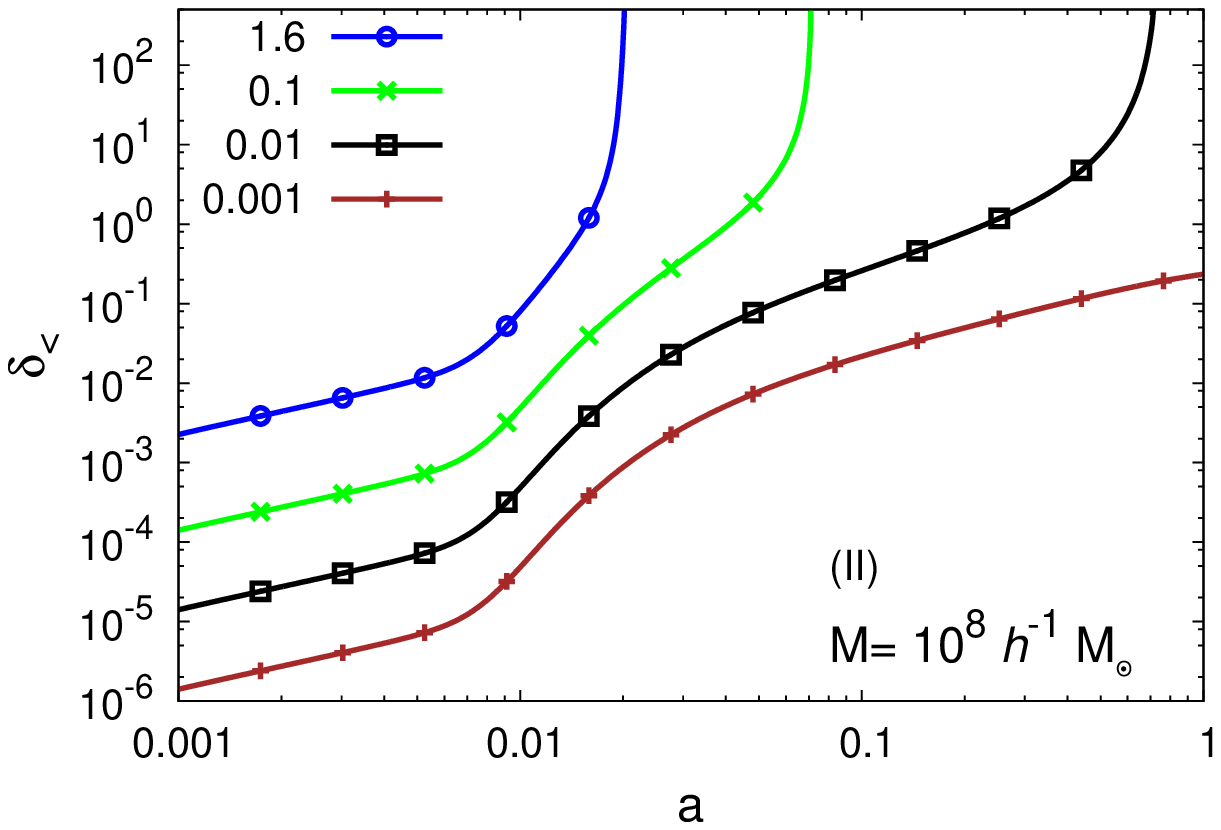}}
\end{center}
\caption{
Time evolution of the nonlinear density contrast $\delta_{<}$ given by the spherical dynamics,
as a function of the scale factor $a$.
{\it Upper panel:} $\delta_{<}(a)$ for several masses, from $M=10^{14}$ to $10^6 h^{-1} M_{\odot}$
from bottom to top, with the same initial condition that corresponds to the $\Lambda$-CM
linear density threshold today $\delta_{<L}^{\Lambda \rm -CDM}=1.6$.
{\it Lower panel:} $\delta_{<}(a)$ for several initial conditions, from $\delta_{<L}^{\Lambda \rm -CDM}=1.6$
to $0.001$ from top to bottom, for the fixed mass $M=10^8 h^{-1} M_{\odot}$.
}
\label{fig_delta_II}
\end{figure}

We show in Fig.~\ref{fig_delta_II} the evolution with time of the nonlinear density contrast within
a shell of mass $M$ given by the spherical dynamics, for the model (II).
As in Fig.~\ref{fig_delta_I}, in the upper panel, we consider the curves obtained for different masses $M$
with a common normalization for the linear density contrast $\delta_{<L}$ at a very high redshift,
$z_i \gtrsim 10^{3}$.
In the case of the $\Lambda$-CDM cosmology, this corresponds to a linear density contrast today,
at $z=0$, of $\delta_{<L}^{\Lambda \rm -CDM}=1.6$, and to a nonlinear density contrast
$\delta_{<} \simeq 200$, hence to a collapsed and just virtualized halo.
In agreement with the results of Sec.~\ref{sec:linear-(II)} and Fig.~\ref{fig_Dp_II},
we find that for large masses, $M \gtrsim 10^{12} h^{-1} M_{\odot}$, which correspond to large scales,
we remain close to the $\Lambda$-CDM behavior (which cannot be distinguished from the
curves for $M \geq 10^{14} h^{-1} M_{\odot}$), whereas the collapse is accelerated for smaller masses
and can occur as soon as $a \sim 0.01$.

In the lower panel of Fig.~\ref{fig_delta_II}, we show the spherical dynamics for the fixed mass
$M=10^8 h^{-1} M_{\odot}$ and several values of the initial linear density contrast, which in
the $\Lambda$-CDM cosmology would give rise today to a linear density contrast of
$\delta_{<L}^{\Lambda \rm -CDM}=1.6$ to $0.001$.
We can clearly see the accelerated growth during the phase, $a_-<a<a_+$, where the fifth force
is important. This implies that linear density contrasts as low as
$\delta_{<L}^{\Lambda \rm -CDM} \simeq 0.05$ can give rise to a collapsed halo today.

\begin{figure}
\begin{center}
\epsfxsize=8. cm \epsfysize=5.5 cm {\epsfbox{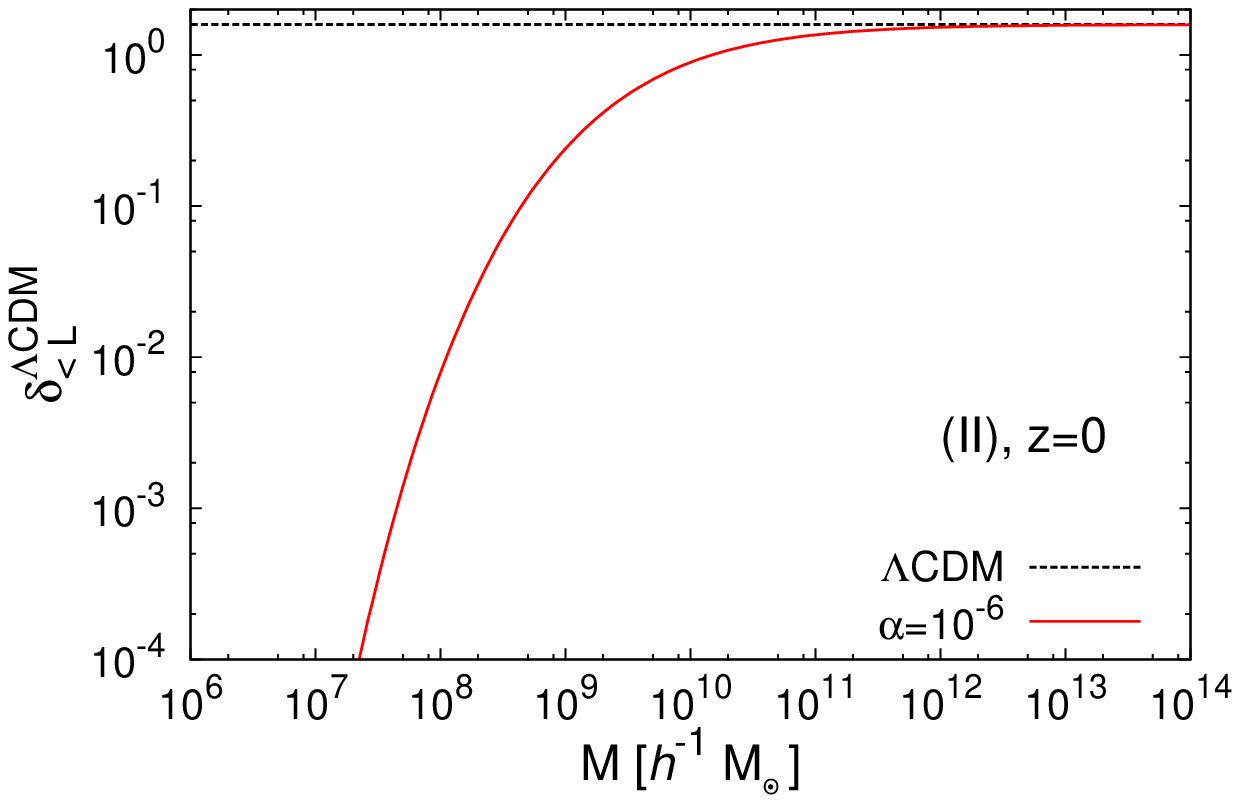}}
\end{center}
\caption{
Initial linear density contrast, as measured by $\delta_{<L}^{\Lambda \rm -CDM}$, that gives rise
to a nonlinear density contrast $\delta_{<}=200$ at $z=0$, as a function of the halo mass $M$.
The upper dashed line is the mass-independent linear density threshold obtained for the
$\Lambda$-CDM cosmology.
}
\label{fig_deltaL_II}
\end{figure}

We show in Fig.~\ref{fig_deltaL_II} the linear density contrast threshold, measured by
$\delta_{<L}^{\Lambda \rm -CDM}$ (i.e., the extrapolation up to $z=0$ of the linear initial
density contrast by the $\Lambda$-CDM growth rate), required to reach a nonlinear density
contrast $\delta_{<}=200$ today.
In  agreement with Fig.~\ref{fig_delta_II}, at large mass we recover the $\Lambda$-CDM
linear density threshold, $\delta_{<L}^{\Lambda \rm -CDM} \simeq 1.6$, whereas at small
mass we obtain a much smaller linear density threshold $\delta_{<L}^{\Lambda \rm -CDM} \ll 1$.
This means that small scales have turned nonlinear at $a_-(M) \lesssim 0.01$, much before
than in the $\Lambda$-CDM cosmology.

\subsection{Model (III)}
\label{sec:spher-III}

\begin{figure}
\begin{center}
\epsfxsize=8. cm \epsfysize=5.5 cm {\epsfbox{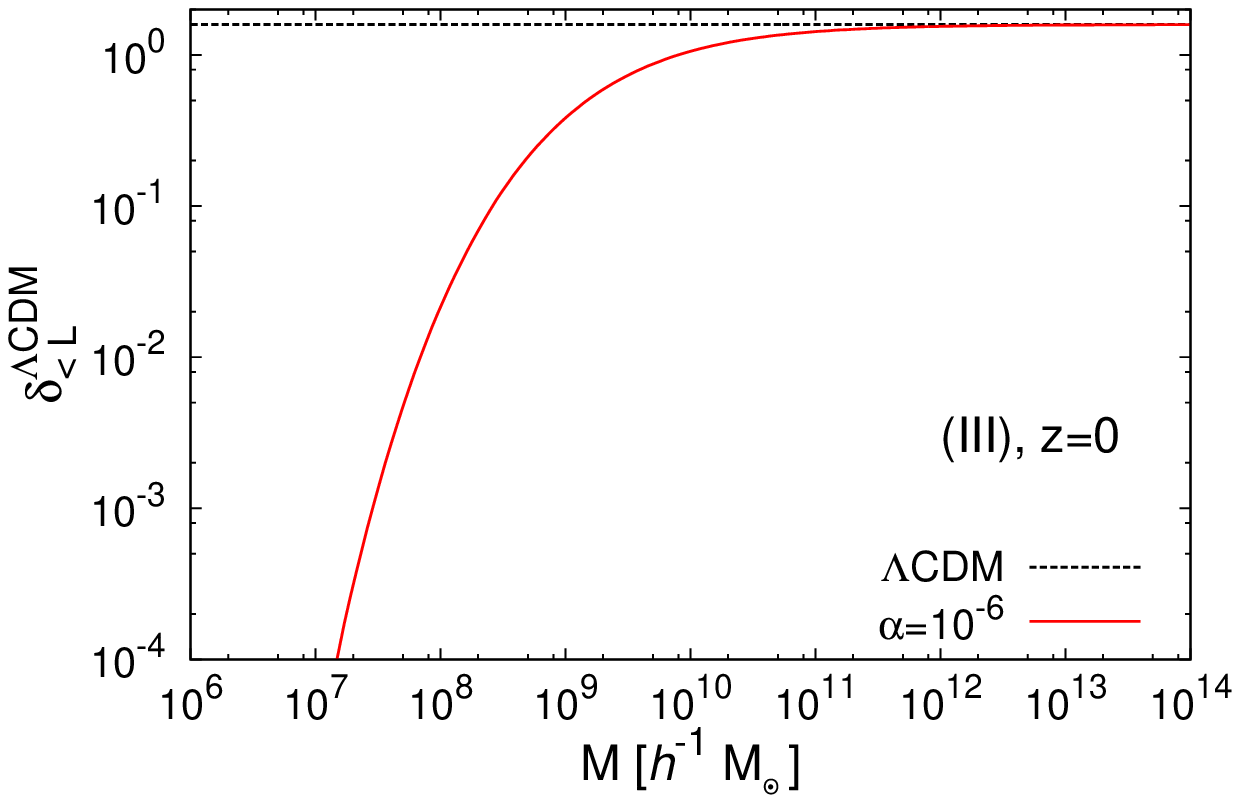}}
\end{center}
\caption{
Initial linear density contrast, as measured by $\delta_{<L}^{\Lambda \rm -CDM}$, that gives rise
to a nonlinear density contrast $\delta_{<}=200$ at $z=0$, as a function of the halo mass $M$.
The upper dashed line is the mass-independent linear density threshold obtained for the
$\Lambda$-CDM cosmology.
}
\label{fig_deltaL_III}
\end{figure}

The model (III) shows a behavior that is very close to the model (II), as was the case
for the linear growing modes studied in Sec.~\ref{sec:linear-(III)}.
Therefore, we only show the linear density threshold $\delta_{<L}^{\Lambda \rm -CDM}$ required
for the nonlinear density contrast $\delta_{<}=200$ at $z=0$, in Fig.~\ref{fig_deltaL_III}.
We can check that this is close to the result displayed in Fig.~\ref{fig_deltaL_II} for the
model (II).
Again, at large mass we recover the standard $\Lambda$-CDM result whereas at small mass
the accelerated growth leads to a much smaller linear threshold $\delta_{<L}^{\Lambda \rm -CDM} \ll 1$.

\section{Halo mass function}
\label{sec:Halo-mass-function}

\subsection{Model (I)}
\label{sec:nM-I}

As for the $\Lambda$-CDM cosmology, we write the comoving halo mass function as \cite{Press1974}
\beq
n(M) \frac{d M}{M} = \frac{\bar\rho_0}{M} f(\nu) \frac{d\nu}{\nu} ,
\label{nM-def}
\eeq
where the scaling variable $\nu(M)$ is defined as
\beq
\nu(M) = \frac{\delta_L^{\Lambda \rm CDM}(M)}{\sigma(M)} ,
\label{nu-def}
\eeq
and $\delta_L^{\Lambda \rm CDM}(M)$ is again the initial linear density contrast
(extrapolated up to $z=0$ by the $\Lambda$-CDM linear growth factor) that is required
to build a collapsed halo (which we define here by a nonlinear density contrast of 200 with
respect to the mean density of the Universe).
The variable $\nu$ measures whether such an initial condition corresponds to a  rare and very high
overdensity in the initial Gaussian field ($\nu \gg 1$) or to a typical fluctuation ($\nu \lesssim 1$).
In the Press-Schechter approach, we have $f(\nu) = \sqrt{2/\pi} \nu e^{-\nu^2/2}$.
Here we use the same function as in \cite{Valageas2009}.
Then, the impact of the modified gravity only arises through the linear threshold
$\delta_L^{\Lambda \rm CDM}(M)$, as we assume the same initial matter density power spectrum
as for the $\Lambda$-CDM reference at high redshift.

\begin{figure}
\begin{center}
\epsfxsize=8. cm \epsfysize=5.5 cm {\epsfbox{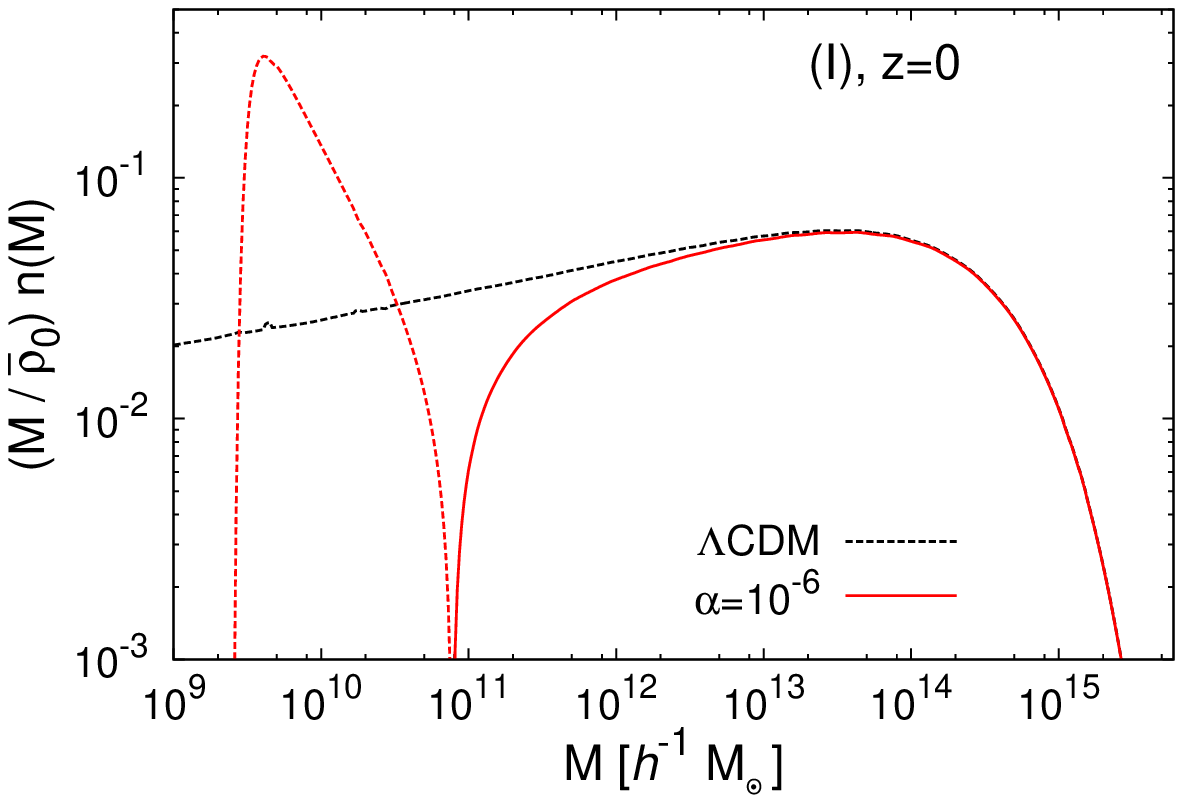}}\\
\epsfxsize=8. cm \epsfysize=5.5 cm {\epsfbox{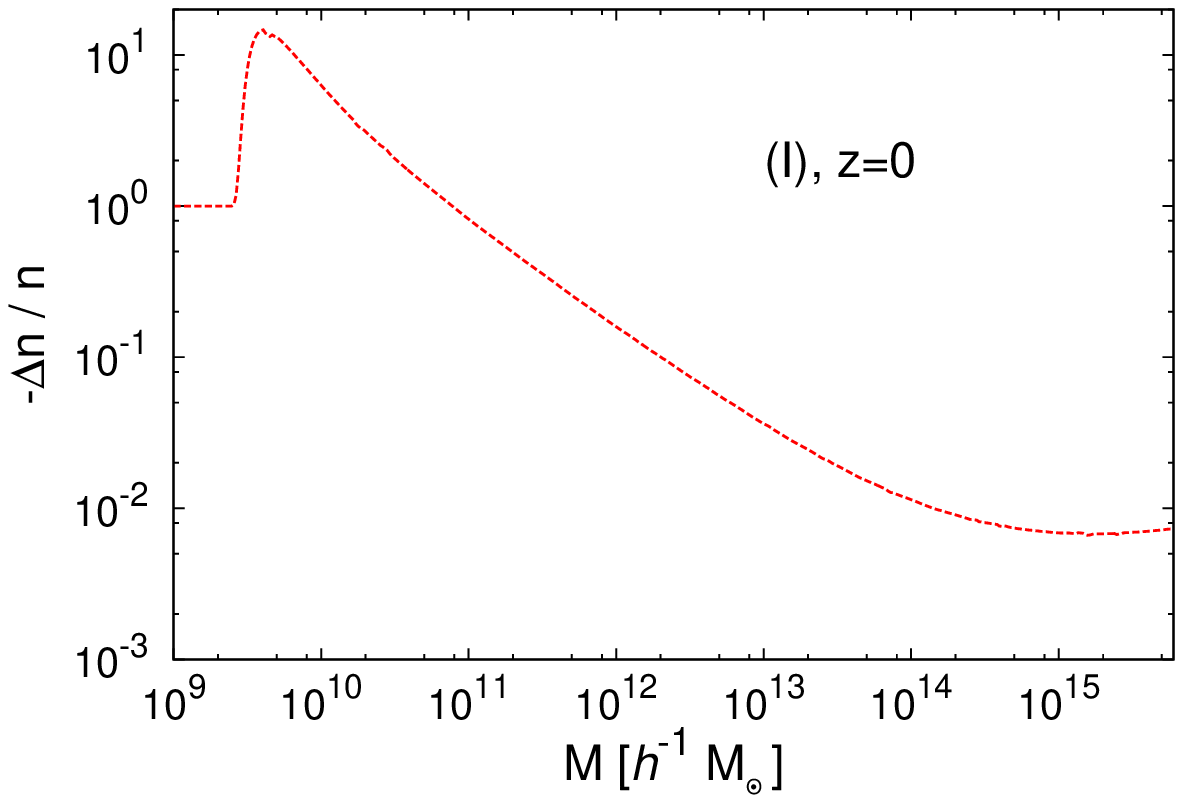}}
\end{center}
\caption{
{\it Upper panel:} halo mass function at $z=0$ for the model (I) (red line with
a downward spike at $M \simeq 10^{11}  h^{-1} M_{\odot}$)
and for the $\Lambda$-CDM reference (smooth black dashed line).
The red solid line shows the range where $n_{\rm (I)}>0$ and the red dashed line the range
where $n_{\rm (I)}<0$.
{\it Lower panel:} relative deviation of the halo mass function from the $\Lambda$-CDM reference,
for the model (I). We show $-\Delta n/n$ as $n_{\rm (I)} < n_{\Lambda\rm CDM}$.
}
\label{fig_nM_I}
\end{figure}

We show our results for the halo mass function obtained for the model (I) in Fig.~\ref{fig_nM_I}.
In agreement with Fig.~\ref{fig_deltaL_I}, at large masses the halo mass function is close to
the $\Lambda$-CDM prediction whereas it is significantly lower at low masses,
$M \sim 10^{11} - 10^{12}  h^{-1} M_{\odot}$, because of the delay of the collapse on small
scales.
In fact, at $M \simeq 10^{11}  h^{-1} M_{\odot}$ the mass function given by
Eq.(\ref{nM-def}) becomes negative.
In the usual $\Lambda$-CDM cosmology $\delta_L^{\Lambda \rm CDM}(M)$
is actually mass independent while $\sigma(M)$ is a monotonic decreasing function of $M$.
Then, $\nu(M)$ is a monotonic increasing function of $M$, which expresses the hierarchical bottom-up
nature of the gravitational clustering: smaller scales and masses collapse first.
As is well known from N-body simulations and semi-analytic modeling, this gives a mass function
that can be described by Eq.(\ref{nM-def}), which is everywhere positive with an almost universal
scaling function $f(\nu)$, and with a low-mass power-law tail and a large-mass
exponential cutoff.

In contrast, in the case of the model (I), the linear threshold $\delta_L^{\Lambda \rm CDM}(M)$
shows a strong mass dependence, as displayed in Fig.~\ref{fig_deltaL_I}.
In particular, it shows a steep increase at lower masses from $M \simeq 10^{11}  h^{-1} M_{\odot}$
down to $M \simeq 2 \times 10^9  h^{-1} M_{\odot}$.
In this range, the variable $\nu(M)$ becomes a decreasing function of mass, so that
the mass function (\ref{nM-def}) becomes negative because of the factor $d\ln\nu/d\ln M$.
At small mass the mass function $n_{\rm (I)}$ becomes very small because of the high values
reached by the linear threshold $\delta_L^{\Lambda \rm CDM}(M)$.
Of course, this negative sign merely signals the breakdown of Eq.(\ref{nM-def}) as the exact mass
function is always positive. The change of sign of $d\ln\nu/d\ln M$ means that at low mass
and small scales gravitational clustering proceeds in an inverse hierarchy: smaller scales and masses
collapse later. This corresponds to a top-down process as in the Hot Dark Matter (HDM)
scenario. In practice, we can expect that small halos form in a very different manner than in the
usual $\Lambda$-CDM cosmology, by the fragmentation of larger-mass halos.
This very different mechanism implies that the halo mass function for low masses cannot be
described by a rescaling of the form (\ref{nM-def}) and one must build a new modeling
suited to this different process. We do not pursue this task here, which would require
comparisons to numerical simulations.

It is interesting to note that this behavior is different from the modelization often used for the
Warm Dark Matter (WDM) scenario, where the formation of low-mass halos is also suppressed
as compared with the CDM scenario.
Indeed, in the WDM case, the main effect comes from a cutoff of the linear power spectrum
at high $k$, due to the free-streaming of the dark matter particles that have a non-negligible
velocity dispersion after recombination. However, at low redshift their velocity dispersion is small
(for typical candidates of particle mass $m \gtrsim 3$ keV) and the collapse proceeds as
in the usual CDM case. Then, the linear threshold $\delta_L^{\Lambda \rm CDM}(M)$
is identical to the $\Lambda$-CDM one and $\nu(M)$ is still a monotonic increasing function
of $M$, but with a smaller decrease at low mass. Typically, $\sigma$ goes to a finite constant
for $M\rightarrow 0$. This pushes Eq.(\ref{nM-def}) to its limits, and the scaling function
$f(\nu)$ may differ from the CDM one, but it remains positive and shows a reasonable shape.
However, numerical simulations suggest that this recipe overestimates the low-mass tail
and this is sometimes cured by using a window function $W(kR)$ [that defines the variance
$\sigma^2$ in Eq.(\ref{nu-def})] that is a top hat in Fourier space instead of configuration
space \cite{Schneider2013}
(but this involves introducing a free parameter to relate the wave number cutoff to the mass
scale, which is fitted to the simulations).
In contrast, in the case of the model (I), the initial linear power spectrum (that defines the initial
conditions, e.g. at $z \sim 1000$) remains the same as in the $\Lambda$-CDM cosmology, but
it is the linear threshold that is modified, because of the different dynamics around $a_-$.
This leads to a dramatic decrease of the halo mass function at low mass, without the need
to change the filter $W(kR)$, and it makes apparent the top-down hierarchy that can be
expected from the analysis of the spherical dynamics.

An alternative modeling, which is closer to the one often used for WDM, would be to define the
initial conditions at sufficiently late time, after $a_+$ when the fifth force is no longer dominant.
Then, the linear power spectrum would be modified from the $\Lambda$-CDM reference,
and given by the lower panel in Fig.~\ref{fig_Dp_I}, whereas the spherical collapse and the
linear density contrast threshold would be the same as for $\Lambda$-CDM.
However, this would hide the inverted hierarchical process [$\nu(M)$ would again be a monotonic
increasing function of mass] and would be likely to underestimate the decrease of the low-mass
tail, as in the WDM case.
In any case, a Press-Schechter-like modeling is unlikely to be meaningful in the low-mass regime
for such scenarios, and obtaining a better match with the numerical simulations by changing
the filter may not amount to much more than coincidence.

On the other hand, at large mass and in the exponential cutoff of the mass function,
where the gravitational clustering proceeds in the usual bottom-up fashion and we probe
rare events governed by the universal tail $e^{-\nu^2/2}$ associated with the Gaussian initial
conditions, we expect our results to be robust.

\subsection{Model (II)}
\label{sec:nM-II}

\begin{figure}
\begin{center}
\epsfxsize=8. cm \epsfysize=5.5 cm {\epsfbox{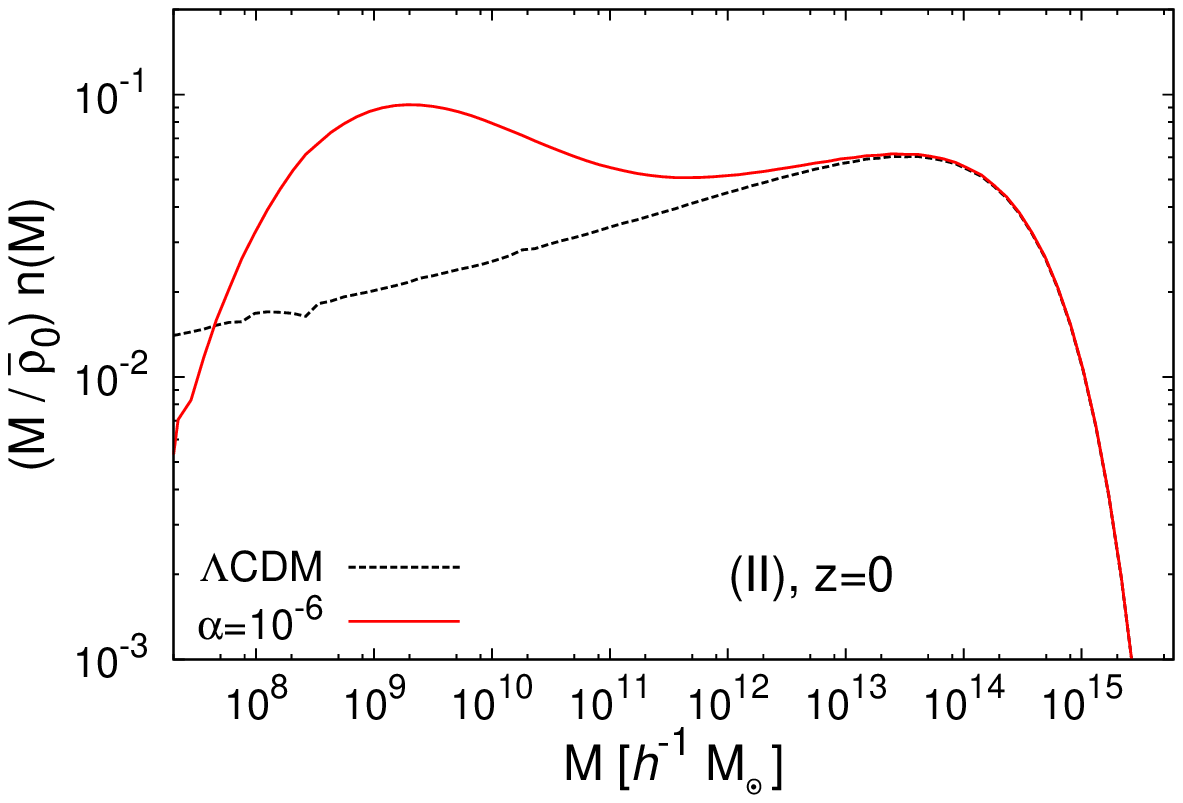}}\\
\epsfxsize=8. cm \epsfysize=5.5 cm {\epsfbox{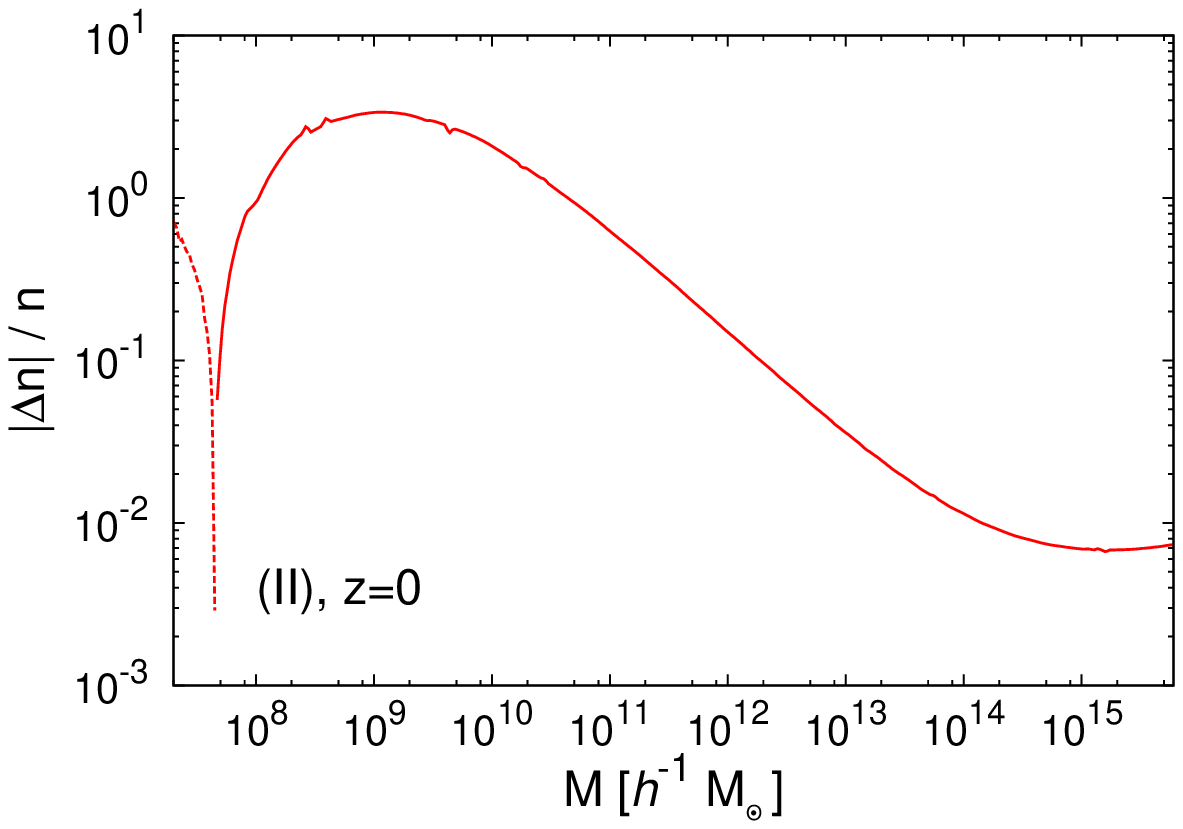}}
\end{center}
\caption{
{\it Upper panel:} halo mass function at $z=0$ for the model (II) (solid line) and the
$\Lambda$-CDM reference (dashed line).
{\it Lower panel:} relative deviation of the halo mass function from the $\Lambda$-CDM reference,
for the model (II). We show the absolute value $|\Delta n|/n$ (with a solid line for
$n_{\rm (II)}>n_{\Lambda\rm CDM}$ and a dashed line otherwise).
}
\label{fig_nM_II}
\end{figure}

We show our results for the halo mass function obtained for model (II) in Fig.~\ref{fig_nM_II}.
In agreement with Fig.~\ref{fig_deltaL_II}, at large masses the halo mass function is close to
the $\Lambda$-CDM prediction whereas it is significantly higher at low masses,
$M \sim 10^8 - 10^{11}  h^{-1} M_{\odot}$, because of the acceleration of the collapse on small
scales.
At low masses the mass function becomes smaller than in the $\Lambda$-CDM cosmology,
because both mass functions are normalized to unity (the sum over all halos cannot give more
matter than the mean matter density).

At large masses, $M > 10^{12}  h^{-1} M_{\odot}$, where the formation of large-scale
structures remains close to the $\Lambda$-CDM case, with only a modest acceleration, and
the mass function is dominated by the Gaussian tail $\sim e^{-\nu^2/2}$, we can expect the
results displayed in Fig.~\ref{fig_nM_II} to be robust.
The relative deviation does not decrease from $10^{14}$ to $10^{15}  h^{-1} M_{\odot}$
because the convergence towards $\Lambda$-CDM is counterbalanced by the Gaussian
tail $e^{-\nu^2/2}$ which increasingly amplifies deviations from $\Lambda$-CDM at high mass.

At low masses, $M < 10^{12}  h^{-1} M_{\odot}$, where the history of gravitational clustering
is significantly different from the $\Lambda$-CDM scenario, as a large range of masses have collapsed
together before a redshift of $100$, and the halo mass function is no longer dominated by its universal
Gaussian tail, these results are unlikely to be accurate. Indeed, there is no reason to expect that the
exponent of the low-$\nu$ power-law tail remains the same as in $\Lambda$-CDM, and because of the
rather different clustering history the mass function may show a significantly different behavior,
even in terms of the scaling variable $\nu$.
Nevertheless, we can still expect the halo mass function to be significantly higher than in the
$\Lambda$-CDM case for masses $M \sim 10^8 - 10^{11}  h^{-1} M_{\odot}$, although it is difficult
to predict the maximum deviation and the transition to a negative deviation at very low masses.

\subsection{Model (III)}
\label{sec:nM-III}

\begin{figure}
\begin{center}
\epsfxsize=8. cm \epsfysize=5.5 cm {\epsfbox{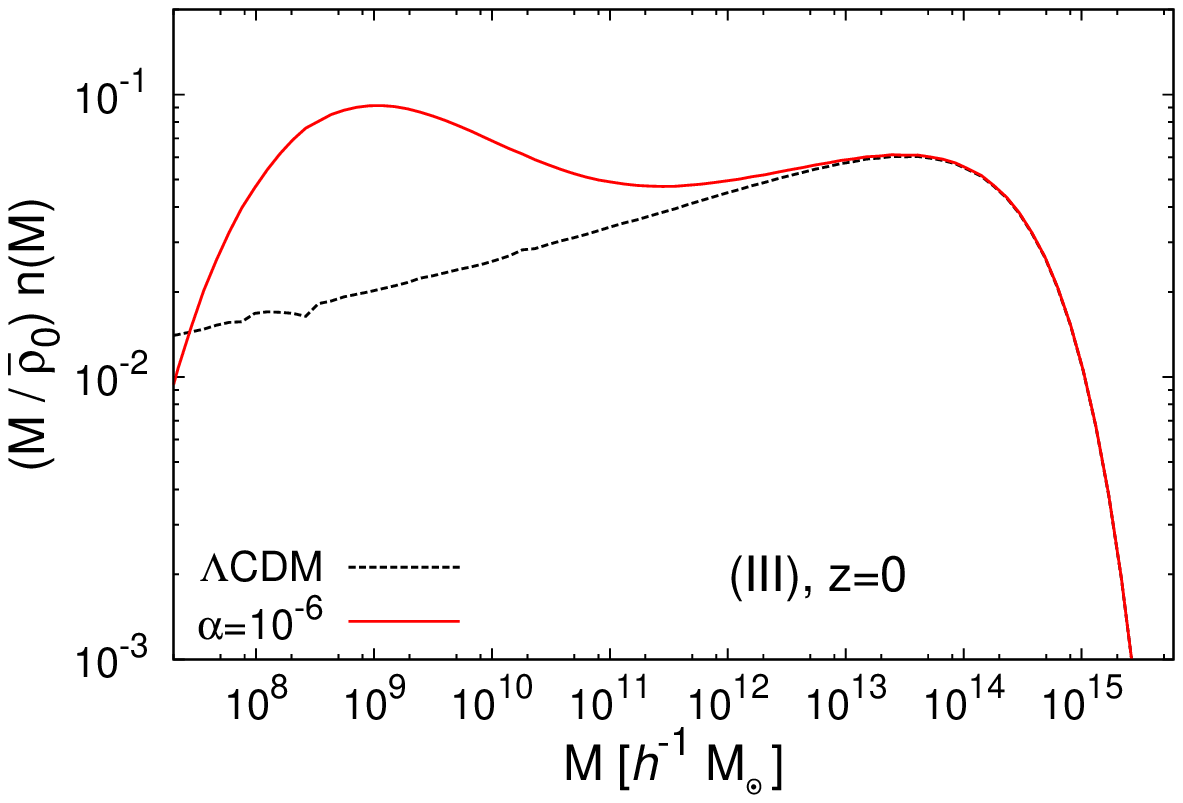}}\\
\epsfxsize=8. cm \epsfysize=5.5 cm {\epsfbox{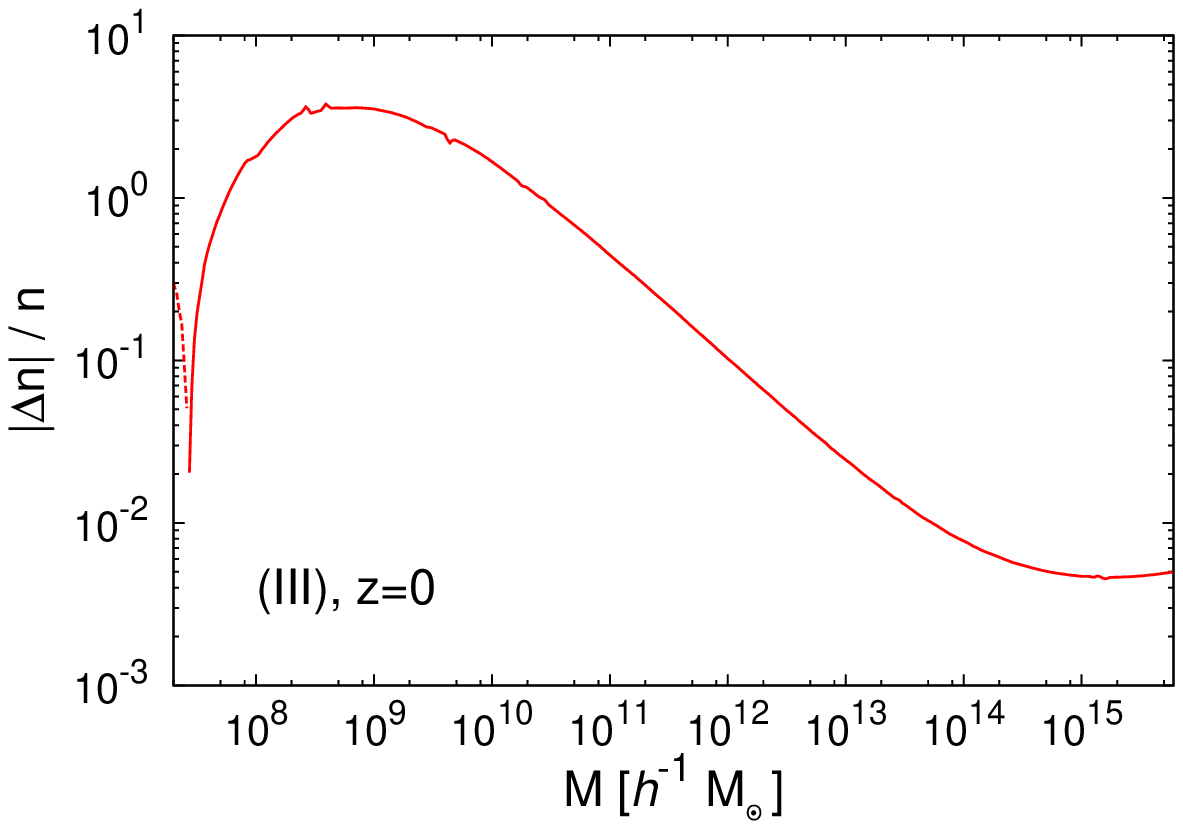}}
\end{center}
\caption{
{\it Upper panel:} halo mass function at $z=0$ for the model (III) (solid line) and the
$\Lambda$-CDM reference (dashed line).
{\it Lower panel:} relative deviation of the halo mass function from the $\Lambda$-CDM reference,
for the model (III). We show the absolute value $|\Delta n|/n$ (with a solid line for
$n_{\rm (III)}>n_{\Lambda\rm CDM}$ and a dashed line otherwise).
}
\label{fig_nM_III}
\end{figure}

We show our results for the halo mass function obtained for model (III) in Fig.~\ref{fig_nM_III}.
In agreement with Sec.~\ref{sec:spher-III}, the results are very close to those obtained for
model (II). The acceleration of the gravitational collapse by the fifth force leads to a higher
halo mass function at moderate and large masses, with an amplification that grows towards
smaller masses, from $M= 10^{13}$ down to $10^9  h^{-1} M_{\odot}$, and a convergence to the
$\Lambda$-CDM falloff around $M \sim 10^{13} - 10^{15} h^{-1} M_{\odot}$.
At very small masses, $M<10^7 h^{-1} M_{\odot}$, the deviation from the $\Lambda$-CDM
halo mass function becomes negative, in agreement with the constraint associated with the
normalization of the halo mass function.

Again, these results should be robust at large mass, $M > 10^{12} h^{-1} M_{\odot}$,
where gravitational collapse remains similar to the usual $\Lambda$-CDM case, whereas
the predictions are unlikely to be accurate at low masses, $M < 10^{12} h^{-1} M_{\odot}$,
where the significant differences in the process of gravitational clustering could change
the shape of the scaling function $f(\nu)$.

\section{Screening of the fifth force in dense environments}
\label{sec:halos}

So far we have focused on the impact of the modification of gravity on the background cosmology
and the large-scale structures, including the linear regime and the formation of collapsed halos.
In practice, we wish to recover General Relativity on small scales, especially in the Solar
System where accurate measurements provide stringent constraints on a possible fifth force.
Therefore, we compare in this section the magnitude of the fifth force with the Newtonian gravity
on a variety of objects, from clusters of galaxies to galaxies and to the Solar System.

\subsection{Screening within clusters or spherical halos}
\label{sec:screening-clusters}

We first consider here how the ratio of the fifth force to Newtonian gravity behaves within
spherical halos with a mean density profile such as the Navarro-Frenk-White (NFW) \cite{Navarro:1996}
density profile, often used to describe massive dark matter halos.
In particular, we wish to find the conditions for the fifth force not to diverge at the center
of the halos and to remain modest at all radii, to be consistent with observations of
X-ray clusters.
Within spherical halos, the Newtonian force reads as
\beq
F_{\rm N} = - \frac{{\cal G}_{\rm N} M(<r)}{r^2} = - \frac{\Omega_{\rm m}}{2} \Delta(<r) r H^2 ,
\label{FN-Delta}
\eeq
where $\Delta(<r)$ is the mean overdensity within radius $r$.
We can also write this as
\beq
F_{\rm N} = - \frac{v_{\rm N}^2(r)}{r} , \;\;\; v_{\rm N}^2 =   \frac{{\cal G}_{\rm N} M(<r)}{r} ,
\label{FN-v2}
\eeq
where $v_{\rm N}^2$ is the circular velocity at radius $r$, which also measures the typical
magnitude of the velocity dispersion when Newtonian gravity is dominant.
The fifth force reads as
\beq
F_A = - c^2 \frac{d\ln A}{d r} = - \frac{c_s^2(r)}{r} = - \frac{c^2}{r} \frac{d\ln A}{d\ln\rho}
\frac{d\ln\rho}{d\ln r} .
\label{FA-def-halo}
\eeq
Therefore, the ratio of the fifth force to the Newtonian force is
\beqa
\eta \equiv \frac{F_A}{F_{\rm N}} & = & \frac{2}{\Omega_{\rm m} \Delta(<r)}
\left( \frac{c}{r H} \right)^{\!2} \frac{d\ln A}{d\ln\rho} \frac{d\ln\rho}{d\ln r} \;\;\;
\label{eta-def-Delta} \\
& = & \frac{c^2}{v_{\rm N}^2} \frac{d\ln A}{d\ln\rho} \frac{d\ln\rho}{d\ln r} .
\label{eta-def-v2}
\eeqa
In agreement with the discussion in Secs.~\ref{sec:Perturbations-J} and \ref{sec:constraints},
the second line (\ref{eta-def-v2}) shows that we need a small amplitude for the coupling
function $\ln A$ to compensate the large factor $c^2/v_{\rm N}^2$, for the ratio $\eta$
not to be much greater than unity in typical astrophysical and cosmological structures.
This is provided by the parameter $\alpha \sim 10^{-6}$.
In agreement with Eq.(\ref{eps-def}) and the analysis of cosmological perturbations
in Sec.~\ref{sec:Linear-perturbations}, the first line (\ref{eta-def-Delta}) shows that
the relative importance of the fifth force typically grows at smaller scales, as $1/r^2$ or
$k^2$, and that the factor $\alpha$ is again needed to ensure that the fifth force does not
greatly exceed Newtonian gravity at scales $\sim 1 h^{-1}\rm{Mpc}$.

From the Euler equation (\ref{Euler-1-J}) or the expression (\ref{FA-def-halo}) of the fifth force,
we can also associate with the fifth force the velocity scale $c_s$, with
\beq
c_s^2 = \left| r F_A \right| = c^2 \left| \frac{d\ln A}{d\ln r} \right| ,
\label{cs2-FA-def}
\eeq
in a fashion similar to $v_{\rm N}^2$ for Newtonian gravity.
Then, the force ratio $\eta$ also reads as
\beq
\left | \eta \right | = \frac{c_s^2}{v_{\rm N}^2} ,
\label{eta-cs-vN}
\eeq
and it also measures the ratio of these two velocity scales.

On very small scales and high densities, the fifth force is also partly screened by the
nonlinearities of the coupling function $\ln A$, as $d\ln A/d\ln\rho$ goes to zero at large
densities (because $\ln A$ is monotonic and bounded).

From Eq.(\ref{dlnAdlnrho-low-density}), we have at moderate densities
\beq
\rho \ll \frac{{\cal M}^4}{\alpha} :  \;\;\; | \eta | \sim \frac{\alpha^2}{a^3} \left( \frac{c}{r H} \right)^2 .
\label{eta-low-density}
\eeq
Thus, at low redshifts the ratio $\eta$ is actually suppressed by a factor $\alpha^2$, for the
models studied in this paper, so that $\eta$ only reaches unity at $r \sim 3 h^{-1} {\rm kpc}$,
i.e. at galaxy scales (see also Sec.~\ref{sec:structures} below).
At higher densities, we obtain for models (I) and (II),
$| d\ln A/d\ln\rho | \sim {\cal M}^8/\alpha\rho^2$ and
\beq
\mbox{(I) and (II)}, \;\;\; \rho \gg \frac{{\cal M}^4}{\alpha} :  \;\;\;
| \eta | \sim \frac{a^6}{\alpha\Delta^3} \left( \frac{c}{r H} \right)^2 ,
\label{eta-high-density-I-II}
\eeq
and for model (III), $| d\ln A/d\ln\rho | \sim \sqrt{\alpha{\cal M}^4/\rho}$ and
\beq
\mbox{(III)}, \;\;\; \rho \gg \frac{{\cal M}^4}{\alpha} :  \;\;\;
| \eta | \sim \sqrt{\frac{\alpha a^3}{\Delta^3}} \left( \frac{c}{r H} \right)^2 .
\label{eta-high-density-III}
\eeq
Let us consider a power-law density profile, of exponent $\gamma>0$ and
critical radius $r_{\alpha}$,
\beq
\rho(r) \sim \frac{\bar\rho_0}{\alpha} \left( \frac{r}{r_{\alpha}} \right)^{-\gamma} .
\label{gamma-def}
\eeq
Since ${\cal M}^4 = \bar\rho_{\rm de 0} \sim \bar\rho_0$, at radii greater than $r_{\alpha}$
we have the behavior (\ref{eta-low-density}),
\beq
r > r_{\alpha} :  \;\;\; | \eta | \sim \frac{\alpha^2}{a^3} \left( \frac{c}{r H} \right)^2 ,
\label{eta-large-r}
\eeq
whereas at smaller radii we have
\beqa
r < r_{\alpha} & : & \mbox{(I) and (II)}, \;\;\;
| \eta | \sim \frac{a^6}{\alpha\Delta^3} \left( \frac{c}{r H} \right)^2 ,
\label{eta-small-r-I-II} \\
&& \mbox{(III)}, \;\;\;
| \eta | \sim \sqrt{\frac{\alpha a^3}{\Delta^3}} \left( \frac{c}{r H} \right)^2 .
\label{eta-small-r-III}
\eeqa
From Eq.(\ref{eta-large-r}) we find that at large radius the relative importance of the fifth
force decreases as $1/r^2$, independently of the shape of the halo profile.
From Eqs.(\ref{eta-small-r-I-II}) and (\ref{eta-small-r-III}) we find that at small radii
the ratio $\eta$ behaves as $r^{3\gamma-2}$ for the models (I) and (II), and as
$r^{3\gamma/2-2}$ for the model (III).
Therefore, the conditions for the ratio to go to zero at the center are:
\beqa
r \rightarrow 0 : \;\;\; \eta \rightarrow 0 \;\; \mbox{if}
&& \gamma> 2/3 \;\; \mbox{for (I) and (II)} , \;\;\;\;\;
\label{gamma-bound-I-II} \\
&& \gamma> 4/3 \;\; \mbox{for (III)} .
\label{gamma-bound-III}
\eeqa
If we consider halos with a mean Navarro-Frenk-White (NFW) density profile,
which has $\gamma=1$, we find that the relative importance of the fifth force vanishes
at the center for the models (I) and (II) but diverges for the model (III).
This means that the model (III) is ruled out, unless the small-scale cutoff $\ell_s$
discussed in Sec.~\ref{sec:cutoff} is of the order of $1 \, h^{-1} \; {\rm kpc}$.
If we do not wish to rely on the small-scale cutoff $\ell_s$, Eq.(\ref{eta-def-Delta})
shows that, to obtain a negligible fifth force at the center of a halo of exponent $\gamma$,
the coupling function must decay at large densities as
\beq
\mbox{for } \eta \rightarrow 0 : \;\;\; \frac{d\ln A}{d\ln\rho} \sim \rho^{-\mu}
\mbox{ with } \mu > \frac{2}{\gamma} -1 .
\label{mu-def}
\eeq
However, we shall come back to this point in section~\ref{sec:halo-centers} and argue
that the divergence of the fifth force at the center could actually disappear because
of the non-linearities of the scalar field dynamics and its ultra-local character.
Indeed, the result (\ref{gamma-bound-III}) was derived from dimensional analysis and
assumes that the density field remains smooth. However, in the non-linear regime
the density field can develop strong inhomogeneities and fragment, because of the
fifth-force instability. This in turn leads to a screening mechanism as isolated subhalos
do not exert a fifth force on each other because of its ultra-local character.

Keeping with the dimensional analysis in this section, the result (\ref{mu-def}) would
suggest that the relative importance of the fifth force
always diverges at the center of halos with a flat core, $\gamma=0$, but this is not
the case as Eq.(\ref{mu-def}) was derived for power-law profiles with $\gamma>0$,
where $d\ln\rho/d\ln r$ in Eq.(\ref{eta-def-Delta}) was assumed to be of order unity.
For halos with a core radius $r_c$, we can write $\rho \simeq \rho_c [ 1 - (r/r_c)^2 ]$
at small radii $r \ll r_c$, hence $|d\ln\rho/d\ln r| \sim (r/r_c)^2$ and
Eq.(\ref{eta-def-Delta}) gives the finite limit
\beq
r \ll r_c : \;\;\; | \eta | \sim \frac{1}{\Delta_c} \left( \frac{c}{r_c H} \right)^{\!2} \,
\left| \frac{d\ln A}{d\ln\rho} \right|_{\rho_c} .
\label{eta-c}
\eeq

\begin{figure}
\begin{center}
\epsfxsize=8.5 cm \epsfysize=6.5 cm {\epsfbox{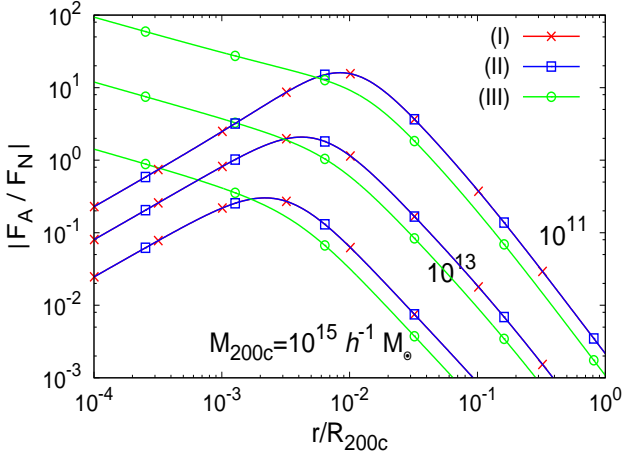}}
\end{center}
\caption{
Absolute value of the ratio $\eta=F_A/F_{\rm N}$, as a function of the radius $r$,
within spherical halos.
We display the halo masses $M_{\rm 200c}= 10^{15}$, $10^{13}$ and
$10^{11} h^{-1} M_{\odot}$, from bottom to top, at $z=0$.
We consider the models (I) (red line with crosses),
(II) (blue line with squares) and (III) (green line with circles);
$\eta <0$ for the model (I); $\eta>0$ for the models (II) and (III).
The absolute values $|\eta|$ of models (I) and (II) are equal.
}
\label{fig_eta_z0}
\end{figure}

We show in Fig.~\ref{fig_eta_z0} the radial profile of the force ratio $\eta$ at $z=0$,
for several halo masses.
Here we consider spherical halos with a mean NFW density profile,
$\rho(r)=\rho_{s}/[(r/r_{s})(1+r/r_{s})^{2}]$, and a concentration parameter,
$c=R/r_s$ given by
$c(M,z) = 11 ( M/10^{12} M_{\odot} )^{-0.1} (1+z)^{-1.5}$.
We define the halo radius $R_{\rm 200c}$ by the mean overdensity threshold
$\Delta_{\rm 200c}=200$ with respect to the critical density $\rho_{\rm crit}$.
In agreement with Eq.(\ref{eta-large-r}) and Eqs.(\ref{eta-small-r-I-II})-(\ref{eta-small-r-III}),
the force ratio decreases as $1/r^2$ at large radii for all three models, it decreases
as $r$ at small radii for the models (I) and (II), while it increases as
$r^{-1/2}$ for the model (III).
The ratio $\eta$ is maximum, for models (I) and (II), or shows a bend between the
small-radius and large-radius regimes, at $r_{\alpha} \sim R_{\rm 200c}/100$
(for the cases considered here).
The overall amplitude of $\eta$ increases for smaller mass (hence smaller
halo radius) because of the characteristic growth on small scale, as $1/r^2$,
of the modification of gravity investigated in this paper.

As noticed above, the steady growth of the ratio $\eta$ towards the center of the halo
for the model (III) suggests that this model would lead to cluster or galaxy halos that
are significantly different from those obtained in the $\Lambda$-CDM scenario.
Then, this model would be ruled out by observations, which show that
$\Lambda$-CDM cosmologies provide a reasonably good agreement with data
for the properties of clusters and galaxies.
The ratio $\eta$ becomes of order $10$ (or greater) around $R_{\rm 200c}/100$ for the
models (I) and (II) for halo masses $M \sim 10^{11} h^{-1} M_{\odot}$ (or lower).
This suggests that these models may also be strongly constrained by observations,
which would provide an upper bound on the model parameter $\alpha$.
However, obtaining a quantitative estimate of this constraint requires a dedicated
study that we leave for future work. We would need to evaluate the impact of the fifth
force on the final halo profile, which may require numerical simulations, and to
estimate the observational accuracy of the halo profiles measured on the
intermediary scale $\sim r_{\alpha}$.
Moreover, as we discuss in section~\ref{sec:halo-centers} below,
the results obtained above may break down in the regime dominated by the fifth force
because it could lead to the formation of strong inhomogeneities that in turn
screen the fifth force in the final configuration of the system.

\subsection{Cosmological and astrophysical structures}
\label{sec:structures}

We now estimate the fifth force to Newtonian gravity ratio $\eta$ for a variety of
astrophysical objects and environments, from clusters of galaxies to the laboratory
on the Earth, at low redshift.

\subsubsection{Clusters of galaxies}
\label{sec:clusters}

In a halo of mass $M$ and radius $R$, the Newtonian potential and the Newtonian
force are of order
\beq
\frac{\Psi_{\rm N}}{c^2} \sim \frac{{\cal G}_{\rm N} M}{c^2 R} , \;\;\;
\frac{F_{\rm N}}{c^2} \sim \frac{{\cal G}_{\rm N} M}{c^2 R^2} .
\label{PsiN-FN-def}
\eeq
As in Eq.(\ref{trajectory-Jordan}), the fifth force
${\bf F}_A=-\nabla \Psi_A= -c^2 \nabla \ln A$ is of order
\beq
\frac{F_A}{c^2} \sim \frac{d\ln A}{d r} \sim \frac{1}{R} \frac{d\ln A}{d\ln \rho} ,
\label{FA-def}
\eeq
where we assumed $d\ln\rho/d\ln r \sim 1$.
As seen from Eq.(\ref{dlnAdlnrho-low-density}), in the low-density regime we have:
\beq
\alpha \Delta \ll 1 : \;\;\; \frac{d\ln A}{d\ln\rho} \sim \alpha^2 \Delta ,
\label{dlnA-dlnrho-low-density}
\eeq
where $\Delta=\rho/\bar{\rho}$ is the typical matter overdensity of the object.
Then, for a cluster of galaxies, with $\Delta \sim 10^3$, $R \sim 1 \, {\rm Mpc}$,
$M \sim 10^{14}M_{\odot}$, we obtain
\beqa
&& \frac{F_{\rm N}}{c^2} \sim 5 \times 10^{-6} {\rm Mpc}^{-1} , \;\;\;
\frac{F_A}{c^2} \sim \alpha^2 10^3 \rm{Mpc}^{-1} , \nonumber \\
&& \frac{F_A}{F_{\rm N}} \sim (10^4 \alpha)^2  \ll 1 .
\label{FA-FN-cluster}
\eeqa
Therefore, the fifth force is negligible on cluster scales.
However, as seen in Sec.~\ref{sec:screening-clusters} and Fig.~\ref{fig_eta_z0},
this is no longer the case far inside the cluster, at $r \lesssim R_{\rm 200c}/100$,
for clusters of mass $M \lesssim 10^{13} h^{-1} M_{\odot}$.

\subsubsection{Galaxies}
\label{sec:galaxies}

We now consider a typical galaxy, such as the Milky Way, with $M \sim 10^{12} M_{\odot}$,
$R \sim 10 \, {\rm kpc}$, and $\Delta \sim 10^6$.
This high value of the density contrast is at the limit of validity of the regime
(\ref{dlnA-dlnrho-low-density}), but this should still provide the order of magnitude of the fifth force.
Then, we obtain
\beqa
&& \frac{F_{\rm N}}{c^2} \sim 5 \times 10^{-4} {\rm Mpc}^{-1} , \;\;\;
\frac{F_A}{c^2} \sim \alpha^2 10^8 \rm{Mpc}^{-1} , \nonumber \\
&& \frac{F_A}{F_{\rm N}} \sim (10^6 \alpha)^2  \sim 1 .
\label{FA-FN-galaxy}
\eeqa
Thus, the fifth force is of the same order as the Newtonian gravity on galaxy scales.
This suggests that interesting phenomena could occur in this regime and that galaxies
could provide a useful probe of such models. On the other hand, since we are at the
border of the regime (\ref{dlnA-dlnrho-low-density}), nonlinear effects may already come
into play and partly screen the fifth force, depending on the details of the coupling
function $A(\tilde\chi)$.

\subsubsection{Solar System}
\label{sec:solar-system}

Many alternative theories to General Relativity are strongly constrained,
or even ruled out, by Solar System tests, based on the trajectories of planets around the Sun
(measurements by the Cassini satellite  \cite{Bertotti:2003rm})
or the motion of the Moon around the Earth
(Lunar Laser Ranging experiment \cite{Williams:2004qba}).
To remain consistent with these data, modified-gravity scenarios often involve nonlinear
screening mechanism that ensure convergence to General Relativity in small-scale
and high-density environments (typically by suppressing the gradients of the scalar field
or its coupling to matter).
In our case, if we consider stars, planets and moons as isolated objects in the vaccum,
the screening is provided by the definition of the model itself and is $100\%$ efficient.
Indeed, because the fifth force is exactly local, as
${\bf F}_A = -c^2 \nabla \ln A(\rho)$ only depends on the local density and its gradient,
the impact of the Sun onto the motion of the Earth through the fifth force is exactly zero,
unless if it creates a distant density gradient by other means (e.g. Newtonian gravity).
However, the impact of the gradient of the
Newtonian force from the Sun onto the matter distribution in the Earth is negligible
and completely superseded by local geophysical sources (the radial structure of the Earth
core and atmosphere and random variations associated with mountains and oceans
 for instance).
Therefore, the Sun is completely ``screened'' as viewed from the Earth by the fifth force,
as well as all planets and moons of the Solar System.
Therefore, the trajectories of astrophysical objects in the Solar System are exactly given
by the usual Newtonian gravity, or more accurately General Relativity,
and all Solar Systems tests of gravity are satisfied, to the same accuracy as
General Relativity.

Here we assumed that the small-scale cutoff $\ell_s$ of the theory,
discussed in Sec.~\ref{sec:cutoff},
is below the Solar System scales. If this is not the case, then one needs to explicitly
consider the small-scale behavior of the complete theory.
If the small-scale regularization is associated with a kinetic term in the scalar-field
Lagrangian, as in Eq.(\ref{L-chi-def-kinetic}), we recover a standard Dilaton model.
Then, high-density regions, or compact objects such as stars, give rise to a long-range
fifth force but the latter is screened in dense environments by the usual Damour-Polyakov
mechanism, as the coupling function $\ln A$ goes to a constant at high-densities
and the coupling strength $d\ln A/d\tilde\chi$ vanishes.
The efficiency of this screening mechanism depends on the details of the model
[the kinetic and potential terms in the original scalar-field Lagrangian
$\tilde{\cal L}_{\varphi}(\varphi)$].

\subsubsection{On the Earth and in the laboratory}
\label{sec:labo}

Even though the fifth force on the Earth is not significantly influenced by the Sun and
other planets, it does not vanish as it is sensitive to the local gradient of the matter density.
Then, we must check that this local force is small enough to have avoided detection
in the laboratory or on the Earth (e.g., at its surface or in the atmosphere).
Here we first assume that the cutoff $\ell_s$ is smaller than the scales we consider.

So far we have assumed that the scalar field is coupled in the same manner to the
dark matter and to the ordinary baryonic matter. For the analysis of cosmological structures,
from the background dynamics down to galaxies, we are dominated by dark matter so
we mostly probed the coupling to the dark matter and it made no difference
whether the coupling to baryons is the same or not.
However, on smaller scales, such as in the Solar System or on Earth, we are dominated by
baryonic matter. Then, a simple manner to ensure that we satisfy observations and experiments
performed in the laboratory or on the Earth is to assume that ordinary matter is not coupled
to the scalar field.

A second alternative is that screening mechanisms are sufficiently efficient to make the
fifth force negligible on the Earth. We now investigate whether this is the case, assuming
dark matter and baryons couple in the same fashion to the scalar field.
As seen in Eq.(\ref{Euler-1-J}), the local nature of the scalar field configuration makes the fifth
force appear as a polytropic pressure $p_A(\rho)$, given by Eq.(\ref{pA-def}), where
$\rho$ is now the baryonic matter density as the dark matter density and its gradient
can be neglected.
Since $\bar{A} \simeq 1$ and ${\cal M}^4 \simeq \bar\rho_{\rm de 0}$, we obtain
for a typical density of 1 g/cm$^{3}$,
\beq
\rho \sim 1 \; {\rm g . cm}^{-3} : \;\;\;
\frac{p_A}{\rho} \sim 3 \times 10^{-13} \; \tilde\chi \; ({\rm m/s})^2 .
\label{pA-rho}
\eeq
For $\tilde\chi \sim 1$, as in the models (I) and (II) where $\tilde\chi$ has a finite range of order unity,
this corresponds to small velocities and motions.
To compare this pressure with the thermal motions found on the Earth or in the laboratory,
we write Eq.(\ref{pA-rho}) as a temperature,
\beq
\frac{m_p p_A}{\rho k_B} \sim 3 \times 10^{-17} \; \tilde\chi \; {\rm K} ,
\label{p-T}
\eeq
where again we chose $\rho \sim 1 \; {\rm g/cm}^{3}$, $m_p$ is the proton mass and $k_B$ the
Boltzmann constant.
For the models (I) and (II) where $\tilde\chi$ has a finite range of order unity, this gives a very
low temperature of order $10^{-17}$ K, which is much smaller than the temperature reached by
cold-atoms experiments in the laboratory, $T \sim 10^{-7}$ K.
Thus, for such models where $\tilde\chi \sim 1$ the fifth force can be neglected in the
laboratory and on the Earth (and in other astrophysical objects).

More generally, Eq.(\ref{p-T}) gives the local upper bound for $|\tilde\chi|$:
\beq
\rho \sim 1 \; {\rm g . cm}^{-3} : \;\;\; | \tilde\chi | < 10^{10} .
\label{bound-chi}
\eeq
For the model (III) where $| \tilde\chi |$ is not bounded, we obtain from
Eq.(\ref{model-III-large-chi-rho-def})
$\tilde\chi \sim - 6\times 10^{14} \alpha^{1/2} \sim - 6\times 10^{11}$, which violates
the upper bound (\ref{bound-chi}).
Therefore, this model would appear to be ruled out by such cold-atoms experiments.
Models where $| \tilde\chi |$ is not bounded are still allowed but their function $\tilde\chi(\rho)$
should be somewhat smaller than Eq.(\ref{model-III-large-chi-rho-def}) for
$\rho \sim 1 \; {\rm g/cm}^{3}$.

However, as noticed in Sec.~\ref{sec:cutoff}, the local model (\ref{L-chi-def}) considered
in this paper is not expected to apply down to arbitrarily small scales, but only above
a small-scale cutoff $\ell_s$. This may arise for instance from a nonzero kinetic term
in the scalar-field Lagrangian. In any case, we should have $\ell_s > 1 \; {\rm m}$
in the cosmological background (i.e., in the intergalactic space).
The cutoff scale $\ell_s$ generically depends on the environment, e.g. on the local
value of the scalar field through the change of variable (\ref{tchi-def}).
On the Earth, the result (\ref{p-T}) suggests that the theory could be valid down to
somewhat smaller scales, as long as we remain above the atomic scale and we can still
define a continuum limit to the density field.
In any case, this small-scale regularization suggests that that the cold-atom bound
(\ref{bound-chi}) can be relaxed and the result (\ref{p-T})
shows that the fifth force is negligible on the Earth and in the laboratory, and hence it is
consistent with local experiments.

\subsection{Fifth-force dominated regime}
\label{sec:fifth-force-regime}

In the previous section, we estimated the fifth force to Newtonian gravity ratio $\eta$
and the impact of the scalar field for a variety of objects and environments.
It is useful to make this analysis more general and to determine the domain of length,
density and mass scales where the fifth force is dominant.
Thus, using for instance Eq.(\ref{eta-def-Delta}) and taking $d\ln\rho/d\ln r \sim 1$,
we write for structures of typical radius $R$, density $\rho$ and mass
$M=4\pi\rho R^3/3$,
\beq
| \eta | \sim \frac{2}{\Omega_{\rm m 0}} \frac{\bar\rho_0}{\rho}
\left( \frac{c}{R H_0} \right)^{\!2} \left| \frac{d\ln A}{d\ln\rho} \right| .
\label{eta-R-rho}
\eeq
Then, the fifth force is greater than Newtonian gravity if we have
\beq
|\eta| \geq 1 : \;\;\; R^2 \leq \left( \frac{c}{H_0} \right)^{\!2}  \frac{2}{\Omega_{\rm m 0}}
\frac{\bar\rho_0}{\rho} \left| \frac{d\ln A}{d\ln\rho} \right| .
\label{R-rho}
\eeq
Although for convenience we write the right-hand side in terms of the cosmological quantities
$H_0$, $\bar\rho_0$ and $\Omega_{\rm m0}$ at $z=0$, this expression does not depend
on redshift nor on cosmology. Moreover, it is only a function of the density $\rho$, as
any coupling function $\ln A(\tilde\chi)$ also defines the functions $\tilde\chi(\rho)$ and
$\ln A(\rho)$ through the scalar-field equation (\ref{KG-A=1}).
Therefore, in a density-radius plane, the domain where $|\eta|\geq 1$ is given by the
area under the curve $R_{\eta}(\rho)$, where $R_\eta(\rho)$ is the density-dependent
radius defined by the right-hand side in Eq.(\ref{R-rho}).

\begin{figure}
\begin{center}
\epsfxsize=9 cm \epsfysize=7. cm {\epsfbox{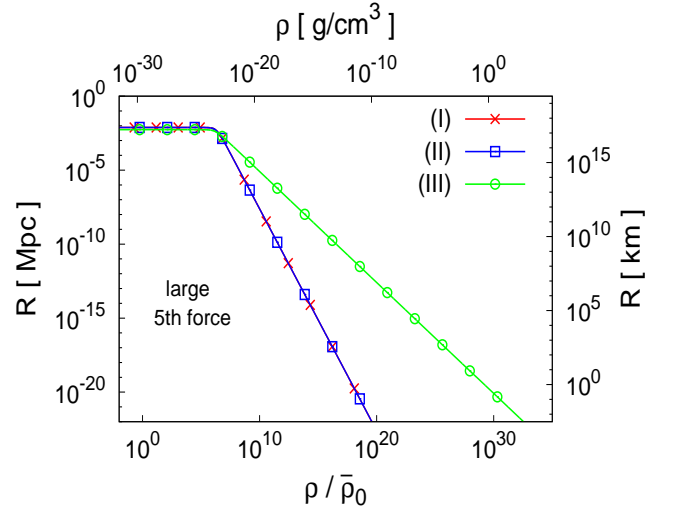}}
\end{center}
\caption{
Domain in the density-radius plane where the fifth force is greater than Newtonian gravity.
This domain is identical for the models (I) and (II), and greater for the model (III).
The horizontal axis is the typical density of the structure, $\rho$, given in units of the mean
matter cosmological density today, $\bar\rho_0$, in the bottom-border scale, and in units
of ${\rm g . cm}^{-3}$ in the top-border scale.
The vertical axis is the typical radius of the structure, $R$, given in Mpc in the left-border
scale and in km in the right-border scale.
}
\label{fig_eta_R_rho}
\end{figure}

We display this domain in the $(\rho,R)$-plane in Fig.~\ref{fig_eta_R_rho}.
At low densities, using Eqs.(\ref{dlnAdlnrho-low-density}) and (\ref{M4-rhode}),
we obtain
\beq
\rho \ll \frac{{\cal M}^4}{\alpha} : \;\; R_{\eta}(\rho) \sim R_{\alpha} \;\;
\mbox{with} \;\; R_{\alpha} \equiv \frac{c}{H_0} \frac{\alpha}{\sqrt{\Omega_{\rm de 0}}} .
\label{Reta-low-density}
\eeq
Thus, at low densities we obtain a constant radius threshold, of order
$R_{\alpha} \sim 0.01 \,{\rm Mpc}$ for $\alpha=10^{-6}$, as we can check in
Fig.~\ref{fig_eta_R_rho}.
At high densities, we have the behaviors
\beqa
\mbox{(I) and (II)} , && \;\; \rho \gg \frac{{\cal M}^4}{\alpha} : \;\;
\left| \frac{d\ln A}{d\ln\rho} \right| \sim \frac{{\cal M}^8}{\alpha\rho^2} , \nonumber \\
&& R_{\eta}(\rho) \sim \frac{c}{H_0} \frac{\Omega_{\rm de 0}}{\sqrt{\alpha\Omega_{\rm m0}^3}}
\frac{\bar\rho_0^{3/2}}{\rho^{3/2}} ,
\label{Reta-high-density-I-II}
\eeqa
and
\beqa
\mbox{(III)} , && \;\; \rho \gg \frac{{\cal M}^4}{\alpha} : \;\;
\frac{d\ln A}{d\ln\rho} \sim \sqrt{ \frac{\alpha{\cal M}^4}{\rho} } , \nonumber \\
&& R_{\eta}(\rho) \sim \frac{c}{H_0}
\left(\frac{\alpha\Omega_{\rm de 0}}{\Omega_{\rm m0}^3} \right)^{1/4}
\frac{\bar\rho_0^{3/4}}{\rho^{3/4}} .
\label{Reta-high-density-III}
\eeqa
Thus, at high densities the upper boundary of the fifth-force domain decreases as
$R_{\eta} \propto \rho^{-3/2}$ for the models (I) and (II) and as $\rho^{-3/4}$ for the
model (III).
As in previous sections, we find that the effects of the fifth force are greater for the
model (III).
This screening of the fifth force at high densities ensures that it becomes negligible
at the center of halos with sufficiently steep density profiles and for astrophysical
objects such as stars and planets.
On the other hand, we find that, independently of the density, the fifth force is always
negligible on scales greater than $R_{\alpha} \sim \alpha c/H_0$, of order
$0.01 \, \rm{Mpc}$.
This confirms again that the fifth force is small on cluster scales and beyond.

\begin{figure}
\begin{center}
\epsfxsize=9 cm \epsfysize=7. cm {\epsfbox{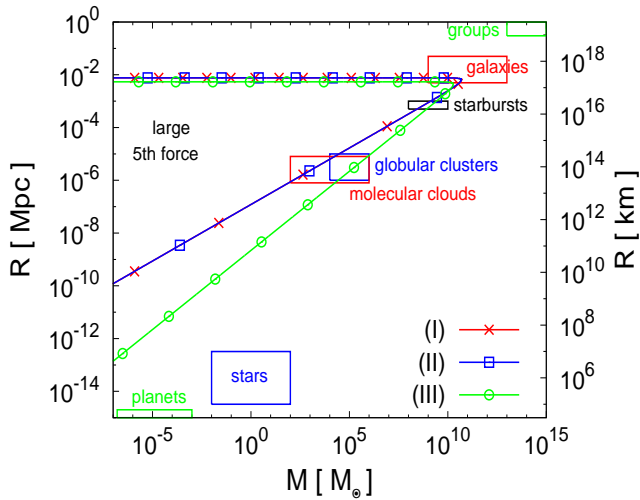}}
\end{center}
\caption{
Domain in the mass-radius plane where the fifth force is greater than Newtonian gravity.
This domain is identical for the models (I) and (II), and greater for the model (III).
The horizontal axis is the typical mass of the structure, $M$, given in units of the solar mass.
The vertical axis is the typical radius of the structure, $R$, given in Mpc in the left-border
scale and in km in the right-border scale.
The rectangles show the typical scales of various astrophysical structures.
}
\label{fig_eta_R_M}
\end{figure}

To facilitate the comparison with astrophysical structures, it is convenient to display
the fifth-force domain (\ref{R-rho}) in the mass-radius plane $(M,R)$.
This is shown in Fig.~\ref{fig_eta_R_M}, as the curve $R_{\eta}(\rho)$ provides
a parametric definition of the boundary $R_{\eta}(M)$, defining the mass
of the structure as $M=4\pi\rho R^3/3$.
We obtain a triangular domain, with a constant-radius upper branch and a lower branch
that goes towards small radius and mass with a slope that depends on the model.
The upper branch corresponds to the regime (\ref{Reta-low-density}), with
\beq
\mbox{upper branch:} \;\;\; R = R_{\alpha} \mbox{ for } M < M_{\alpha} ,
\label{R-M-upper}
\eeq
and
\beq
M_{\alpha} \equiv \frac{\alpha^2\bar\rho_0}{\Omega_{\rm m0} \sqrt{\Omega_{\rm de 0}}}
\left( \frac{c}{H_0} \right)^3 ,
\label{Malpha-def}
\eeq
where $M_{\alpha} = \rho_{\alpha} R_{\alpha}^3$ with $\rho_{\alpha}={\cal M}^4/\alpha$.
For $\alpha=10^{-6}$ this yields $M_{\alpha} \sim 10^{10} M_{\odot}$.
The lower branch corresponds to the regimes (\ref{Reta-high-density-I-II}) and
(\ref{Reta-high-density-III}), which yield
\[
\mbox{lower branch for } M < M_{\alpha} :
\]
\beq
\mbox{(I) and (II) :} \;\; R = R_{\alpha} \left( \frac{M}{M_{\alpha}} \right)^{3/7} ,
\label{R-M-lower-I-II}
\eeq
\beq
\mbox{(III) :} \;\; R = R_{\alpha} \left( \frac{M}{M_{\alpha}} \right)^{3/5} .
\label{R-M-lower-III}
\eeq
We also show in Fig.~\ref{fig_eta_R_M} the regions in this $(M,R)$-plane occupied
by various astrophysical objects.
From left-bottom to right-top, we show planets, stars, molecular clouds, globular clusters,
extended starburst regions, galaxies and groups of galaxies.
In agreement with Secs.~\ref{sec:screening-clusters} and \ref{sec:structures},
we find that the fifth force is negligible for clusters and groups (at their global scale)
and Solar-System objects, while it is of the same order as Newtonian gravity for galaxies.
In particular, it appears that various galactic structures, from the molecular clouds and
extended starburst regions, where star formation takes place, to the overall extent of
low-mass galaxies, as well as the small old globular clusters, all lie close to the boundary of the
fifth-force domain.
Therefore, they may provide strong constraints on the models considered in this paper.
In fact, the model (III) might be ruled out by galaxy observations, independently
of the issue found in Sec.~\ref{sec:screening-clusters} with the divergence of the fifth
force at the center of NFW halos (\ref{gamma-bound-III}).
However, we leave a detailed study of molecular clouds and globular clusters to future
works to check the quantitative constraints they can provide on the scalar-field theories
(\ref{L-phi-def}).

\section{History and properties of the formation of cosmological structures}
\label{sec:structures-thermo-analysis}

In the previous sections we have studied the evolution of the linear perturbations and
of the spherical collapse by assuming that the density field remains smooth and that
the fifth force on cosmological scales $x$ is set by the density gradient smoothed
on these large scales.
However, in the ultra-local models that we study in this paper the fifth force is
directly sensitive to the local density gradient, as $\nabla\ln A = (d\ln A/d\rho) \nabla\rho$.
As compared with the $\Lambda$-CDM cosmology, the models of the type (II) and (III)
accelerate the growth of small-scale perturbations, and increasingly so on smaller scales
because of the $k^2$ term in Eq.(\ref{eps-def}), as seen in Figs.~\ref{fig_Dp_II}
and \ref{fig_Dp_III} of the linear growing mode. This suggests that very small scales
can develop strong inhomogeneities at early times and the local density gradient
could always be set by such very small scales (actually the small-scale cutoff of the theory)
rather than by the cosmological scales of interest.
Then, the fifth force would be screened as in the Solar System, see the discussion
in section~\ref{sec:solar-system}, because of this ultra-local property, and there would
be no effect left on cosmological scales. In this case, the universe would be made of small
high-density clumps (set by the cutoff of the theory), built at high redshift, while
perturbations on cosmological scales would evolve according to General Relativity.
To address this issue, we need to go beyond perturbation theory and spherical
dynamics, as this is a highly non-linear and inhomogeneous problem.
In this article, we consider a thermodynamic analysis that provides
a simple analytic framework, which we present in section~\ref{sec:thermo-equilibrium}
below.

However, before we tackle this problem, we first describe in section~\ref{sec:cosmo-trajectory}
the evolution with redshift of the scales that enter the non-linear regime.
This allows us to distinguish various regimes: while at high redshift the non-linear transition
is set by the fifth force, more precisely by the pressure-like term $\propto \nabla^2\delta$
in Eq.(\ref{delta-evol}) associated with the ultra-local potential $\ln A$, at low redshift it is set
by the standard Newtonian gravity [the right-hand side in Eq.(\ref{delta-evol})].

In this section we focus on models (II) and (III), because model (I) actually damps small-scale
perturbations, so that the issue of a possible sensitivity to small scales does not arise.
Moreover, we have seen in section~\ref{sec:quasi-static-background} that such scenarios
are disfavored on theoretical grounds because they are not stable with respect to a small
kinetic term.

\subsection{Evolution of the cosmological non-linear transition for the model (II)}
\label{sec:cosmo-trajectory}

\begin{figure}
\begin{center}
\epsfxsize=8. cm \epsfysize=5.5 cm {\epsfbox{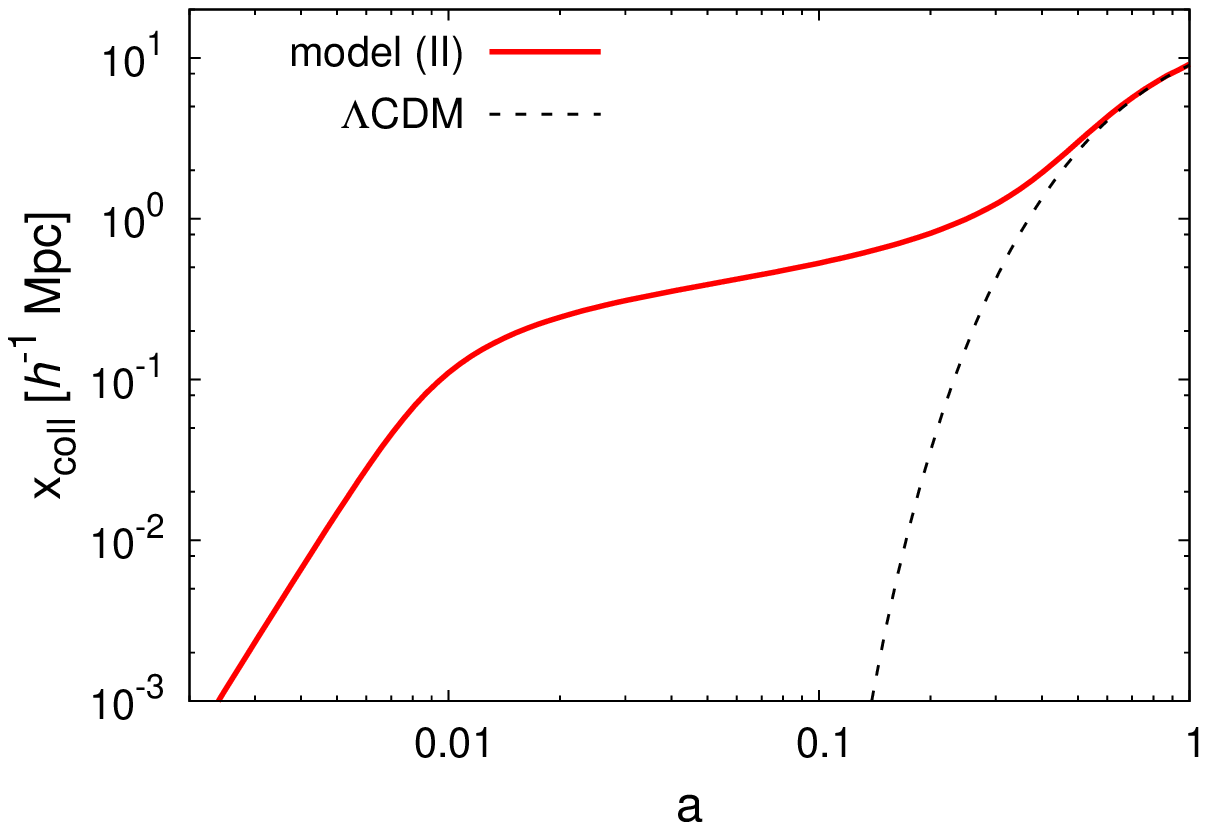}}\\
\epsfxsize=8. cm \epsfysize=5.5 cm {\epsfbox{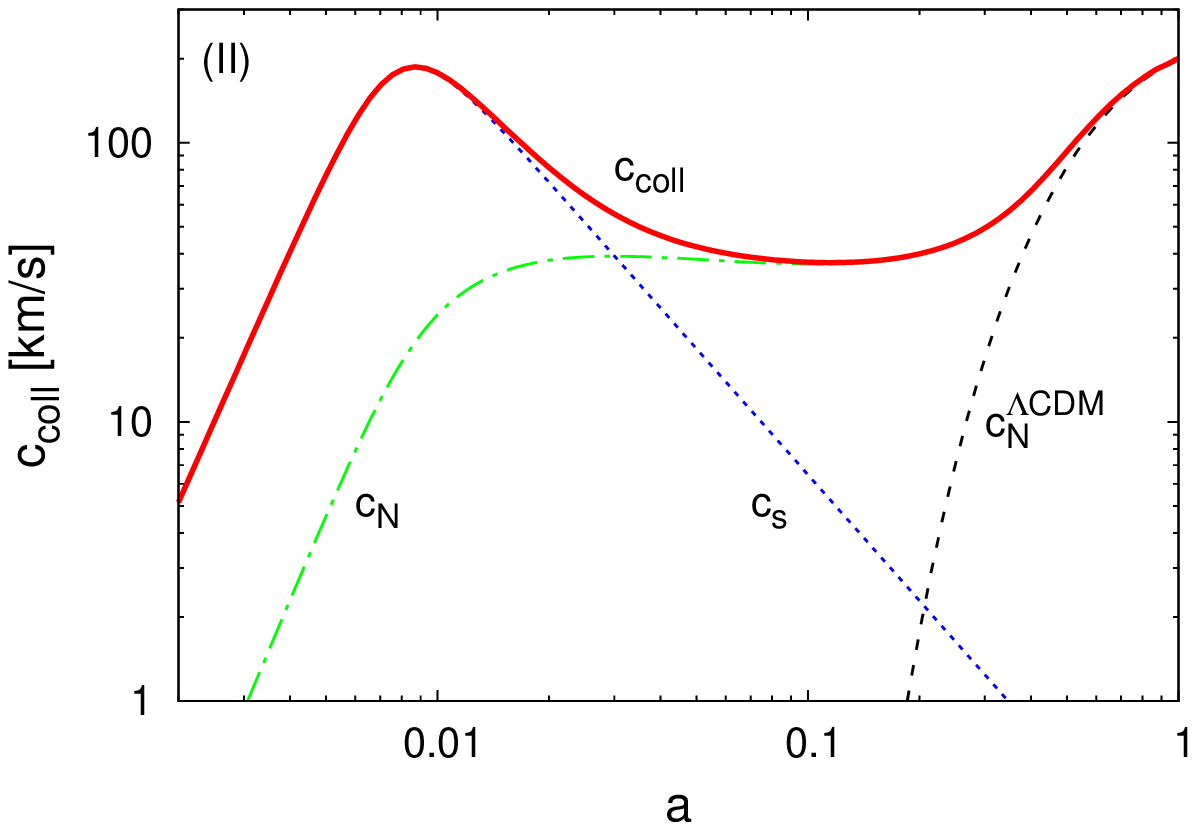}}
\end{center}
\caption{
{\it Upper panel:} collapse radius $x_{\rm coll}(z)$ (in comoving coordinates)
as a function of the scale factor $a$. The solid line is for the model (II) while the dashed line
is for the $\Lambda$-CDM cosmology.
{\it Lower panel:} collapse velocity scale $c_{\rm coll}(z)$ (solid line) as a function of the
scale factor $a$ for the model (II). The dotted and dot-dashed lines are $c_s$ and
$c_{\rm N}$ whereas the dashed line on the right is the result
$c_{\rm coll}^{\Lambda \rm -CDM} = c_{\rm N}^{\Lambda \rm -CDM}$
in the case of the $\Lambda$-CDM cosmology.
}
\label{fig:r_c-v_c-II}
\end{figure}

As explained in previous sections for models (II) and (III), at high redshift the fifth force amplifies
the growth of structures and the non-linear transition $x_{\rm coll}(z)$ is much greater than for the
$\Lambda$-CDM cosmology, as seen from the linear power spectrum in
Fig.~\ref{fig_Dp_II}. Using comoving coordinates, we define this non-linear scale by
\beq
\Delta^2_L(\pi/x_{\rm coll},z) = 1.5
\label{condition-collapse-radius}
\eeq
and we show $x_{\rm coll}(z)$ in the upper panel of Fig.~\ref{fig:r_c-v_c-II}.
The factor $1.5$, which should be order unity, is chosen to give a scale of order
$8 h^{-1}{\rm Mpc}$ at $z=0$, when the Newtonian gravity dominates and
we recover the usual $\Lambda$-CDM behavior.
We define the non-linear scale $x_{\rm coll}(z)$ by the condition
(\ref{condition-collapse-radius}) on the Fourier-space power spectrum $\Delta^2_L(k)$ rather
than the real-space linear variance $\sigma^2_L(x)$ because of the steep growth of the
linear growing mode $D_+(k,t)$ at high $k$.
This makes the linear variance $\sigma^2_L$ divergent or ill-defined,
dominated by a small-scale cutoff, but this is not physical because the linear theory
cannot be trusted in the non-linear regime. Using $\Delta^2_L(k)$ allows us to avoid this
problem (in contrast, for the $\Lambda$-CDM cosmology, where the slope of the linear
power spectrum decreases at higher $k$, using either $\Delta^2_L(k)$ or
$\sigma^2_L(x)$ gives similar results).

To perform the thermodynamic analysis presented in section~\ref{sec:thermo-equilibrium}
below, we shall need the initial kinetic energy or typical velocity of the collapsing region.
From the evolution equation (\ref{delta-evol}) of the linear density, we define an effective
velocity scale $c_{\rm coll}$ by
\beq
c_{\rm coll}^2(z) = c_s^2 + c_{\rm N}^2 ,
\label{ceff-def}
\eeq
with
\beq
c_s^2 = \epsilon_1 c^2 , \;\;\;\;
c_{\rm N}^2 = (1+\epsilon_1) \frac{3\Omega_{\rm m}}{2\pi^2}
\left( H a x_{\rm coll} \right)^2 .
\label{cs-cN-def}
\eeq
The factor $c_s^2$ comes from the pressure-like term $\epsilon_1 c^2 \nabla^2\delta$
in Eq.(\ref{delta-evol}) while the term $c_{\rm N}^2$ comes from the right-hand
side, associated with the usual gravitational force (where the Newton constant is amplified
by the negligible factor $\epsilon_1 \ll 1$).
We show our results in the lower panel of Fig.~\ref{fig:r_c-v_c-II}.
We also display the case of the $\Lambda$-CDM cosmology where
$c_{\rm coll}^{\Lambda \rm -CDM} = c_{\rm N}^{\Lambda \rm -CDM}$
as there is no pressure-like term.
It gives $c_{\rm coll}^{\Lambda \rm -CDM} \sim 200 \, {\rm km/s}$ at $z=0$,
which is indeed of the order of the velocities associated with collapsed structures today.
It is a bit low, by a factor two if we compare with large clusters of galaxies, which is not
surprising as the relation (\ref{cs-cN-def}) is only an order of magnitude estimate,
but this is sufficient for our purposes.

The component $c_s$, associated with the pressure-like term associated with the
fifth-force potential $\ln A$, dominates at high redshift. Its amplitude follows the rise and
fall of $\epsilon_1(z)$ displayed in Fig.~\ref{fig_Aa_1}, with a peak at
$z_{\alpha} \sim \alpha^{-1/3} \sim 100$.
The component $c_{\rm N}$, associated with the Newtonian gravity, explicitly depends
on the scale $r_{\rm coll}(z)$. It grows with time, along with $r_{\rm coll}(z)$, and
dominates at late times, $a \gtrsim 0.03$. The plateau for $0.01 \lesssim a \lesssim 0.2$
follows from the very slow growth of $r_{\rm coll}(z)$ found in the upper panel in this
redshift range.
This can be understood from the peak at $z_{\alpha} \sim 100$ of the fifth-force
characteristic amplitude $\epsilon_1$ and from the analysis of the linear growing modes
and of the spherical collapse shown in Figs.~\ref{fig_Dp_II} and \ref{fig_delta_II}.
As seen in the previous sections, the fifth force amplifies the growth of structures with a peak
at $z_{\alpha}$ and a strong dependence on scales, following the $k^2$ factor in
Eq.(\ref{eps-def}).
As can be seen in Fig.~\ref{fig_delta_II}, the main effect is that wave numbers
higher than the characteristic value $k_{\alpha}^{\rm min} \sim 3 h {\rm Mpc}^{-1}$ of
Eq.(\ref{k-alpha-min}) become strongly amplified and reach the non-linear regime at
$z_{\alpha}$, with a steep scale dependence of $D_+(k)$.
This leads to the steady rise of $r_{\rm coll}(z)$ and $c_{\rm coll}(z)$ until $z_{\alpha}$ and its
subsequent stop as the fifth force declines and the steep scale dependence imprinted on the
linear perturbations implies that it requires a very long time for the usual gravitational instability
to push the non-linear regime towards greater scales. We recover the standard
$\Lambda$-CDM behavior at late times, $a > a_{\Lambda\rm -CDM} \simeq 0.2$, when the
Newtonian gravity dominates and the scales that turn non-linear had not been significantly
impacted by the fifth force at $z_{\alpha}$ (i.e. $x \gtrsim 1/k_{\alpha}^{\rm min}$).

Thus, we can distinguish three regimes from Fig.~\ref{fig:r_c-v_c-II},
defining $a_{c_s/c_{\rm N}} \simeq 0.03$ as the transition where $c_s = c_{\rm N}$
and $a_{\Lambda\rm -CDM} \sim 0.2$ as the time when we recover the
$\Lambda$-CDM behavior.
At early times, $a<a_{\alpha}=0.01$, the fifth force dominates and increasingly large scales
enter the non-linear regime. This is the period when the thermodynamic analysis of
section~\ref{sec:thermo-equilibrium} below applies and allows us to estimate the behavior
of the system in the non-linear regime.
For $a_{\alpha} < a < a_{c_s/c_{\rm N}}$, the fifth force remains dominant but
$r_{\rm coll}(z)$ does not significantly grow so that no new structures form.
For $a_{c_s/c_{\rm N}} < a < a_{\Lambda\rm -CDM}$, the Newtonian gravity becomes
dominant but again $r_{\rm coll}(z)$ does not significantly grow so that no larger structures
form. However, some top-down structure formation might occur (in the range where gravity
remains dominant), as in hot dark matter scenarios.
Finally, for $a_{\Lambda\rm -CDM} < a < 1$, we recover the $\Lambda$-CDM
behavior as Newtonian gravity is dominant and the linear power spectrum on the
large scales that now turn non-linear has not been strongly modified by earlier fifth-force
effects.

We can note that this history singles out a characteristic mass and velocity scale, associated
with the plateau found in Fig.~\ref{fig:r_c-v_c-II} over $0.02 \lesssim a \lesssim 0.2$.
This yields
\beqa
&& x_* \sim 0.355 \; h^{-1} {\rm Mpc} , \;\;\;
M_* \sim 2 \times 10^{10} \; h^{-1} M_{\odot} , \nonumber \\
&& c_* \sim 50 \; {\rm km/s} .
\label{x*-M*-c*-def}
\eeqa
As in Fig.~\ref{fig_eta_R_M}, we recover galaxy scales, more precisely here the scales
associated with small galaxies.
Again, it is tempting to wonder whether this could help alleviate some of the problems
encountered on galaxy scales by the standard $\Lambda$-CDM scenario.
However, this would require detailed numerical studies that are beyond the scope of this
paper.

\subsection{Thermodynamic equilibrium in the fifth-force regime for the model (II)}
\label{sec:thermo-equilibrium}

As explained above, we have so far implicitly assumed that during the initial phase
$a<a_{\alpha}$ of structure formation governed by the fifth force the
density field remains smooth on cosmological scales.
In other words, we assumed for the computation of the fifth force in linear theory and
for the spherical collapse dynamics that the gradient of the fifth force potential,
$\nabla \ln A$, is set by the density field smoothed on cosmological scales.
This is not obvious because small scales, $x\leq x_{\rm coll}(z)$, have already turned
non-linear at high redshift, $z > z_{\alpha}$, as seen in the upper panel in
Fig.~\ref{fig:r_c-v_c-II}.
Then, the density field could have become strongly inhomogeneous, made of objects
of mass $M_{\rm coll}(z_{\rm cutoff})$ formed at a high cutoff redshift $z_{\rm cutoff}$
amid empty space. Then, the gradient of the fifth force potential $\nabla\ln A$ at a given
location in space would be unrelated with the gradient of the density field smoothed
on cosmological scales. This strong sensitivity to the small-scale distribution of the density
field does not arise for the Newtonian gravitational force, because the force at a distance
$d$ explicitly depends on the density smoothed over a size of the same order, through the
integral ${\bf F} = {\cal G}_{\rm N} \int d^3r \rho({\bf r}) {\bf r}/r^3$.
This comes from the fact that the Newtonian potential is given by the Poisson equation
(\ref{tPsi-N-J}), $\Psi_{\rm N} \propto \nabla^{-2} \rho$, which regularizes the
density field, whereas the fifth force potential $\ln A$ is a direct function of the local
density through Eq.(\ref{KG-A=1}).
Thus, this issue only arises in the first stage $a<a_{\alpha}$ found in
Fig.~\ref{fig:r_c-v_c-II}, where new scales enter the non-linear regime and are
dominated by the fifth force.

To address this question we need to go beyond perturbation theory and spherical
dynamics, as this is a highly non-linear and inhomogeneous problem.
We use a thermodynamic analysis, which provides a simple analytic framework, and
we leave dedicated numerical studies for future works.
Assuming that the scales that turn non-linear because of the fifth force at high redshift
reach a statistical equilibrium through the rapidly changing effects of the fluctuating
potential, in a fashion somewhat similar to the violent relaxation that takes
place for gravitational systems \cite{Lynden-Bell1967},
we investigate the properties of this thermodynamic equilibrium.
This first requires the study of the phase transitions and of the phase
diagram associated with the potential $\ln A(\rho)$ that defines our models.
Because this issue arises from the behavior of the fifth force in the regime where it
dominates over Newtonian gravity, we can neglect the latter to investigate this problem.
Note that contrary to the usual gravitational case, the
potential $\ln A$ is both bounded and short-ranged , so that we cannot build infinitely
large negative (or positive) potential energies and a stable thermodynamic equilibrium
always exists, and it is possible to work with either micro-canonical, canonical or
grand-canonical ensembles.
In this respect, a thermodynamic analysis is better suited for such systems than for
standard 3D gravitational systems, where the potential energy is unbounded
from below and stable equilibria do not always exist, and different statistical
ensembles are not equivalent \cite{Padmanabhan1990}.

\subsubsection{Thermodynamic phase transition and phase diagram}
\label{sec:thermo-diagram}

We work in the grand-canonical ensemble, where the dark matter particles are
confined in a box of size $x$ (the scale that reaches the non-linear regime at a given
redshift) with a mean temperature $T=1/\beta$ and chemical potential $\mu$.
These two thermodynamic quantities will be set by the initial energy and density
at the nonlinear transition $x_{\rm coll}(z)$.
If the potential $\ln A(\rho)$ were constant, there would be no fifth force and as usual
the potential would disappear as an irrelevant constant in the statistical analysis.
Then, we would recover the homogeneous equilibrium of the perfect gas, without
interactions.
However, because of the variations of $\ln A$ we expect inhomogeneities to develop.
For the models (II) and (III), where the potential $\ln A(\rho)$ decreases at higher
density, see Eqs.(\ref{lnA-rho-model-II}) and
(\ref{lnA-chi-model-IIIa})-(\ref{lnA-rho-model-III-b}), the fifth force generates
instabilities, as already seen from the behavior of linear perturbations, and the medium
 can be expected to become strongly inhomogeneous, with small high-density clumps
amid large voids. However, this outcome depends on the temperature $1/\beta$.
At high temperature, we are dominated by the kinetic energy and the potential
energy is negligible as $\ln A$ is bounded. Then, we recover the perfect gas
with an homogeneous distribution. At low temperature, the potential becomes
important and we expect the system to present strong inhomogeneities.
As for the thermodynamics of many standard systems, we shall find
that there is a phase transition between the homogeneous and the inhomogeneous
phases at a critical temperature $T_c=1/\beta_c$.
We do not need to consider cases such as model (I), where $\ln A(\rho)$ increases
at higher density and the fifth force has a stabilizing influence that prevents the
formation of small-scale inhomogeneities, as already seen from the behavior of linear
perturbations.

In the continuum limit, where the mass $m$ of the dark matter particles goes to zero,
we describe the system by the smooth phase-space distribution function
$f(\vx,\vv)$. The mass $M$, the energy $E$ and the entropy $S$ of the system read as
\cite{Padmanabhan1990,Chavanis2005,Valageas2006}
\beqa
M & = &\int d^3 x d^3 v \; f(\vx,\vv) ,
\label{mass-thermo} \\
E & = & \int d^3 x d^3 v \; f(\vx,\vv) \left( \frac{v^2}{2} + c^2 \ln A[\rho(\vx)] \right) ,
\hspace{0.5cm}
\label{energy-thermo} \\
S & = & -\int d^3 x d^3 v \; f(\vx,\vv) \; \ln \frac{f(\vx,\vv)}{f_0} ,
\label{entropy-thermo}
\eeqa
where $f_0$ is a normalization constant and we used the fact that the potential
$\ln A$ is a function of the local density.
In the grand-canonical ensemble the statistical equilibrium is obtained by minimizing
the grand-canonical potential $\Omega$, which is given by
\beq
\Omega = E - S/\beta - \mu M ,
\label{grand-potential}
\eeq
where $\beta$ and $\mu$ are the inverse temperature and the chemical potential.
With our notations $\beta$ has units of inverse squared velocity and $\mu$ has units
of squared velocity.
The equilibrium phase-space distribution is given by the minimum of the grand potential,
${\cal D} \Omega / {\cal D} f = 0$. This yields
\beqa
f(\vx,\vv) = f_0 \; e^{-\beta \left(v^2/2 + c^2 \ln A + c^2 d \ln A / d \ln \rho \right)
+ \beta\mu - 1 } .
\label{phase-space-eq}
\eeqa
Since $\ln A$ only depends on the positions of the particles but not on their velocities,
we recover as expected the Maxwellian distribution over velocities,
$f(\vx,\vv) \propto \rho(\vx) e^{-\beta v^2/2}$.
The proportionality factor is obtained by integrating over velocities, which gives
the usual result
\beq
f(\vx,\vv) = \left( \frac{\beta}{2\pi} \right)^{3/2} \rho(\vx) \; e^{-\beta v^2/2} ,
\label{f-rho-beta}
\ee
and Eq.(\ref{phase-space-eq}) yields
\beq
\rho(\vx) = f_0  \left( \frac{2\pi}{\beta} \right)^{3/2}
e^{-\beta c^2 \left( \ln A + d \ln A / d \ln \rho \right)+ \beta\mu - 1 } .
\label{rho-lnA-eq}
\eeq

Because of the specific form of the potential $\ln A$, which is local and only depends
on the local density $\rho(\vx)$, the thermodynamic equilibrium condition
(\ref{rho-lnA-eq}) factorizes over different positions $\vx$. The different space locations
are thus decoupled and we can omit the space coordinate $\vx$: the equilibrium
condition (\ref{rho-lnA-eq}), which was a functional equation over the field $\rho(\vx)$,
simplifies to an ordinary function of the local density $\rho$.
As noticed in section~\ref{sec:rhoc}, it is convenient to introduce the rescaled
dimensionless potential and density $\lambda$ and $\hat\rho$, from
Eqs.(\ref{lambda-def}) and (\ref{rho-hat-def}).
Defining also the rescaled dimensionless inverse temperature $\hat\beta$ and
chemical potential $\hat\mu$,
\beq
\hat\beta = \alpha c^2 \beta ,
\label{beta-hat-def}
\eeq
\beq
\hat\mu = \ln \left[ \frac{\alpha f_0 c^3}{{\cal M}^4}
\left( \frac{2\pi}{\beta c^2} \right)^{3/2} \right] + \beta\mu -1 ,
 \label{mu-hat-def}
\eeq
the equilibrium condition (\ref{rho-lnA-eq}) reads as
\beq
\hat\mu = \theta + \hat\beta \, \nu(\theta) ,
\label{mu-theta-nu-eq}
\eeq
where we introduced
\beq
\theta = \ln\hat\rho , \;\;\; \nu(\theta) = \lambda + \frac{d\lambda}{d\theta} .
\label{theta-nu-def}
\eeq
For a given value of the rescaled inverse temperature $\hat\beta$ and chemical
potential $\hat\mu$, this gives the equilibrium density $\theta$ as the solution
of the implicit equation (\ref{mu-theta-nu-eq}).
In terms of these dimensionless variables, the grand-canonical potential
(\ref{grand-potential}) reads as
\beq
\Omega = \frac{{\cal M}^4 c^2 V}{\hat{\beta}} \hat\Omega \;\;\; \mbox{with} \;\;\;
\hat\Omega = e^{\theta} \left[ \hat\beta \lambda - \hat\mu-1 + \theta \right] ,
\label{hat-Omega-def}
\eeq
where $V$ is the total volume of the system. Thus, the equilibrium equation
(\ref{mu-theta-nu-eq}) is the condition $d\hat\Omega/d\theta = 0$, as the
thermodynamic equilibrium corresponds to the minimization of the grand-potential.

\begin{figure}
\begin{center}
\epsfxsize=8. cm \epsfysize=5.5 cm {\epsfbox{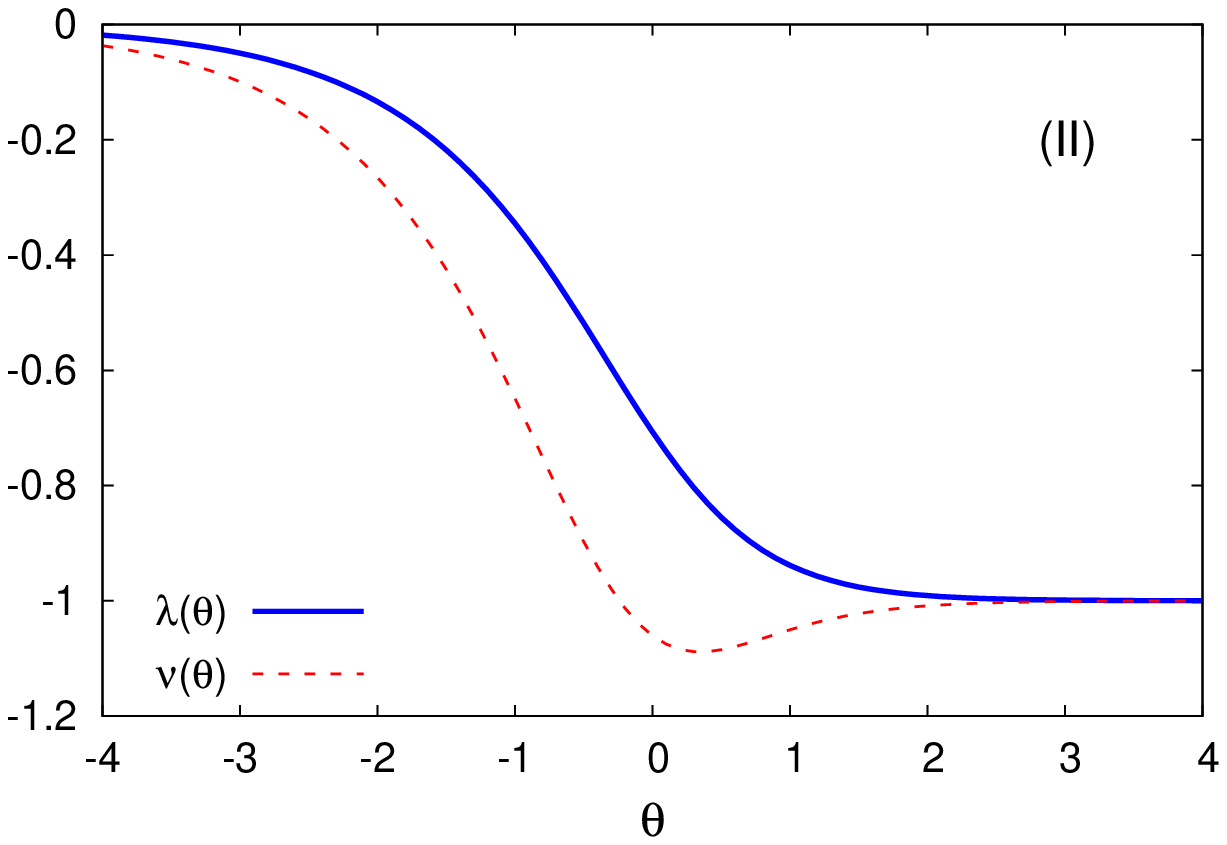}}\\
\epsfxsize=8. cm \epsfysize=5.5 cm {\epsfbox{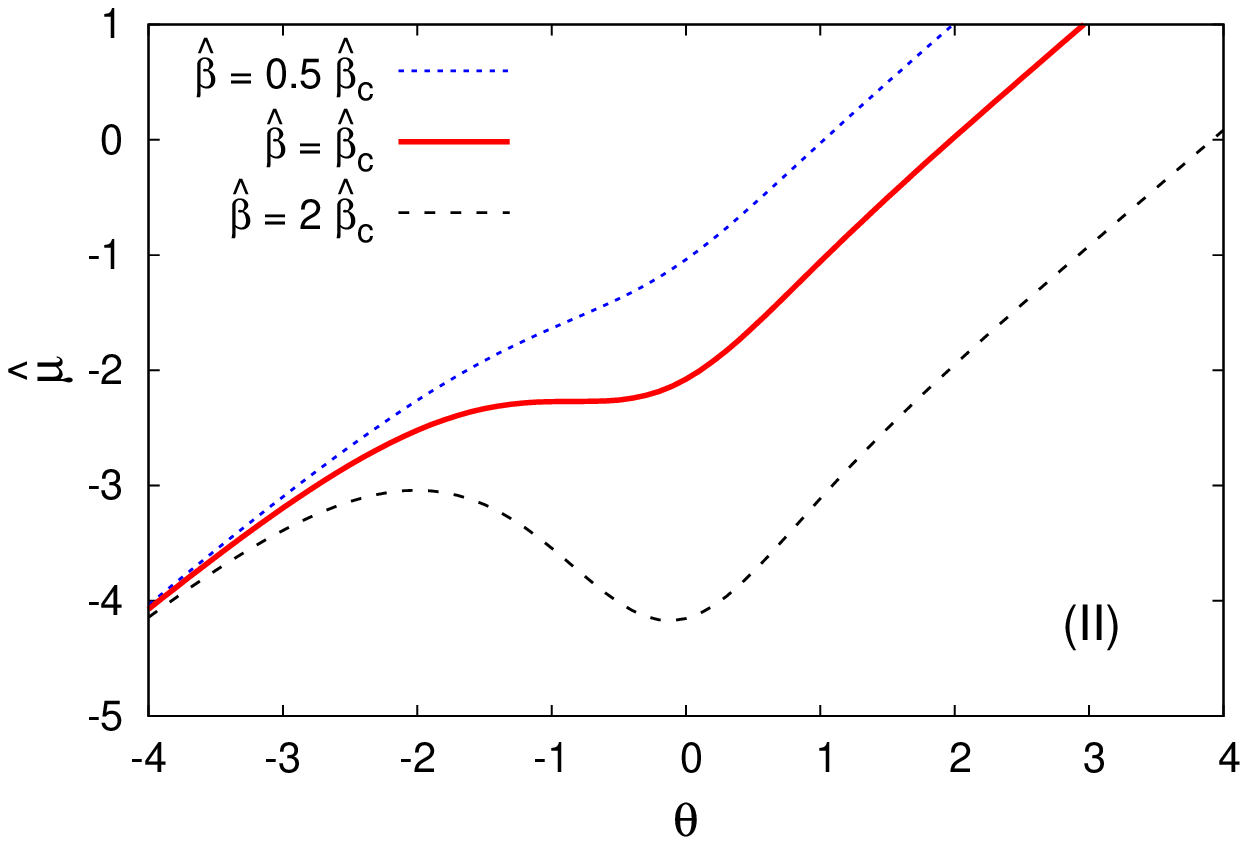} }
\end{center}
\caption{{\it Upper panel:} fifth-force potential functions $\lambda(\theta)$ and
$\nu(\theta)$ for the model (II).
{\it Lower panel:} thermodynamic equilibrium relation
$\hat{\mu} = \hat{\mu}(\theta,\hat{\beta})$ as a function of $\theta$,
fixing $\hat{\beta}=0.5 \hat{\beta}_c, \hat{\beta}_c$ and $2 \hat{\beta}_c$.}
\label{fig:nu-mu-II}
\end{figure}

It is convenient to analyse the system at a fixed temperature, which corresponds to
a given initial velocity dispersion, as a function of the chemical potential $\hat\mu$
or of the density $\hat\rho$, seen as conjugate variables.
At high temperature, $\hat\beta\rightarrow 0$, Eq.(\ref{mu-theta-nu-eq}) becomes
$\hat\mu=\theta$ and there is a unique density for each $\hat\mu$. This corresponds
to the high-temperature homogeneous phase where we recover the perfect gas as
the potential energy is negligible.
At low temperature, $\hat\beta\rightarrow \infty$, the right hand side of
Eq.(\ref{mu-theta-nu-eq}) can become non-monotonic so that for some values of
the chemical potential $\hat\mu$ there are several solutions $\theta_i$.
This corresponds to the inhomogeneous phase, where the system splits over several
regions of different densities $\theta_i$, with an admixture such that the mean
density over the large scale $x=V^{1/3}$ is the initial density $\bar\rho$,
see \cite{Balian2007} for an analysis of such phase transitions.

We first consider the model (II) defined in Eq.(\ref{model-II-def}). From
Eq.(\ref{lnA-rho-model-II}), with again $\chi_*=-2$, we obtain
\beq
{\rm (II) :} \;\;\; \lambda(\theta) = - \frac{1}{\sqrt{1+e^{-2\theta}}} , \;\;\;
\nu(\theta) = - \frac{1+2 e^{-2\theta}}{(1+e^{-2\theta})^{3/2}} .
\label{nu-theta-II}
\eeq
We show these two functions in the upper panel of Fig.~\ref{fig:nu-mu-II}.
From Eq.(\ref{mu-theta-nu-eq}), the function $\hat\mu(\theta)$, at fixed inverse
temperature $\hat\beta$, is strictly monotonic if
$d\hat\mu/d\theta=1+\hat\beta d\nu/d\theta > 0$.
Therefore, the function $\hat\mu(\theta)$ becomes non-monotonic below the temperature
$1/\hat\beta_c$, where $\hat\beta_c$ is given by the most negative value of
$d\nu/d\theta$,
\beq
\hat\beta_c = \frac{-1}{\min (d\nu/d\theta)} =
\frac{(15+\sqrt{105})^{5/2}}{16(51+5\sqrt{105})} \simeq 1.96
\label{betac-II}
\eeq
We display in the lower panel of Fig.~\ref{fig:nu-mu-II} the function $\hat\mu(\theta)$ for
three values of $\hat\beta$. As explained above, for low $\hat\beta$ (i.e. high temperature)
the function $\hat\mu(\theta)$ is monotonic while for high $\hat\beta$
(i.e. low temperature) it is non-monotonic over some range of densities,
with a first-order phase transition at $\hat\beta_c$.
Then, for $\hat\beta < \hat\beta_c$, we always have a single solution $\theta(\hat\mu)$
for any chemical potential $\hat\mu$.
For $\hat\beta > \hat\beta_c$, in a finite range $[\hat\mu_1,\hat\mu_2]$ and
$[\theta_1,\theta_2]$, we have three solutions, $\theta_- < \theta_{\rm m} < \theta_+$,
for a given chemical potential $\hat\mu$.
Both $\theta_-$ and $\theta_+$ are local minima of the grand-potential $\hat\Omega$
whereas $\theta_{\rm m}$ is a local maximum. Then, the physical solution
$\theta(\hat\mu)$ is the global minimum among $\{\theta_-,\theta_+\}$ (i.e. the deepest
minimum).
For $\hat\mu \simeq \hat\mu_1$, where we are close to the bottom left monotonic
branch in the lower panel of Fig.~\ref{fig:nu-mu-II} (i.e. the low-density branch), this global
minimum is the lowest-density one $\theta_-$.
For $\hat\mu \simeq \hat\mu_2$, where we are close to the upper right monotonic branch
(i.e. the high-density branch), this global minimum is the highest-density one $\theta_+$.
Then, there is a critical value $\hat\mu_s$ in between,
$\hat\mu_1<\hat\mu_s<\hat\mu_2$,
where we make the transition from $\theta_-$ to $\theta_+$. This happens at the crossing
of their values of the grand-potential, when
$\hat\Omega(\theta_-;\hat\mu_s) = \hat\Omega(\theta_+;\mu_s)$
\cite{Balian2007}.
This condition allows us to compute $\hat\mu_s$, as a function of $\hat\beta$, from
Eqs.(\ref{mu-theta-nu-eq}) and (\ref{hat-Omega-def}).
At leading order for large $\hat\beta$ we obtain
\beq
\hat\beta \rightarrow \infty : \;\; \hat\mu_s \sim - \hat\beta , \;\;\;
\theta_- \sim - \hat\beta , \;\;\; \theta_+ \sim \frac{\ln\hat\beta}{2} .
\label{mus-large-beta}
\eeq
This means that the transition occurs close to the low-density boundary
$(\theta_1,\hat\mu_1)$ of the multi-valued region.
Thus, we have a first-order phase transition, as the density of the system jumps from
$\theta_-(\hat\mu_s)$ to $\theta_+(\hat\mu_s)$ when the chemical potential goes
through $\hat\mu_s$. At $\hat\mu_s$, where $\hat\Omega_-=\hat\Omega_+$,
there is a coexistence of the two phases. One part of the volume $V$ is at the low
density $\theta_-$ and the other part at the high density $\theta_+$.
The relative fraction between the two phases is set by the mean density $\bar\theta$
of the full volume, $\theta_- \leq \bar\theta \leq \theta_+$, which is given by the initial
condition of the system (the constraint on the average density of the full system).

\begin{figure}
\begin{center}
\epsfxsize=8. cm \epsfysize=5.5 cm {\epsfbox{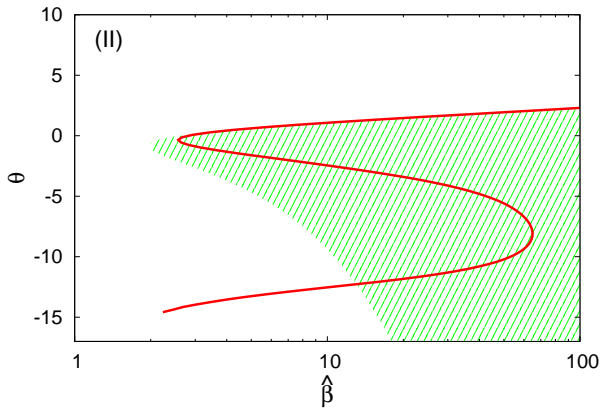}}
\end{center}
\caption{
Thermodynamic phase diagram of model (II). The shaded area is the region
of initial inverse temperature $\hat\beta$ and density $\theta$ where the system
reaches an inhomogeneous thermodynamic equilibrium. The white area corresponds to
the homogeneous phase.
The solid line is the cosmological trajectory $(\hat\beta_{\rm coll}(z),\theta_{\rm coll}(z))$.
}
\label{fig:cosmo-tra-II}
\end{figure}

The thermodynamic phase diagram of the system, in the inverse temperature - density
plane, is shown by the shaded area in Fig.~\ref{fig:cosmo-tra-II}.
This domain is limited at low $\hat\beta$ by the critical temperature
$\hat\beta_c$. The lower and upper limits of the domain are the curves
$\theta_-(\hat\beta) \equiv \theta_-(\hat\mu_s(\hat\beta),\hat\beta)$ and
$\theta_+(\hat\beta) \equiv \theta_+(\hat\mu_s(\hat\beta),\hat\beta)$, which obey the
asymptotes (\ref{mus-large-beta}).
We choose the $(\hat\beta,\theta)$ plane to display the phase diagram, rather
than $(\hat\beta,\hat\mu)$ for instance, because the density is a more direct physical
variable than the chemical potential, while the temperature $1/\hat\beta$ is also
directly related to the initial kinetic energy.
Whereas in the $(\hat\beta,\hat\mu)$ plane the transition appears as a critical line
$\hat\mu_s(\hat\beta)$, in the $(\hat\beta,\theta)$ plane it appears as an extended
domain, because the critical line $\hat\mu_s(\hat\beta)$ corresponds to the jump
from $\theta_-$ to $\theta_+$ over the density.
The meaning of the diagram in Fig.~\ref{fig:cosmo-tra-II} is the following.
If the average initial temperature and density, $(1/\hat\beta,\theta)$, fall outside of
the shaded region, the system remains in the homogeneous phase.
If the initial condition falls inside the shaded region, the system becomes inhomogeneous
and splits over domains with density $\theta_-$ or $\theta_+$, with a proportion such that
the total mass over the full volume is conserved.

\subsubsection{Cosmological trajectory in the phase diagram}
\label{sec:thermo-diagram-cosmo}

Using the phase diagram of Fig.~\ref{fig:cosmo-tra-II}, we can now consider the
behavior of the collapsing scales $r_{\rm coll}(z)$ obtained in Fig.~\ref{fig:r_c-v_c-II},
in the time interval $a<a_{\alpha}$ where the new structures that reach the non-linear
regime are governed by the fifth-force potential $\ln A$.
For the typical density associated with the non-linearity transition we simply
take $\rho_{\rm coll}(z) = \bar\rho(z)$, as the transition corresponds to density contrasts
of order unity, hence
\beq
\rho_{\rm coll}(z) = \bar\rho(z) , \;\;\;
\theta_{\rm coll}(z) = \ln \left[ \frac{\alpha\bar\rho(z)}{{\cal M}^4} \right]  .
\label{theta-coll-def}
\eeq
At the thermodynamic equilibrium (\ref{f-rho-beta}) the kinetic energy reads
as $E_{\rm kin} = 3 M T/2 = 3 M / 2 \beta$. From the typical velocity scale
$c_{\rm coll}(z)$ of Eq.(\ref{ceff-def}) we use the simple estimate
\beq
\beta_{\rm coll}(z) = \frac{1}{c^2_{\rm coll}(z)} \;\;\; \mbox{hence} \;\;\;
\hat\beta_{\rm coll}(z) = \frac{\alpha c^2}{c^2_{\rm coll}(z)} .
\label{beta-coll-def}
\eeq
We show in Fig.~\ref{fig:cosmo-tra-II} the cosmological trajectory
$(\hat\beta_{\rm coll}(z),\theta_{\rm coll}(z))$ over the phase space diagram of the system
defined by the fifth-force potential $\ln A$ of the model (II).
The curve runs downwards to lower densities $\theta_{\rm coll}$ as cosmic time grows.
In agreement with the lower panel of Fig.~\ref{fig:r_c-v_c-II},
the inverse temperature $\hat\beta_{\rm coll}$ first decreases until $a_{\alpha}$,
as the velocity $c_{\rm coll}(z)$ grows.
Next, $\hat\beta_{\rm coll}$ increases while $c_{\rm coll}(z)$ decreases until
$a_{\Lambda\rm -CDM}$, when we recover the $\Lambda\rm -CDM$ behavior,
and $\hat\beta_{\rm coll}$ decreases again thereafter.
We are interested in the first era, $a<a_{\alpha}$, and we find that the cosmological
trajectory is almost indistinguishable from the upper boundary $\theta_+(\hat\beta)$
of the inhomogeneous thermodynamic phase.
Indeed, from Eq.(\ref{ceff-def}) and Fig.~\ref{fig:r_c-v_c-II} we have at early times
$c_{\rm coll} \simeq c_s$, hence $\hat\beta_{\rm coll} \simeq \alpha/\epsilon_1$.
Using Eq.(\ref{eps1-2-II}) we have at high densities, which also correspond
to $a<a_{\alpha}$, $\epsilon_1 \simeq \alpha \hat\rho^{-2} = \alpha e^{-2\theta}$,
hence
\beq
a \ll a_{\alpha} : \;\;\; \theta_{\rm coll} \sim \frac{1}{2} \ln \hat\beta_{\rm coll} ,
\label{theta-beta-coll-II}
\eeq
and we recover the asymptote (\ref{mus-large-beta}) of $\theta_+(\hat\beta)$.
Depending on the choice of some numerical factors, e.g. whether we modify
Eq.(\ref{beta-coll-def}) as $\beta_{\rm coll}(z) = 2/c^2_{\rm coll}(z)$ or
$\beta_{\rm coll}(z) = 1/2c^2_{\rm coll}(z)$, we can push $\theta_{\rm coll}$ slightly
above or below $\theta_+$.
If $\theta_{\rm coll}>\theta_+$ we are in the homogeneous phase and the system
remains at the initial density $\bar\rho$.
If $\theta_{\rm coll}<\theta_+$ we are in the inhomogeneous phase and the system
splits over regions of densities $\theta_+$ and $\theta_-$. However, as we remain close
to $\theta_+$ most of the volume is at the density $\theta_+ \simeq \theta_{\rm coll}$
and only a small fraction of the volume is at the low density $\theta_-$.
Neglecting these small regions, we can consider that in both cases the system remains
approximately homogeneous.
This means that, according to this thermodynamic analysis, the cosmological density
field does not develop strong inhomogeneities that are set by the cutoff scale
of the theory when it enters the fifth-force non-linear regime.
Therefore, density gradients remain set by the large-scale cosmological density
gradients and the analysis of the linear growing modes in
section~\ref{sec:Linear-perturbations} and of the spherical collapse
in section~\ref{sec:Spherical-collapse} are valid.
Of course, on small non-linear scales and at late times, where Newtonian gravity
becomes dominant, we recover the usual gravitational instability that we neglected in
this analysis and structure formation proceeds as in the standard $\Lambda$-CDM
case.

\subsection{Cosmological trajectory in the phase diagram for the model (III)}
\label{sec:cosmo-trajectory-III}

\begin{figure}
\begin{center}
\epsfxsize=8. cm \epsfysize=5.5 cm {\epsfbox{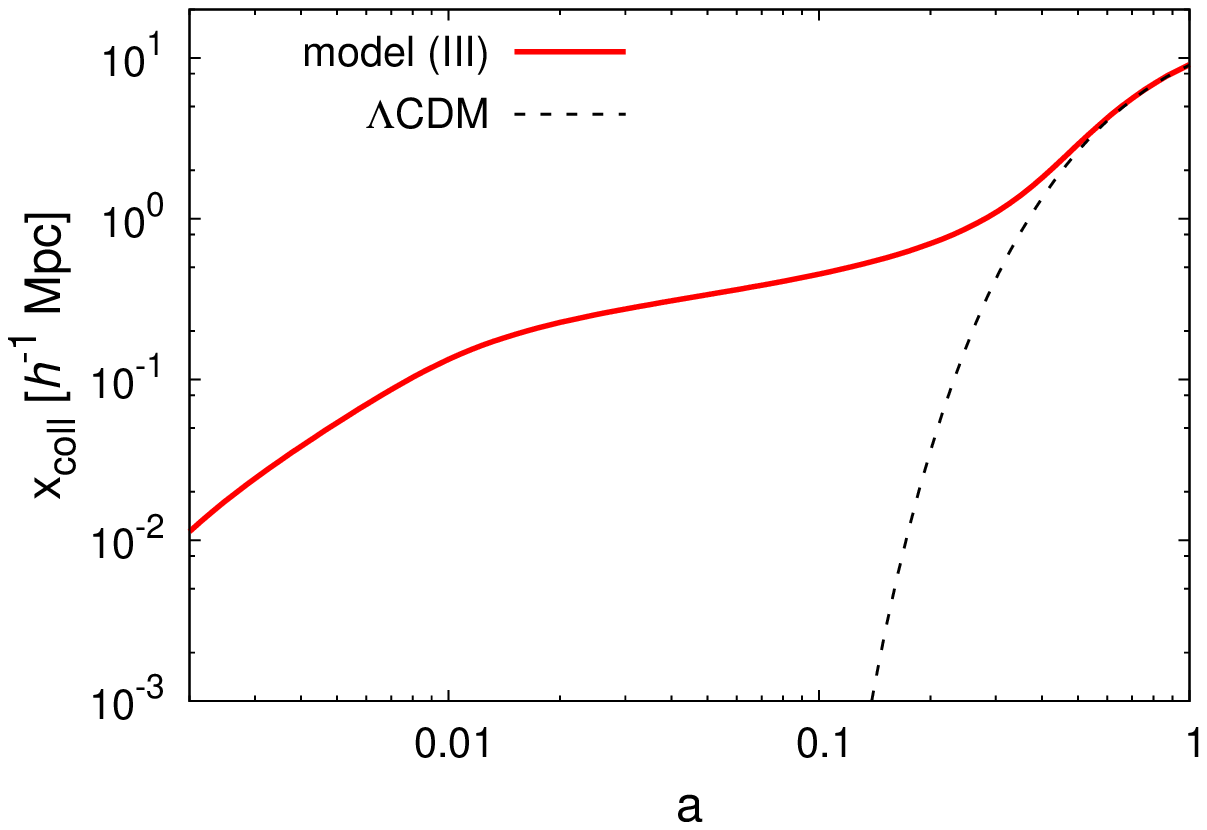}}\\
\epsfxsize=8. cm \epsfysize=5.5 cm {\epsfbox{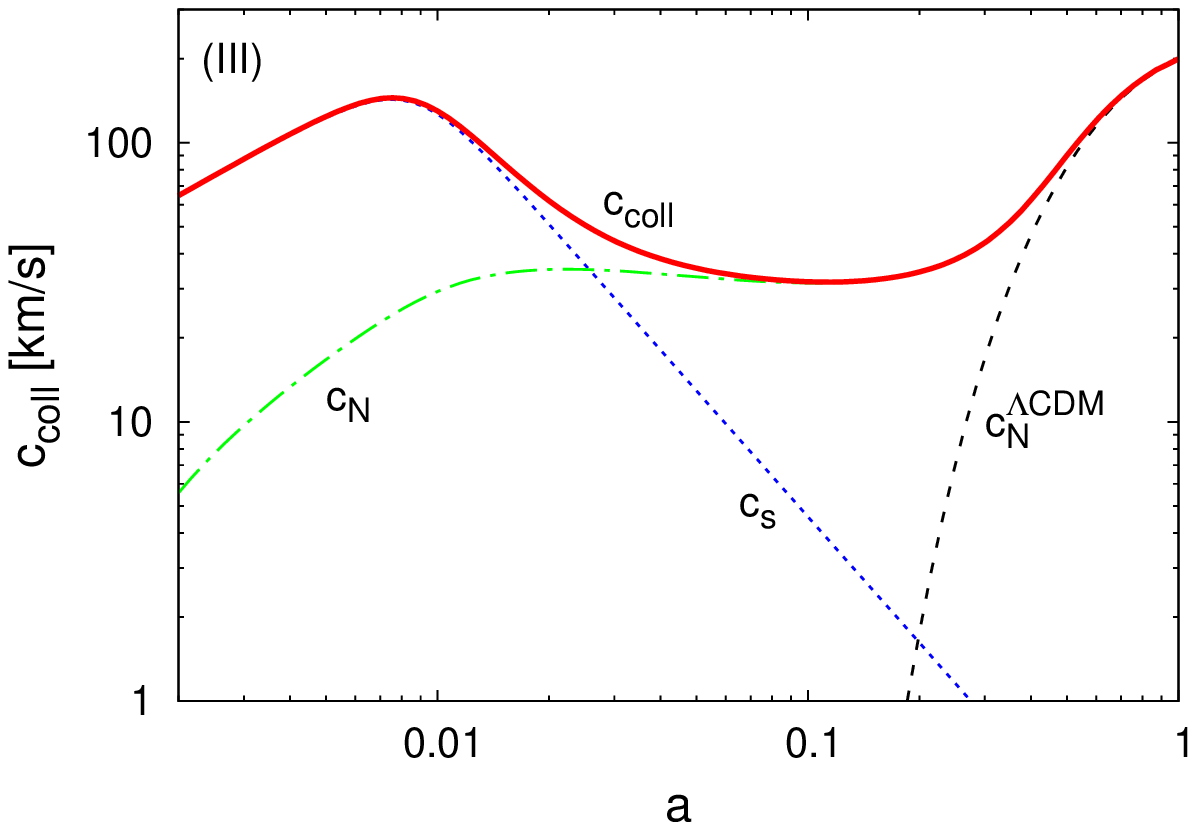}}
\end{center}
\caption{
{\it Upper panel:} collapse radius $x_{\rm coll}(z)$ (in comoving coordinates)
as a function of the scale factor $a$. The solid line is for the model (II) while the dashed line
is for the $\Lambda$-CDM cosmology.
{\it Lower panel:} collapse velocity scale $c_{\rm coll}(z)$ (solid line) as a function of the
scale factor $a$ for the model (II). The dotted and dashed lines are $c_s$ and $c_{\rm N}$
whereas the dashed line on the right is the result
$c_{\rm coll}^{\Lambda \rm -CDM} = c_{\rm N}^{\Lambda \rm -CDM}$
in the case of the $\Lambda$-CDM cosmology.
}
\label{fig:r_c-v_c-III}
\end{figure}

We can repeat the previous analysis for the model (III), which also amplifies density
perturbations and is similar to the model (II) in many respects.
We show the evolution of the non-linearity scale $r_{\rm coll}(z)$ and of the velocity
scale $c_{\rm coll}(z)$ in Fig.~\ref{fig:r_c-v_c-III}.
We can see that the behavior is similar to the one obtained in Fig.~\ref{fig:r_c-v_c-II}
for the model (II), except that $r_{\rm coll}(z)$ and $c_{\rm coll}(z)$ decrease more
slowly at high redshift, $z \gg z_{\alpha}$.
This is because the amplitude of the fifth force, as measured by $\epsilon_1$,
decreases more slowly at high $z$ for this model, as found in Fig.~\ref{fig_Aa_1}
and explained in section~\ref{sec:background-quantities}.
At lower redshifts, $z<z_{\alpha}$, the models behave in the same fashion,
as was also seen in Fig.~\ref{fig_Aa_1}. This leads to the same characteristic
mass and velocity scales (\ref{x*-M*-c*-def}), associated with the intermediate redshift
plateau, $z_{\Lambda\rm -CDM} \ll z \ll z_{\alpha}$.

The thermodynamic behavior is similar to the one obtained for the model (II) in
section~\ref{sec:thermo-diagram}.
As in the upper panel of Fig.~\ref{fig:nu-mu-II}, the fifth-force potential functions
$\lambda(\theta)$ and $\nu(\theta)$ again decrease from $0$ at low density to $-1$
at high density, except that $\nu(\theta)$ is now strictly decreasing and does not show
a local minimum at $\theta \simeq 0$ (which did not play a significant role anyway).
We again obtain a first-order phase transition as described in the lower panel
of Fig.~\ref{fig:nu-mu-II}.
The inverse critical temperature is now
\beq
\hat\beta_c \simeq 3.53 ,
\label{betac-III}
\eeq
and at low temperature we obtain the asymptotic behaviors
\beq
\hat\beta \rightarrow \infty : \;\; \hat\mu_s \sim - \hat\beta , \;\;\;
\theta_- \sim - \hat\beta , \;\;\; \theta_+ \sim 2 \ln\hat\beta .
\label{mus-large-beta-III}
\eeq

\begin{figure}
\begin{center}
\epsfxsize=8. cm \epsfysize=5.5 cm {\epsfbox{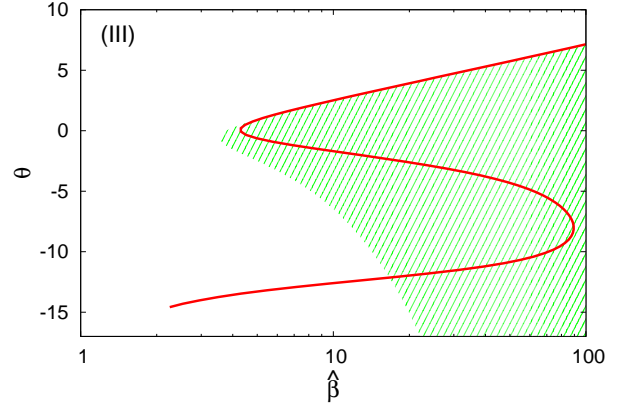}}
\end{center}
\caption{
Thermodynamic phase diagram of model (III). The shaded area is the region
of initial inverse temperature $\hat\beta$ and density $\theta$ where the system
reaches an inhomogeneous thermodynamic equilibrium. The white area corresponds to
the homogeneous phase.
The solid line is the cosmological trajectory $(\hat\beta_{\rm coll}(z),\theta_{\rm coll}(z))$.
}
\label{fig:cosmo-tra-III}
\end{figure}

We show the thermodynamic phase diagram of the model (III) in
Fig.~\ref{fig:cosmo-tra-III}. We recover the same features as for the model (II) shown in
Fig.~\ref{fig:cosmo-tra-II}, with a somewhat higher inverse critical temperature
$\hat\beta_c$ and upper boundary $\theta_+$ of the inhomogeneous phase.
The cosmological trajectory $(\hat\beta_{\rm coll}(z),\theta_{\rm coll}(z))$ again
roughly follows the upper boundary $\theta_+$ at high redshift, $z>z_{\alpha}$.
Indeed, using again $\hat\beta_{\rm coll} \simeq \alpha c^2/c_s^2 = \alpha/\epsilon_1$
and Eq.(\ref{eps12-model-III-b}), we obtain at high densities and redshifts
$\epsilon_1 \simeq \alpha/\sqrt{8\hat\rho} = \alpha e^{-\theta/2}/\sqrt{8}$,
hence
\beq
a \ll a_{\alpha} : \;\;\; \theta_{\rm coll} \sim 2 \ln \hat\beta_{\rm coll} ,
\label{theta-beta-coll-III}
\eeq
and we again recover the asymptote (\ref{mus-large-beta-III}) of $\theta_+(\hat\beta)$.
Therefore, as for model (II), we can conclude that during the fifth-force era
of structure formation, $a<a_{\alpha}$, density gradients up to the linear transition
remain set by large scales and do not suffer from cutoff-scale dependence, so that
the analysis of the linear growing modes in section~\ref{sec:Linear-perturbations}
and of the spherical collapse in section~\ref{sec:Spherical-collapse} are valid.

\subsection{Halo centers}
\label{sec:halo-centers}

It is interesting to apply the thermodynamic analysis presented above to the
inner radii of clusters and galaxies.
Indeed, we have seen in section~\ref{sec:screening-clusters} that
the fifth force can become large inside spherical halos and the ratio
$F_A/F_{\rm N}$ can actually diverge at the center for shallow density
profiles, see Fig.~\ref{fig_eta_z0} and
Eqs.(\ref{gamma-bound-I-II})-(\ref{gamma-bound-III}).
However, this analysis was based on dimensional and scaling arguments and it fails
if the density field becomes strongly inhomogeneous so that the typical density
inside the halo is very different from the global average density.
The thermodynamic analysis presented in section~\ref{sec:thermo-diagram}
neglected Newtonian gravity. However, we can also apply its conclusions to a regime
dominated by Newtonian gravity where at radius $r$ inside the halo the structures
built by gravity and the density gradients are on scale $r$.
Then, we can ask whether at this radius $r$ fifth-force effects may lead to a fragmentation
of the system on much smaller scales $\ell \ll r$. To study this small-scale behavior
we can neglect the larger-scale gravitational gradients $r$ and discard gravitational
forces.

\begin{figure}
\begin{center}
\epsfxsize=8. cm \epsfysize=5.5 cm {\epsfbox{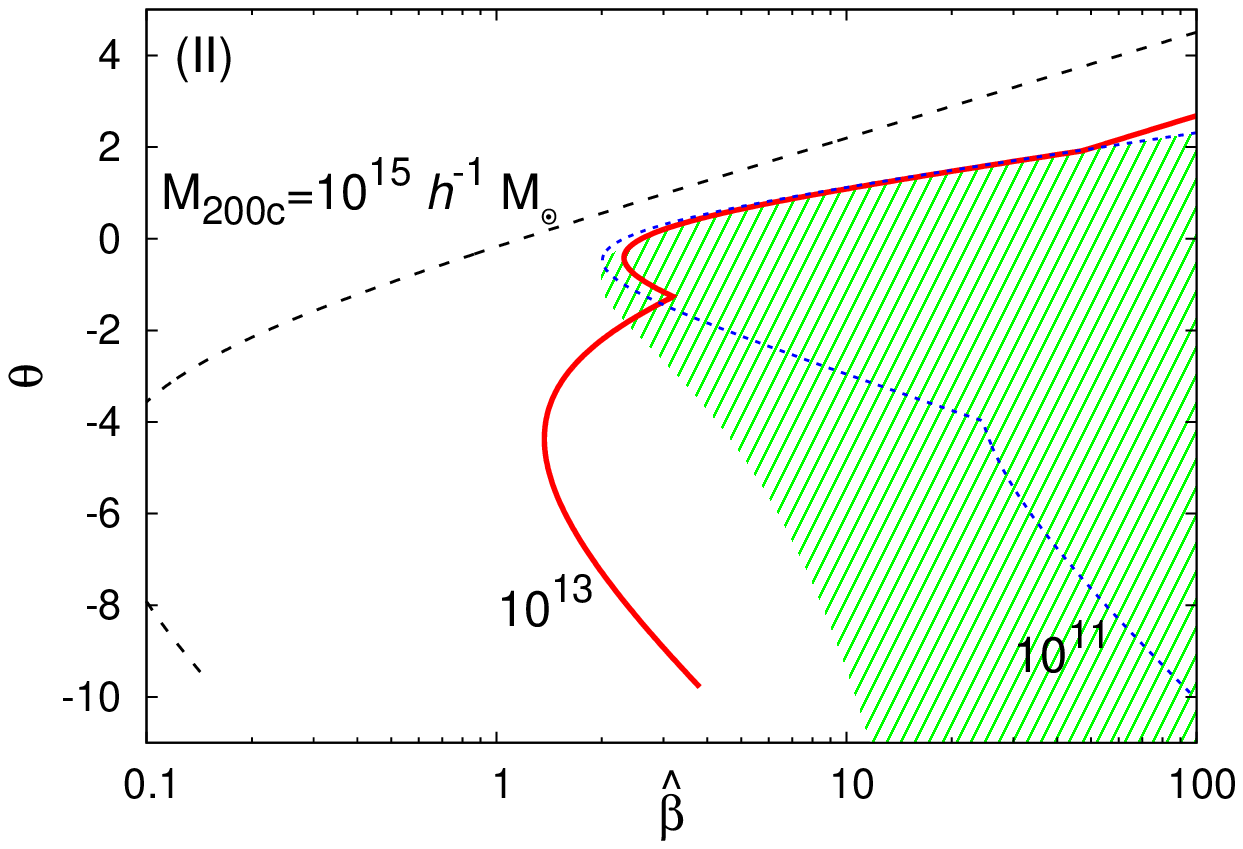}}\\
\epsfxsize=8. cm \epsfysize=5.5 cm {\epsfbox{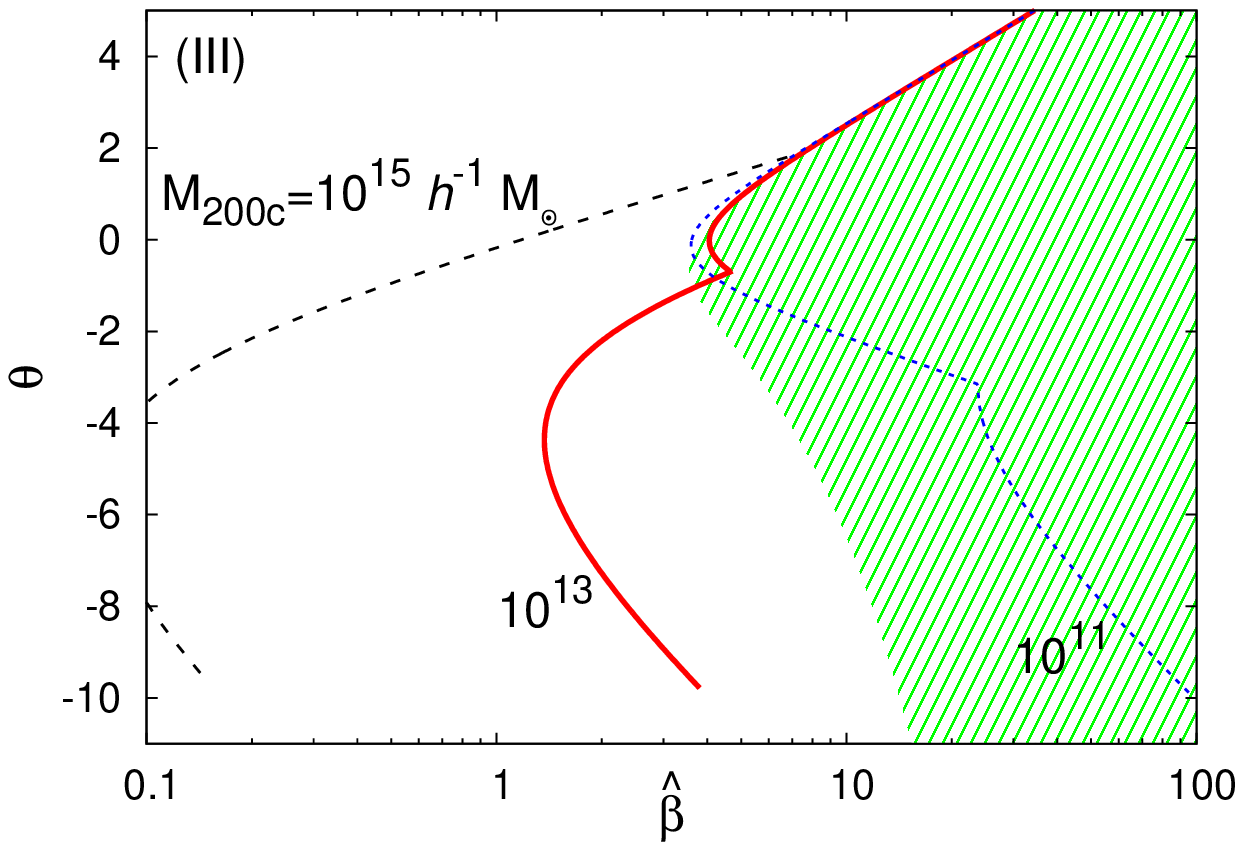}}
\end{center}
\caption{
Radial trajectory $(\hat\beta_r,\theta_r)$ over the thermodynamic phase diagram
inside halos of mass $M_{\rm 200c}= 10^{15}$, $10^{13}$ and
$10^{11} h^{-1} M_{\odot}$, at $z=0$.
We show our results for the models (II) (upper panel) and (III) (lower panel).
}
\label{fig:diagram_Halo_II_III}
\end{figure}

Within a radius $r$ inside the halo the averaged reduced density is
\beq
\theta_r = \ln \left[ \frac{\alpha \rho(<r)}{{\cal M}^4} \right]
= \ln\left[ \frac{\alpha 3 M(<r)}{4\pi r^3 {\cal M}^4} \right] .
\label{theta-halo-def}
\eeq
We write the reduced inverse temperature as
\beq
\hat\beta_r = \frac{\alpha c^2}{{\rm Max}(c_s^2,v_{\rm N}^2)} ,
\label{beta-halo-def}
\eeq
where $v_{\rm N}$ is the circular velocity (\ref{FN-v2}) associated with the Newtonian gravity
while $c_s$ is the velocity scale (\ref{cs2-FA-def}) associated with the fifth force.
As noticed in Eq.(\ref{eta-cs-vN}), the maximum ${\rm Max}(c_s^2,v_{\rm N}^2)$
shifts from one velocity scale to the other when the associated force becomes dominant.
Here we choose the non-analytic interpolation ${\rm Max}(c_s^2,v_{\rm N}^2)$
instead of the smooth interpolation $c_s^2+v_{\rm N}^2$ that we used
in Eq.(\ref{ceff-def}) for the cosmological analysis for illustrative convenience.
Indeed, the discontinuous changes of slope in Fig.~\ref{fig:diagram_Halo_II_III}
below will show at once the location of the transition $|\eta|=1$ between the
fifth-force and Newtonian gravity regimes.

If the density grows at small radii as a power law, $\rho \propto r^{-\gamma}$,
we have seen in Eqs.(\ref{eta-small-r-I-II}) and (\ref{eta-small-r-III})
that the fifth-force to gravity ratio $\eta$ behaves as
$\eta_{\rm (II)} \sim r^{3\gamma-2}$ for the model (II) and
$\eta_{\rm (III)} \sim r^{3\gamma/2-2}$ for the model (III). This led to the bounds
(\ref{gamma-bound-I-II}) and (\ref{gamma-bound-III}) over $\gamma$ for the fifth force
to become negligible at the center.
From Eqs.(\ref{theta-halo-def}) and (\ref{beta-halo-def}) we obtain in this power-law
regime $\theta_r \sim - \gamma \ln r$ and
\beq
v_{\rm N}^2 \sim r^{2-\gamma} ,  \;\;\; c_{s{\rm (II)}}^2 \sim r^{2\gamma} , \;\;\;
c_{s{\rm (III)}}^2 \sim r^{\gamma/2} ,
\label{cs-vN-II-III}
\eeq
where we used Eqs.(\ref{lnA-rho-model-II}) and (\ref{lnA-rho-model-III-b}).
In the Newtonian gravity regime this gives for both models
\beq
| \eta | < 1 : \;\;\; \theta_r \sim \frac{\gamma}{2-\gamma} \ln\hat\beta_r ,
\label{theta-r-beta-r-eta-small}
\eeq
and in the fifth-force regime
\beq
| \eta | > 1 : \;\;\; \theta_{r{\rm (II)}} \sim \frac{1}{2} \ln\hat\beta_r , \;\;\;
\theta_{r{\rm (III)}} \sim 2 \ln\hat\beta_r .
\label{theta-r-beta-r-eta-large}
\eeq
For $\gamma>2$ we are in the Newtonian regime for both models and
$v^2_{\rm N} \rightarrow \infty$, $\hat\beta_r \rightarrow 0$, so that
we are in the homogeneous phase of the thermodynamic phase diagram
as $\hat\beta_r < \hat\beta_c$.
Let us now consider the case $\gamma<2$.
For model (II), Newtonian gravity dominates at small radii if $\gamma>2/3$ from
Eq.(\ref{gamma-bound-I-II}). In this regime Eq.(\ref{theta-r-beta-r-eta-small})
yields $\theta_r > (1/2) \ln\hat\beta_r$, so that we are above the upper boundary
$\theta_+$ of the inhomogeneous phase obtained in Eq.(\ref{mus-large-beta}).
For shallower density profiles, $\gamma<2/3$, the fifth force dominates
and we obtain $\theta_r \sim \theta_+ \sim (1/2) \ln\hat\beta_r$.
The model (III) shows a similar behavior. Newtonian gravity now dominates for
$\gamma>4/3$ from Eq.(\ref{gamma-bound-III}), this gives $\theta_r > 2 \ln\hat\beta_r$
hence $\theta_r>\theta_+$. In the fifth-force regime, $\gamma<4/3$, we obtain
$\theta_r \sim \theta_+ \sim 2\ln\hat\beta_r$.
Therefore, in both models in the Newtonian gravity regime we are far in the
homogeneous phase of the thermodynamic diagram whereas in the fifth-force
regime we are along the upper boundary of the inhomogeneous phase domain.
This means that the dimensional analysis of section~\ref{sec:screening-clusters}
is valid as the fifth force does not push towards a fragmentation of the system
down to very small scales.

The previous results were obtained in the small-radius limit $r\rightarrow 0$.
In Fig.~\ref{fig:diagram_Halo_II_III} we show the full radial trajectories
$(\hat\beta_r,\theta_r)$ over the thermodynamic phase diagram, from $R_{200\rm c}$
inward, for the NFW halos that were displayed in Fig.~\ref{fig_eta_z0}
at $z=0$.
As we move inside the halo, towards smaller radii $r$, the density $\theta_r$ grows.
The turn-around of $\hat\beta_r$ at $\theta_r \simeq -4$ corresponds to the NFW
radius $r_s$ where the local slope of the density goes through $\gamma=2$
and the circular velocity is maximum.
At smaller radii, $r \ll r_s$, the NFW profile goes to $\rho \propto r^{-1}$, hence
$\gamma = 1$.
For model (II) (upper panel) this corresponds to the Newtonian regime and we
move farther away above the inhomogeneous phase.
However, for low-mass halos, $M \lesssim 10^{13} h^{-1} M_{\odot}$,
at intermediate radii we are in the fifth force regime, as seen in Fig.~\ref{fig_eta_z0},
and the trajectory converges towards the upper boundary of the inhomogeneous
phase. These behaviors agree with the discussion above and
Eqs.(\ref{theta-r-beta-r-eta-small})-(\ref{theta-r-beta-r-eta-large}).
The transitions between the Newtonian-gravity and fifth-force regimes correspond
to the discontinuous changes of slope in the figure.
For $M = 10^{15} h^{-1} M_{\odot}$ there is no intermediate fifth-force regime,
for $M = 10^{13} h^{-1} M_{\odot}$ it corresponds to $-1 \lesssim \theta_r \lesssim 2$,
while for $M = 10^{11} h^{-1} M_{\odot}$ the low-radius boundary of the intermediate
fifth-force regime is beyond the scales shown in the figure.
For model (III) (lower panel) the small-radius density slope $\gamma=1$
is in the fifth-force domain and we can see that for the three masses the trajectory
converges to the upper boundary $\theta_+$ of the inhomogeneous domain,
in agreement with Eqs.(\ref{theta-r-beta-r-eta-small})-(\ref{theta-r-beta-r-eta-large}).

The results found in Fig.~\ref{fig:diagram_Halo_II_III} suggest that for large-mass
halos, $M \gtrsim10^{13} h^{-1} M_{\odot}$ at $z=0$, the dimensional analysis
of section~\ref{sec:screening-clusters} is valid. In the case of model (III)
this would lead to an increasingly dominant fifth force at small radii and
characteristic velocities that are higher than the Newtonian circular velocity.
This is likely to rule out this scenario.
For low-mass halos, $M \lesssim10^{11} h^{-1} M_{\odot}$ at $z=0$,
we find that a significant part of the halo is within the inhomogeneous thermodynamic
phase for both models II and III.
This may leave some signature as a possible fragmentation of the system
on these intermediate scales into higher-density structures.
This process would next lead to a screening of the fifth force,
as discussed for the Solar System and the Earth in sections~\ref{sec:solar-system}
and \ref{sec:labo}, because of the ultra-local character of the fifth force.
Indeed, because it is set by the local density gradients, the fragmentation
of the system leads to a disappearance of large-scale collective effects and the
fifth force behaves like a surface tension at the boundaries of different domains.
Such a process may also happen in the case of massive halo at earlier stages
of their formation, which could effectively screen the fifth force in the case of model (III)
where a simple static analysis leads to a dominant fifth force at small radii.
However, a more precise analysis to follow such evolutionary tracks and check
the final outcomes of the system requires numerical studies that are beyond the scope
of this paper.

\section{Dependence on the parameter $\alpha$}
\label{sec:alpha}

\begin{figure*}
\begin{center}
\epsfxsize=8 cm \epsfysize=6. cm {\epsfbox{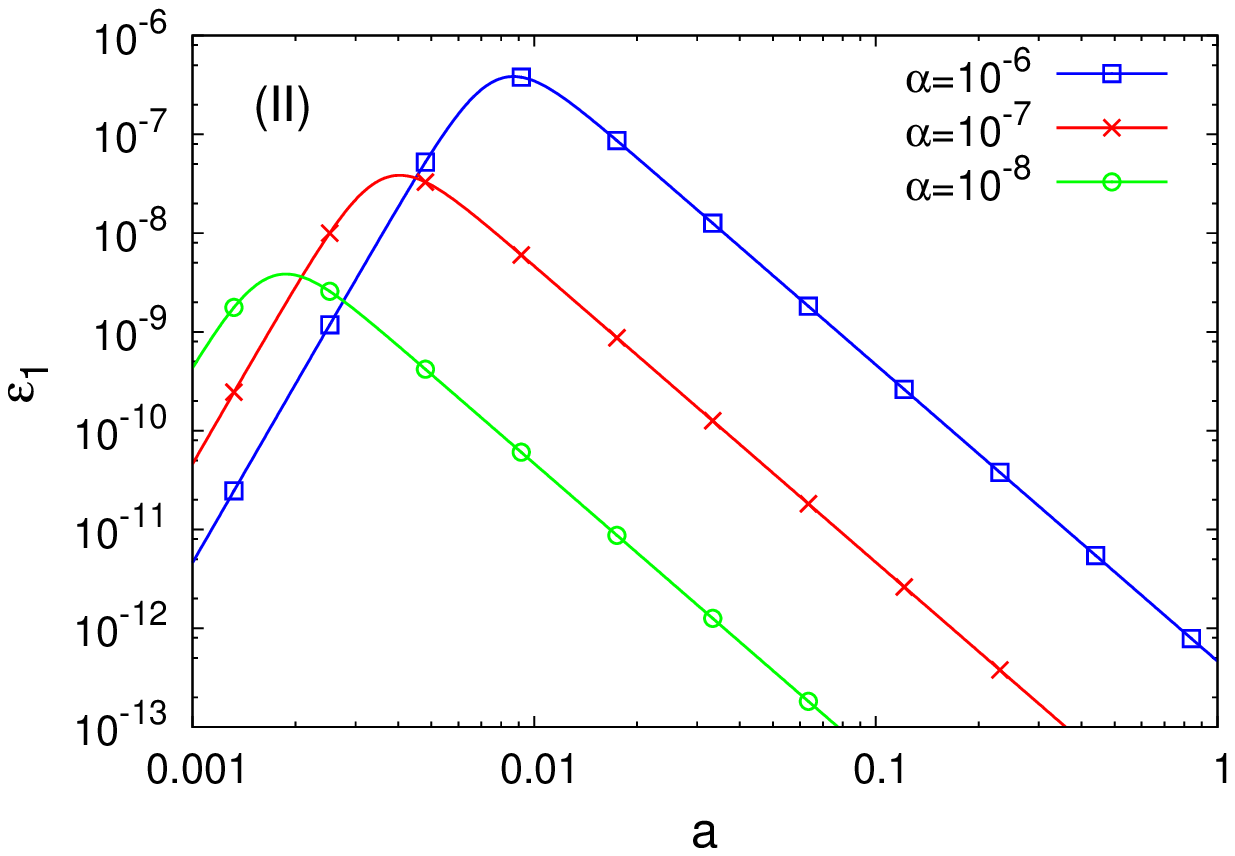}}
\epsfxsize=8 cm \epsfysize=6. cm {\epsfbox{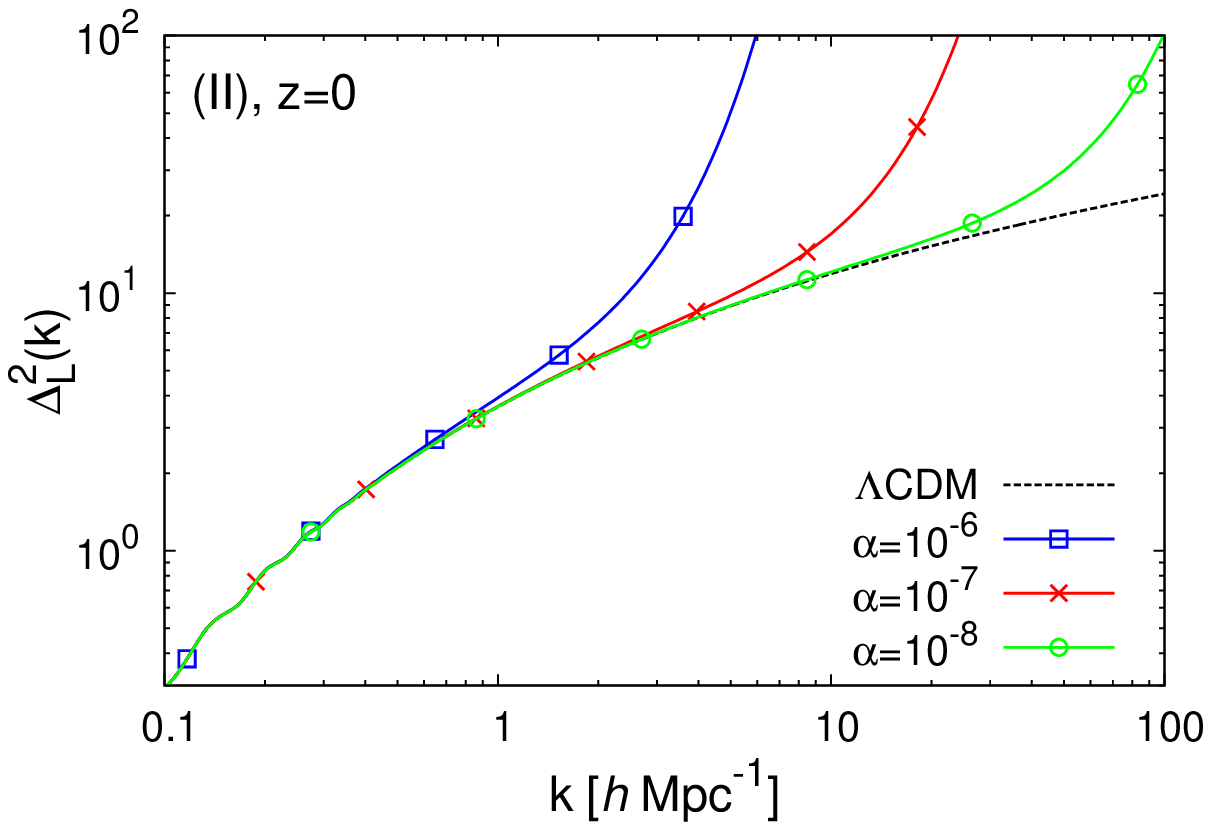}}\\
\epsfxsize=8 cm \epsfysize=6.5 cm {\epsfbox{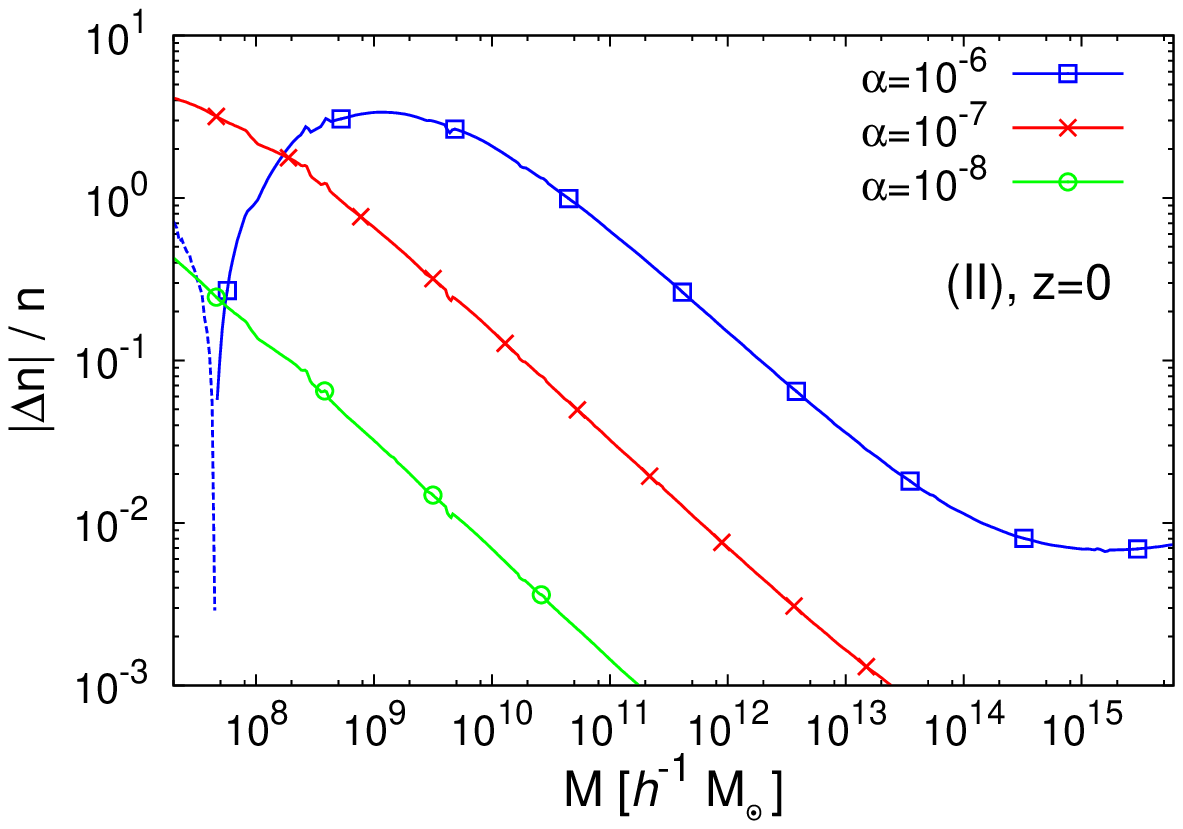}}
\epsfxsize=9 cm \epsfysize=6.5 cm {\epsfbox{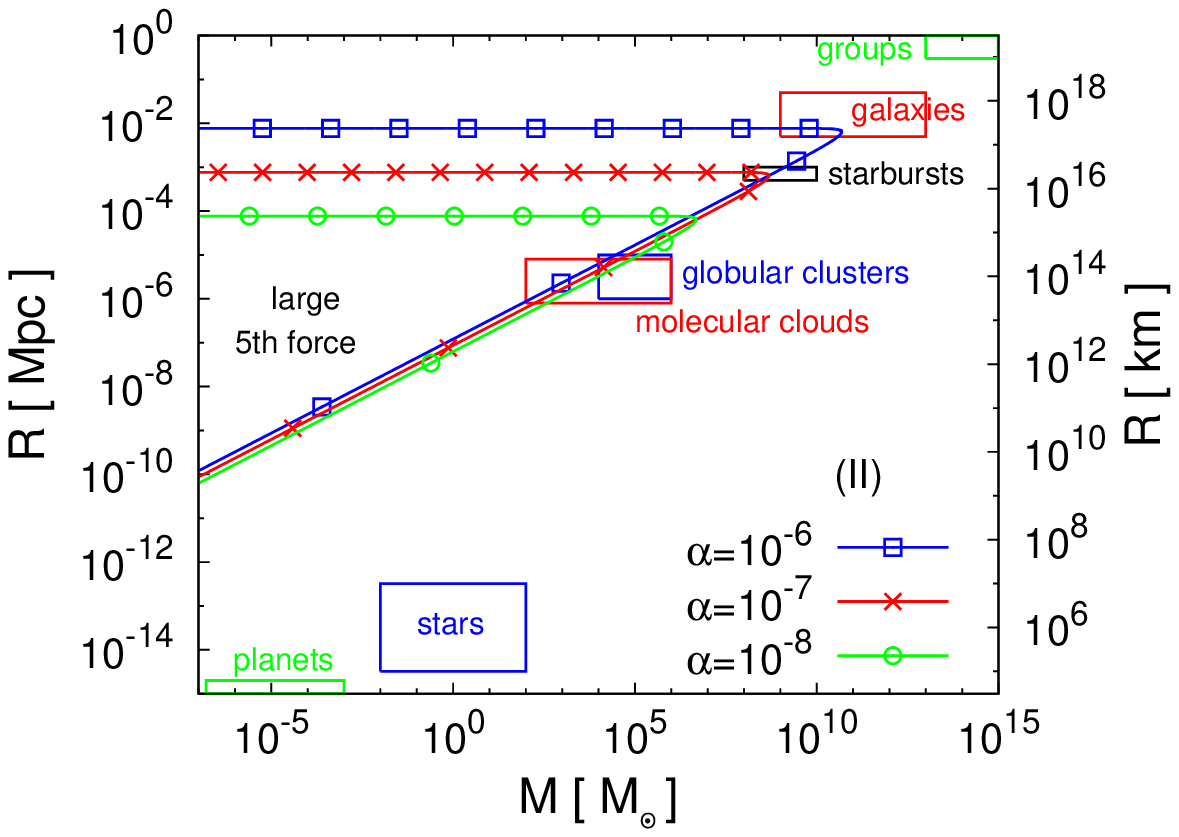}}
\end{center}
\caption{
Dependence on the parameter $\alpha$ of the deviations from the $\Lambda$-CDM
predictions. We plot models of the type (II) with $\alpha=10^{-6}$, $10^{-7}$ and
$10^{-8}$.
{\it Upper left panel:} $\epsilon_1(a)$ as a function of the scale factor, as in
Fig.~\ref{fig_Aa_1}.
{\it Upper right panel:} linear logarithmic power spectrum $\Delta_L^2(k)$ at redshift
$z=0$, as in Fig.~\ref{fig_Dp_II}.
{\it Lower left panel:} absolute value $\Delta n|/n$ of the relative deviation of the halo
mass function from the $\Lambda$-CDM result, as in Fig.~\ref{fig_nM_II}.
{\it Lower right panel:} domain in the mass-radius plane where the fifth force is greater than
Newtonian gravity, as in Fig.~\ref{fig_eta_R_M}.
}
\label{fig_alpha}
\end{figure*}

It is interesting to investigate how the results obtained in the previous sections change
when we vary the parameter $\alpha$ that measures the amplitude of the modification
to General Relativity.
For illustration, we consider the model (II) defined by Eq.(\ref{model-II-def}),
keeping $\tilde\chi_*=-2$.
We show our results in Fig.~\ref{fig_alpha}, where we compare the case $\alpha=10^{-6}$
considered in the previous sections with the two cases $\alpha=10^{-7}$ and
$\alpha=10^{-8}$.

In agreement with the discussion in Sec.~\ref{sec:evolution-back-fields},
the factor $\epsilon_1$ shown in the upper left panel, which measures the amplitude
of the modification of gravity at linear order over field fluctuations, decreases linearly
with $\alpha$ while its peak is pushed towards higher redshift as
$z_{\alpha} \sim \alpha^{-1/3}$.

The smaller value of $\epsilon_1$ implies that the effect of the scalar field on gravitational
clustering is pushed to smaller scales, as $k_{\alpha} \propto \epsilon_1^{-1/2}$
from Eq.(\ref{k-alpha-def}), and hence $k_{\alpha} \propto \alpha^{-1/2}$.
We can check in the upper right panel that the deviation from the $\Lambda$-CDM linear
power spectrum is indeed pushed towards smaller scales as $\alpha$ decreases.
This also means that the deviation of the halo mass function is repelled to smaller
masses, as we can see in the lower left panel.
At a given mass, the relative deviation $\Delta n/n$ decreases with $\alpha$, but
one can still reach deviations of order unity by going to small enough masses.

As expected, the area in the $(M,R)$ plane where the fifth force is greater than
Newtonian gravity shrinks as $\alpha$ decreases, as we can see in the lower right panel.
The upper branch at constant radius is pushed towards smaller scales, as
$R_{\alpha} \propto \alpha$ from Eq.(\ref{Reta-low-density}).
The lower branch keeps the same slope and goes down at the very slow rate
$R \propto \alpha^{1/7}$ at fixed mass [as can be seen from Eq.(\ref{R-M-lower-I-II})
and the expressions of $R_{\alpha}$ and $M_{\alpha}$].
Because the lower branch is almost insensitive to $\alpha$, the various galactic structures
shown in the figure remain along the border of the fifth-force dominated region.
They only progressively leave this region, starting from the largest and most massive
objects, as the upper branch is pushed downward.
Therefore, globular clusters and molecular clouds remain sensitive to the modification
of gravity until $\alpha$ becomes smaller than about $10^{-10}$.

\section{Comparison with scalar-field models with a kinetic term and tomographic reconstruction}
\label{sec:Comparison}

The ultra-local models introduced in this paper can be easily compared to models of
modified gravity of the chameleon type. These models are defined by two functions,
the potential $V(\phi)$ and the conformal coupling $A(\phi)$ of a scalar field $\phi$.
They can be reconstructed from two functions $m^2(\rho)$ and $\beta(\rho)$,
which are respectively the mass squared and the coupling to matter in an environment of
density $\rho$, using the tomographic mapping \cite{Brax2012a,BraxPV2013}
\be
\frac{\phi(\rho)}{M_{\rm Pl}} = \frac{\phi_{\rm BBN}}{M_{\rm Pl}}
-\int_{\rho_{\rm BBN}}^\rho d\rho \, \frac{\beta(\rho)}{M^2_{\rm Pl} m^2(\rho)} ,
\label{phi-tomography}
\ee
and we have
\be
\ln A (\rho)= -\int_{\rho_{\rm BBN}}^\rho d\rho \,
\frac{\beta^2(\rho)}{M^2_{\rm Pl}m^2(\rho)}
\label{lna}
\ee
where we assumed that $A_{\rm BBN}(\rho)$ is close to one, and
\be
V(\rho)=V_{\rm BBN}+\int_{\rho_{\rm BBN}}^\rho d\rho \,
\frac{\beta^2(\rho)\rho }{M^2_{\rm Pl}m^2(\rho)} .
\label{VV}
\ee
This parametric mapping  defines all the models of the chameleon-type such as $f(R)$
models, chameleons, dilatons and symmetrons.

In the case of the ultra-local models, as the rescaling $A(\chi)$ between the Einstein and
the Jordan frames is constrained to vary cosmologically by less than $10^{-6}$, the
dynamics of the models can be equally understood in the Einstein frame.
Then, we can write the ultra-local model in the same form as
Eqs.(\ref{phi-tomography})-(\ref{VV}), where $\tilde\chi$ plays the role of the
reduced scalar field $\phi/M_{\rm Pl}$.
The effective potential reads
\be
V_{\rm eff}(\tilde\chi)= -\cM^4 \tilde\chi + \rho \, \ln A (\tilde\chi) ,
\ee
where $\rho$ is the conserved matter density, and the equation of motion (\ref{KG-A=1})
corresponds to the minimum of the effective potential,
\be
\left. \frac{\partial V_{\rm eff}(\tilde\chi)}{\partial \tilde\chi} \right\vert_{\tilde\chi_{\rm min} (\rho)}=0 .
\ee
Thus, we recover the behavior of models of the chameleon type, where the field is stuck
at the minimum of the effective potential since Big Bang Nucleosynthesis.
At this minimum one can define the effective coupling to matter
\be
\beta_1(\rho) \equiv \left. \frac{d\ln A}{d\tilde\chi} \right\vert_{\tilde\chi_{\rm min}(\rho)}
\label{beta1-tomo-def}
\ee
and the effective mass
\be
m^2(\rho) \equiv  \left. \frac{1}{M^2_{\rm Pl}}
\frac{\partial^2 V_{\rm eff}}{\partial\tilde\chi^2} \right\vert_{\tilde\chi_{\rm min}(\rho)}
= \frac{\rho \beta_2(\rho)}{M_{\rm Pl}^2} .
\label{m2-tomo-def}
\ee
From $\beta_2 \equiv d\beta_1/d\tilde\chi=-\beta_1 d\ln\rho/d\tilde\chi$, where we used the
equation of motion $\beta_1={\cal M}^4/\rho$, we obtain $d\tilde\chi=-(\beta_1/\beta_2)d\ln\rho$.
With Eq.(\ref{m2-tomo-def}) this yields
\be
\tilde\chi (\rho)= \tilde\chi_{\rm BBN} - \int_{\rho_{\rm BBN}}^\rho d\rho \,
\frac{\beta_1(\rho)}{M^2_{\rm Pl} m^2(\rho)} .
\label{chi-tomo}
\ee
Thus, we recover the same tomographic mapping as for chameleon-type models,
where $\tilde\chi$ plays the role of the rescaled field $\phi/M_{\rm Pl}$ and
$\beta_1$ that of $\beta$ in Eq.(\ref{phi-tomography}).
We can also write $d\ln A/d\rho = \beta_1 d\tilde\chi/d\rho = - \beta_1^2/\beta_2\rho$,
which yields
\be
\ln A(\rho) =  - \int_{\rho_{\rm BBN}}^\rho d\rho \,
\frac{\beta_1^2(\rho)}{M^2_{\rm Pl} m^2 (\rho)} ,
\label{lnA-tomo}
\eeq
which also coincides with Eq.(\ref{lna}).
Finally, writing $V=-{\cal M}^4\tilde\chi$ and using ${\cal M}^4=\rho\beta_1$, we recover
Eq.(\ref{VV}). This completes the equivalence, at the background level, of the ultra-local models
with a subclass of the chameleon-type models. Thus, the ultra-local models are defined by
the specific choice
\be
\mbox{ultra-local} \sim \mbox{chameleon with} \;\; \beta(\rho) = \frac{\cM^4}{\rho} ,
\label{beta1-rho-def}
\ee
while the squared-mass $m^2(\rho)$, or equivalently the coupling $\beta_2(\rho)$, remains
a free function.
We recover the fact that all ultra-local models can be defined by a single function of the matter
density, as was already seen in section~\ref{sec:models}.

At the linear perturbation level, the chameleon-type models modify the growth of structure
as Newton's constant becomes space and time dependent
\cite{Brax2012a,BraxPV2012}
\be
{\cal G}_{\rm eff} = {\cal G}_{\rm N} \, (1+ \epsilon(k,t)) ,
\ee
with
\be
\epsilon_{\rm cham}(k,t) = \frac{2 \beta^2 (a)}{1+ \frac{a^2 m^2 (a)}{k^2}} .
\label{epsilon-chameleon-def}
\ee
On large scales beyond the Compton radius (but still below the horizon) we have
\be
H \ll \frac{k}{a} \ll m : \;\; \epsilon_{\rm cham}(k,t) = \frac{2 \beta^2(a) k^2}{a^2 m^2(a)} .
\label{epsilon-low-k-chameleon}
\ee
On the other hand, from Eq.(\ref{eps-def}) we find that on sub-horizon scales the ultra-local
models also exhibit a modified Newton constant with
\beq
H \ll \frac{k}{a} : \;\; \epsilon_{\rm ultra}(k,t) =  \epsilon_1 (a)
\frac{2k^2}{3\Omega_{\rm m} a^2 H^2} 
= \frac{2 \beta_1^2 k^2}{a^2 m^2} ,
\label{epsilon-ultra-local-subhorizon}
\ee
where in the second equality we used the definition (\ref{eps1-def}), 
$\epsilon_1 = \beta_1^2/\beta_2$, and the identification (\ref{m2-tomo-def}),
$\beta_2 = M_{\rm Pl}^2 m^2/\bar\rho$.
Thus, we recover the result (\ref{epsilon-low-k-chameleon}) of the chameleon models,
over the intermediate scales $H \ll k/a \ll m$.

So we find that the ultra-local models can be seen as chameleon-type models
when their mass terms are much larger than the kinetic energy outside the Compton
wavelength of the scalar field.
We will give an explicit model with such a large mass in a companion paper where we
discuss the supersymmetric chameleons.
However, we should note that the correspondence found in 
Eq.(\ref{epsilon-ultra-local-subhorizon}) is not complete as it breaks down
inside the Compton wave-length. From Eq.(\ref{m2-tomo-def})
and the estimate (\ref{beta2-a-esp1-a}) we obtain at low redshift
\beq
z \lesssim 1 : \;\; m^2 \sim \frac{\bar\rho}{\alpha^2 M^2_{\rm Pl}} 
\sim \frac{H^2}{\alpha^2}  .
\eeq
This means that the correspondence with the chameleon models, in the low-$k$ regime
(\ref{epsilon-low-k-chameleon}), applies up to $m \sim H/\alpha$.
Since $\alpha \ll 1$ this means that it holds down to scales that are much below the
horizon, $1/m \sim 3 h^{-1} \, \rm{kpc}$ for $\alpha \sim 10^{-6}$.
However, for the ultra-local models that we consider in this paper the rise with $k$
of $\epsilon$ in Eq.(\ref{eps-def}) goes on to much higher $k$, until we reach the cutoff of the
theory. Therefore, ultra-local models go beyond chameleon models with a relatively
large squared-mass $m^2$; taking the kinetic terms in the Lagrangian or the unit factor
in Eq.(\ref{epsilon-chameleon-def}) to zero is not exactly the same as taking
$m$ large in Eq.(\ref{epsilon-chameleon-def}).
This is also clear from the phenomenology presented in this paper, which shows many
different qualitative features from usual chameleon models at short enough scale inside the
Compton wave-length of the chameleon scalar field.

\section{Conclusions}
\label{sec:Conclusions}

We have introduced in this paper ultra-local models, a class of modified gravity theories
where we add a scalar field  with a negligible kinetic term to the Einstein-Hilbert action
and a conformal coupling to matter.
This gives rise to a new screening mechanism, which is not mainly due to the non-linearity
of the scalar field potential or coupling function but to the absence of kinetic term.
Indeed, it is this feature that leads to the ultra-local character of the model, where
the fifth force potential only depends on the local density. This removes any fifth force
between isolated objects in vacuum.
Another property of this class of models is that the scalar field potential and coupling
function are degenerate, so that predictions only depend on a single free function.
We have then presented a cosmological analysis
of these scenarios.

We have shown the ultra-local models recover the  $\Lambda$-CDM expansion history
at a level of accuracy which is set by a free parameter $\alpha$ of the theory but is
always smaller than $\alpha \lesssim 10^{-6}$.
Moreover we have demonstrated that, for some of the models considered
in this paper, the results obtained for the expansion history are stable if we add a small initial
kinetic term to the Lagrangian. We have also checked that the non-linearities of the models
do not lead to strong back-reaction effects on the cosmological background.
In addition to the dark energy density today, $\bar\rho_{\rm de0}$, these models
single out a characteristic density $\rho_{\alpha} \sim \bar\rho_{\rm de0}/\alpha$
and redshift $z_{\alpha} \sim \alpha^{-1/3} \gtrsim 100$ where the fifth force is the greatest.

At the linear level of cosmological perturbations, the presence of the ultra-local scalar field
has a major impact on the growth rate of structures at small scales, enhancing or diminishing
it, even though the last case corresponds to a model that is found to be unstable if we add
a small initial kinetic term to its Lagrangian.

We have studied the spherical collapse in this framework showing that, due to the
modification of the growth rate at small scales, the halo mass function is substantially
modified in the low mass tail. However, it must be taken into account that we have used
a Press-Schechter-like approach without considering qualitative modifications to the spherical
collapse, which we may be taken into account in future studies.

We have  shown that due to the ultra-local behavior of the theory, very dense environments
such as the Solar System are completely screened but on the other hand the importance of
the fifth force in astrophysical systems with a continuous distribution of matter, such as
galaxies or clusters of galaxies, may or may not diminish going towards the center of the
objects depending on the shape of the coupling function.
This could provide very stringent constraints on the latter, which may require a better
understanding of the possible modifications of the halo profile for this theory and/or the use of
numerical simulations.

To study the non-linear and inhomogeneous regime of the fifth force, which requires
to go beyond perturbation theory or the spherical collapse, we have presented
a thermodynamic analysis. This leads to a phase diagram with a first-order phase
transition. At at low temperature (i.e. low initial kinetic energy) and intermediate density,
the system becomes inhomogeneous and splits over domains of either larger or smaller
density. We have checked that this inhomogeneous transition does not invalidate
our cosmological analysis.
On the other hand, for small masses $M\lesssim 10^{11} M_\odot$, the ultra-local force
may alter significantly the landscape of inhomogeneities inside the object.
The study of this effect requires numerical methods beyond the present work.

Then, we have briefly considered the dependence of our results on the main free parameter
$\alpha$ of these models. As it decreases we slowly converge to the $\Lambda$-CDM
scenario. However, from $\alpha \sim 10^{-6}$ down to $\alpha \sim 10^{-8}$
we expect some signatures on galactic or subgalactic scales.
Indeed, it is a peculiar feature of these modified gravity scenarios that the fifth force
appears to be most significant on galactic scales, $1 \, {\rm pc} -- 10 \, {\rm kpc}$,
whereas clusters of galaxies and astrophysical compact objects (stars or planets)
show no significant fifth force or are screened.

In the last section we have compared the ultra-local models to chameleon-type models
with a  mass term that is much greater than the potential one. Both scenarios are similar
outside the Compton wave-length of the scalar but differ otherwise.
We shall discuss a supersymmetric implementation of ultra-local models in a companion
paper \cite{Brax:2012mq,Brax:2013yja}.

\begin{acknowledgments}

This work is supported in part by the French Agence Nationale de la Recherche
under Grant ANR-12-BS05-0002. This project has received funding from the European Union’s Horizon 2020 research and innovation programme under the Marie Skłodowska-Curie grant agreement No 690575.

\end{acknowledgments}

\bibliography{ref1_submit}   

\end{document}